\documentclass[aps,prd,twocolumn,groupedaddress,nofootinbib,amssymb]{revtex4-1}

\usepackage[english]{babel} 
\usepackage[utf8]{inputenc}
\usepackage[T1]{fontenc}
\usepackage{lmodern}
\usepackage{amsmath}
\usepackage{amsfonts}
\usepackage{amssymb}
\usepackage{slashed}
\usepackage{bbm}
\usepackage{graphicx}
\usepackage{color}
\usepackage{multirow}
\usepackage{hyperref}
\usepackage{enumitem}

\newcommand{\I}{\mathrm{i}}
\newcommand{\E}{\mathrm{e}}
\newcommand{\tr}{\mathrm{tr}}

\newcommand{\diag}{\mathrm{diag}}

\newcommand{\pd}{\partial}
\newcommand{\pdm}{\pd_{\mu}}

\newcommand{\im}{\mathrm{Im}}
\newcommand{\re}{\mathrm{Re}}

\newcommand{\m}{{\mu}}

\newcommand{\SU}{\mathrm{SU}}
\newcommand{\U}{\mathrm{U}}

\newcommand{\mh}{m_{\mathrm{h}}}

\newcommand{\mA}{m_{\mathrm{A}}}
\newcommand{\MA}{M_{\mathrm{A}}}
\newcommand{\bma}{\begin{pmatrix}}
\newcommand{\ema}{\end{pmatrix}}

\begin{document}
\title{The spectrum of an SU(3) gauge theory with a fundamental Higgs field}

\author{Axel~Maas}
\email{axel.maas@uni-graz.at}
\affiliation{Institute of Physics, NAWI Graz, University of Graz, Universit\"atsplatz 5, 8010 Graz, Austria}

\author{Pascal~T\"orek}
\email{pascal.toerek@uni-graz.at}
\affiliation{Institute of Physics, NAWI Graz, University of Graz, Universit\"atsplatz 5, 8010 Graz, Austria}

\begin{abstract}
In gauge theories, the physical, experimentally observable spectrum consists only of gauge-invariant states. This spectrum can be different from the elementary spectrum even at weak coupling and in the presence of the Brout-Englert-Higgs effect.

We demonstrate this for an $\SU(3)$ gauge theory with a single fundamental Higgs, a toy theory for grand-unified theories. The manifestly gauge-invariant approach of lattice gauge theory is used to determine the spectrum in four different channels. It is found to be qualitatively different from the elementary one, and especially from the one predicted by standard perturbation theory.

The result can be understood in terms of the Fr\"ohlich-Morchio-Strocchi mechanism. In fact, we find that analytic methods based on this mechanism, a gauge-invariant extension of perturbation theory, correctly determines the spectrum, and gives already at leading order a reasonably good quantitative description. Together with previous results this supports that this approach is the analytic method of choice for theories with a Brout-Englert-Higgs effect.
\end{abstract}

\maketitle

\section{Introduction}\label{sec:intro}
Non-Abelian gauge theories in combination with scalars are compelling theories 
to study. Of special interest is the case of an $\SU(2)$ gauge group with a single 
scalar field in the fundamental representation, since this is the gauge-Higgs sector of the 
standard model.

The physical spectrum of these kind of theories needs to be gauge invariant.  
This, almost tautological, insight has a realization which is far from obvious in the 
standard model. In QCD confinement 
takes care of this issue \cite{Weinberg:1996kr}, while for
QED dressings by Dirac phases create observable states \cite{Haag:1992hx}. 
In the weak sector the same necessity 
applies \cite{'tHooft:1979bj,Osterwalder:1977pc,Banks:1979fi,Frohlich:1980gj,Frohlich:1981yi}. 
At first sight, this seems surprising, as a perturbative description using the BRST-invariant, 
but still gauge-dependent, elementary states of the Lagrangian, the $W$, the $Z$, the Higgs, 
and the fermion fields, describes experimental results remarkably well \cite{pdg}. 

The, subtle, reason for this is the mechanism described by Fr\"ohlich, Morchio, and Strocchi 
(FMS) \cite{Frohlich:1980gj,Frohlich:1981yi}: Under certain conditions, realized in the 
standard model, the properties of the physical states can be mapped to the gauge-dependent 
states which appear in the Lagrangian. 
This FMS mechanism has been confirmed in lattice calculations for the scalar 
sector \cite{Maas:2012tj,Maas:2013aia}. An extensive review on this (and more concerning 
field theories with scalars) can be found in \cite{Maas:2017wzi}. 

However, the conditions mentioned are quite specific, and the standard model is special to 
fulfill them. Especially, the weak gauge group is the same as the global custodial symmetry group. In general beyond-standard-model theories this is not the case, and they therefore potentially not meet these requirements \cite{Maas:2015gma}. Then, discrepancies between the actual physical spectrum and the elementary one, and thus the one described 
by perturbation theory, may arise. Investigations of explicit examples have found both types of 
behaviors \cite{Maas:2016qpu,Maas:2017xzh,Maas:2017wzi}. In particular, this can imply that the 
low-lying observable spectrum is different from the standard model, even if a model 
features perturbatively the $W$ and $Z$ bosons and a light Higgs. Such theories would 
therefore not be suitable extensions of the standard model.

The FMS mechanism can be used to create the analytic tool of gauge-invariant perturbation theory (GIPT) \cite{Maas:2017wzi}. This tool has been applied to $\SU(N)$ gauge theories 
with scalars in different representations in \cite{Maas:2017xzh}. Such theories are of particular interest as their structures is typical for so-called grand-unified theories (GUTs), which are one of the candidates for beyond-standard-model theories. This lead to analytical predictions of the spectrum, which generically disagree with the elementary one.

The primary aim of this work is to check these analytical predictions. This requires a manifestly gauge-invariant approach which is capable of treating non-perturbative physics. We choose this method to be the lattice. The required resources forced us to concentrate on a particular case, an $\SU(3)$ gauge theory with a fundamental scalar. The technicalities and details of the lattice simulations can be found in Section \ref{sec:tech}. This includes how the spectroscopy of gauge-invariant operators 
is performed, and how gauge-variant quantities, like the propagators of elementary fields and the running gauge coupling, are obtained. This section can be skipped entirely if only the results are of interest.

In Section \ref{sec:physics} we concentrate on the properties of the theory. We present the phase diagram of the theory, and show in which 
regions Brout-Englert-Higgs (BEH) physics or QCD-like physics takes place. We then compute the 
spectrum of gauge-invariant states as well as the spectrum of elementary fields 
in the Higgs-like region of the phase diagram. 

Finally, in Section \ref{sec:testfms} the primary aim of this work will be achieved, the test of FMS mechanism and GIPT. To this end, we first show that standard perturbation theory is not able to describe the physics of the theory even qualitatively. Then, to be self-contained, we first rehearse the predictions of the spectrum \cite{Maas:2017xzh}. Finally, we compare the results of the lattice simulations to the predictions of GIPT. We find that, already at leading order, all channels are qualitatively correctly described, and even the quantitative agreement is good in all channels where the lattice results are reasonably reliable. This strongly supports GIPT as the analytic tool for this type of theories. This agrees with all other available results, especially in the standard model \cite{Maas:2017wzi}.

Preliminary and related results of this work can be found in \cite{Maas:2017xzh,Maas:2017pcw,Torek:2016ede,Maas:2016ngo}.

\section{Technicalities}\label{sec:tech}

We consider an $\SU(3)$ gauge theory with a single scalar in the fundamental representation of the gauge group. The theory has therefore a global U(1) custodial symmetry \cite{Maas:2017xzh}.

\subsection{Preliminaries}\label{ssec:pre}
The action of the theory on a $4$-dimensional, Euclidean, isotropic, hypercubic lattice with 
lattice constant $a$, and volume $V=L^4$, is given by \cite{Montvay:1994cy}
\begin{align}
\begin{split}
 S[U,\phi] = &\sum_x
 \Bigg( \phi(x)^\dagger\phi(x) 
 +\lambda~\big(\phi(x)^\dagger\phi(x)-1\big)^2  \\
 &-\kappa\sum_{\mu=\pm 1}^{\pm 4}\phi(x)^\dagger~U_\mu(x)~\phi(x+\hat{\mu}) \\
&+ \frac{\beta}{3}\sum_{\mu<\nu}\mathrm{Re}~\mathrm{tr}
 \big[\mathbbm{1}-U_{\mu\nu}(x)\big]  
 \Bigg)\;,
\end{split}
\label{eq:lattice_action}
\end{align}
The first sum runs 
over all lattice sites $x=(x_1,x_2,x_3,x_4)$, $x_i=0,1,\dots,L-1$ and $\hat{\mu}$ denotes the unit 
vector in the $\mu$-direction. 
The first term of the action is the Wilson gauge action with the plaquette variable $U_{\mu\nu}(x)$, which 
is a product of four link variables $U_\mu(x)$ forming a closed loop, i.e., 
\begin{align}
U_{\mu\nu}(x) &= U_\mu(x)~U_\nu(x+\hat{\mu})~U_\mu(x+\hat{\nu})^\dagger~U_\nu(x)^\dagger\;,
\label{eq:plaq}
\end{align}
and is essentially the field-strength tensor squared plus $\mathcal{O}(a^2)$-corrections in the naive 
continuum limit $a\to 0$.
The links are related to the gauge fields by $U_\mu(x) = \exp(\I aA^c_\mu(x)T^c)$, with $2T^c$ 
being the Gell-Mann matrices. Thus, the links are elements of the gauge group $\mathrm{SU}(3)$. Note, 
that $U_{-\mu}(x) \equiv U_\mu(x-\hat{\mu})^\dagger$.

Both, the scalar field as well as the links, obey periodic boundary conditions, i.e., 
\begin{align}
\phi(x+\hat{\nu}L) &= \phi(x) \;,\; U_\mu(x) = U_\mu(x+\hat{\nu}L)\;.
\end{align}

Under gauge transformations the scalar field and the gauge links transform as
\begin{align}
\begin{split}
\phi(x) &\to g(x)~\phi(x)\;,\;\\ 
U_\mu(x) &\to g(x)~U_\mu(x)~g(x+\hat{\mu})^\dagger\;,
\end{split}
\label{eq:gaugetrafo}
\end{align}
with $g(x)\in \mathrm{SU}(3)$. The scalar field also transforms under global custodial U(1) transformations $\exp(i\alpha)\in\U(1)$. The Equation \eqref{eq:lattice_action} is invariant under these transformations, making the action gauge and custodial invariant. 

In total three parameters appear in the action \eqref{eq:lattice_action}:
$\beta$ is the inverse gauge coupling, $\lambda$ is the coupling for the self-interaction of 
the scalar fields, 
and $\kappa$ is related to the square of the inverse bare mass. 
Those lattice parameters are related to the continuum ones by
\begin{align}
\beta &= \frac{6}{g^2}\quad,\quad a^2 m_0^2 = \frac{1-2\lambda}{\kappa}-8
\quad,\quad \lambda_\mathrm{c} = \frac{\lambda}{\kappa^2}\;,
\label{eq:params}
\end{align}
where $m_0$ is the bare mass and $\lambda_\mathrm{c}$ the bare self-interaction of the corresponding 
continuum theory. 

\par
In order to generate configurations we use one multi-hit Metropolis sweep for the links, 
where $5$ attempts are made to update one link by 
standard techniques \cite{Gattringer:2010zz} before moving to the next link, 
and one subsequent Metropolis sweep for the scalar field using a Gau{\ss}ian proposal. 
We tuned the widths of the proposals adaptively to achieve a $50\%$ acceptance rate for both updates. 
After every $5$ sweeps through the lattice, a projection step of the gauge links to $\SU(3)$ matrices is 
performed by a standard Gram-Schmidt procedure \cite{Gattringer:2010zz} in order to keep 
rounding errors under control. 

\par
A list of all lattice parameter sets with the corresponding lattice volumes and number of 
configurations is given in Appendix \ref{app:datafittables} in Table \ref{tab:numvals}.

\subsection{Techniques for gauge-invariant quantities}\label{ssec:gi_tech}
For the spectroscopy we use the zero-momentum projected 
interpolators listed in Table \ref{tab:interpol} with distinct $J^{PC}_{\U(1)}$ quantum numbers, where the 
lower index is the quantum number of the 
global custodial group $\U(1)$, which only acts on the scalar field. 
\begin{table*}[tbh!]
\begin{center}
\caption{List of interpolators used for our spectroscopic analysis. Definitions of the objects 
$D_\mu$, $L_\mu^{(1)}$, $L_\mu^{(2)}$, and $L_\mu^{(3)}$ can be found in the main text. We 
perform a zero-momentum projection for all interpolators. 
We use the notation $\pmb x =(x_1,x_2,x_3)$, and $t=x_4$.}
\begin{tabular}{@{}clc@{}}

\hline

\hline
																												
Name\qquad\qquad & Interpolator\quad	& $J^{\mathrm{PC}}_{\U(1)}$  \cr
\hline	
																							
$O^{0^{++}_0}_1(t)$\qquad\qquad & $\frac{1}{L^3}\sum\limits_{\pmb x} \phi(\pmb x,t)^\dagger\phi(\pmb x,t)$ 
\quad& $0^{++}_0$\cr

$O^{0^{++}_0}_2(t)$\qquad\qquad & $O^{0^{++}_0}_1(t)~O^{0^{++}_0}_1(t)$\quad & $0^{++}_0$ \cr

$O^{0^{++}_0}_3(t)$\qquad\qquad & $\frac{1}{L^3}\sum\limits_{\pmb x}\sum\limits_{\mu,\nu=1,\mu<\nu}^{3}
\textnormal{Re tr}\big[U_{\mu\nu}(\pmb x,t)\big]$\quad & $0^{++}_0$ \cr

$O^{1^{--}_0}_{1,\mu}(t)$\qquad\qquad & $\frac{\I}{L^3}~\sum\limits_{\pmb x}\phi(\pmb x,t)^\dagger~
D_{\mu}~\phi(\pmb x,t)$ \quad& $1^{--}_0$ \cr

$O^{1^{--}_0}_{2,\mu}(t)$\qquad\qquad & $O^{0^{++}_0}_1(t)~O^{1^{--}_0}_{1,\mu}(t)$ \quad& $1^{--}_0$ \cr

$O^{1^{--}_0}_{3,\mu}(t)$\qquad\qquad & $\sum\limits_{\nu=1}^{3}\Big(O^{1^{--}_0}_{1,\nu}(t)~
O^{1^{--}_0}_{1,\nu}(t)\Big)~O^{1^{--}_0}_{1,\mu}(t)$\quad & $1^{--}_0$ \cr

$O^{1^{--}_0}_{4,\mu}(t)$\qquad\qquad & $\frac{1}{L^3}\sum\limits_{\pmb x}\text{Im}~L_\mu^{(1)}(\pmb x,t)$ 
\quad& $1^{--}_0$ \cr

$O^{1^{--}_0}_{5,\mu}(t)$\qquad\qquad & $\frac{1}{L^3}\sum\limits_{\pmb x}\text{Im}~L_\mu^{(2)}(\pmb x,t)$ 
\quad& $1^{--}_0$ \cr

$O^{1^{--}_0}_{6,\mu}(t)$\qquad\qquad & $\frac{1}{L^3}\sum\limits_{\pmb x}\text{Im}~L_\mu^{(3)}(\pmb x,t)$ 
\quad& $1^{--}_0$ \cr

$O^{0^{++}_0}_4(t)$\qquad\qquad & $\sum\limits_{\mu,=1}^{3}O^{1^{--}_0}_{1,\mu}(t)~
O^{1^{--}_0}_{1,\mu}(t)$\quad & $0^{++}_0$ \cr

$O^{2^{++}_0}(t)$\qquad\qquad & $\frac{1}{L^3}\sum\limits_{\pmb x}\textnormal{Re tr}\big[U_{12}(\pmb x,t)+
U_{23}(\pmb x,t)-2~U_{13}(\pmb x,t)\big]$\quad & $2^{++}_0$ \cr

$O^{0^{-+}_0}(t)$\qquad\qquad & $\frac{1}{L^3}\sum\limits_{\pmb x}\sum\limits_{\mu\neq \nu\neq \gamma
\neq \rho=1}^3\textnormal{tr}\big[U_{\mu\nu}(\pmb x,t)~U_{\gamma\rho}(\pmb x,t)\big]$ 
& $0^{-+}_0$ \cr

$O^{0^{++}_1}(t)$\qquad\qquad & $\frac{1}{L^3}\sum\limits_{\pmb x}\sum\limits_{\mu,\nu=1}^3 \epsilon_{ijk}~
\Big[\phi_i~(D_{\mu}\phi)_j~(D_{\mu}D_{\nu}D_{\nu}\phi)_k\Big]
(\pmb x,t)$ \quad
& $0^{++}_1$ \cr

$O^{1^{--}_{1}}_{\mu}(t)$\qquad\qquad & $\frac{1}{L^3}\sum\limits_{\pmb x}\sum\limits_{\nu=1}^3 
\epsilon_{ijk}~\Big[\phi_i~(D_{\mu}\phi)_j~(D_{\nu}D_{\nu}\phi)_k\Big](\pmb x,t)$ \quad
& $1^{--}_1$ \cr

\hline

\hline																							

\end{tabular}
\label{tab:interpol}
\end{center}
\end{table*}
The parity $P$, the charge parity $C$, and the total angular 
momentum $J$,  are assigned to the interpolators by their transformation properties under the octahedral 
symmetry group $O_\mathrm{h}$, which is the discrete symmetry group of an isotropic lattice. A 
method on how these quantum numbers are assigned to the interpolators according to the irreducible 
representations of the octahedral group can be found in, e.g., \cite{Wurtz:2013ova} and 
\cite{Berg:1982kp}.

Note, that we only use the spatial-directions $\mu=1,2,3$ for the operators, since we are 
interested in the propagation of the state in Euclidean time-direction $\mu=4$.

\par
In Table \ref{tab:interpol} several of the interpolators can be viewed as bound states of the scalar and the 
gauge bosons in the language of a naive constituent interpretation (we only discuss the 'atomic' 
interpolators here):
\begin{itemize}[leftmargin=*]
\item $O^{0^{++}_0}_1$ describes a two-scalar bound state.
\item $O^{1^{--}_0}_{1,\mu}$ is a gauge boson dressed with two scalar fields. The gauge bosons appear 
in the lattice version of the covariant derivative defined as 
\begin{align}
D_\mu\phi(x) &= \frac{U_{\mu}(x)\phi(x+\hat{\mu})-U_{\mu}(x-\hat{\mu})^\dagger 
\phi(x-\hat{\mu})}{2}\;.
\label{eq:latcovderiv}
\end{align}
\item $O^{0^{++}_0}_3$ is a scalar gaugeball.

\item $O^{1^{--}_0}_{4,5,6,\mu}$ are vector gaugeball interpolators as defined in \cite{Wurtz:2013ova}. 
Several definitions are needed to define the spatially summed quantities $L^{(1,2,3)}_\mu$ 
from Table \ref{tab:interpol}. 
These quantities are built from the Wilson loop operator
\begin{align}
\begin{split}
W_{\mu\nu\rho}&(x) = 
\\
\text{tr}\Big[&U_\mu(x)U_\mu(x+\hat{\mu})U_\nu(x+2\hat{\mu})
U_\mu(x+\hat{\mu}+\hat{\nu})^\dagger \\
\times &U_\rho(x+\hat{\mu}+\hat{\nu})U_\mu(x+\hat{\nu}+\hat{\rho})^\dagger
U_\rho(x+\hat{\nu})^\dagger U_\nu(x)^\dagger\Big]\,,
\end{split}
\end{align}
and linear combinations thereof,
\begin{align}
\begin{split}
L^{(1)}_{\mu\nu\rho} =\; &W_{+\mu+\nu+\rho}+W_{+\mu+\nu-\rho}+W_{+\mu-\nu+\rho}
+W_{+\mu-\nu-\rho}\\
-&W_{-\mu+\nu+\rho}-W_{-\mu+\nu-\rho}-W_{-\mu-\nu+\rho}-W_{-\mu-\nu-\rho}\;,\\
L^{(2)}_{\mu\nu\rho} =\; &W_{+\mu+\nu+\rho}+W_{+\mu+\nu-\rho}+W_{+\mu-\nu+\rho}
-W_{+\mu-\nu-\rho}\\\
+&W_{-\mu+\nu+\rho}+W_{-\mu+\nu-\rho}-W_{-\mu-\nu+\rho}-W_{-\mu-\nu-\rho}\;,\\
L^{(3)}_{\mu\nu\rho} =\; &W_{+\mu+\nu+\rho}-W_{+\mu+\nu-\rho}+W_{+\mu-\nu+\rho}
-W_{+\mu-\nu-\rho}\\
+&W_{-\mu+\nu+\rho}-W_{-\mu+\nu-\rho}+W_{-\mu-\nu+\rho}-W_{-\mu-\nu-\rho}\;,
\end{split}
\label{eq:vectorglue_lincomb}
\end{align}
where we skipped the spacetime argument $x$ for brevity. The last step is to build the following linear 
combinations of Equation \eqref{eq:vectorglue_lincomb} to build the interpolators that give the 
vector representation, $J=1$, and negative parity $P$:
\begin{align}
\begin{split}
L^{(1)} &= \Big(L^{(1)}_{123}+L^{(1)}_{132}~,~L^{(1)}_{231}+L^{(1)}_{213}
~,~L^{(1)}_{312}+L^{(1)}_{321}\Big)\;,\\
L^{(2)} &= \Big(L^{(2)}_{123}+L^{(2)}_{321}~,~L^{(2)}_{231}+L^{(2)}_{132}
~,~L^{(2)}_{312}+L^{(2)}_{213}\Big)\;,\\
L^{(3)} &= \Big(L^{(3)}_{123}+L^{(3)}_{213}~,~L^{(3)}_{231}+L^{(3)}_{321}
~,~L^{(3)}_{312}+L^{(3)}_{132}\Big)\,.
\end{split}
\end{align}
Taking the imaginary parts of these quantities yields interpolators with negative charge parity $C$. 
Vector gaugeball interpolator with other $P$ and $C$ quantum numbers could be constructed from 
the definitions given above \cite{Wurtz:2013ova}. However, we are particularly 
interested in the $1^{--}_0$ gaugeball for reasons illustrated in the next subsection where this 
quantum number channel is discussed.

\item $O^{0^{-+}_0}$, and $O^{2^{++}_0}$ are a pseudo-scalar gaugeball, and a tensor gaugeball, 
respectively, see \cite{Maas:2014pba}. 

\item $O^{0^{++}_1}$ and $O^{1^{--}_1}_{\mu}$ are the only interpolators with an open 
$\U(1)$-quantum number. We assigned a $\U(1)$ charge of $1/3$ to the scalar field $\phi$. 
The continuum versions are discussed in \cite{Maas:2017xzh} and 
the corresponding lattice versions are 
\begin{widetext}
\begin{align}
O^{0^{++}_1}(t) &= \frac{1}{L^3}\sum\limits_{\pmb x}\sum\limits_{\mu,\nu=1}^3 \epsilon_{ijk}~
\Big[\phi_i~(D_{\mu}\phi)_j~(D_{\mu}D_{\nu}D_{\nu}\phi)_k\Big](\pmb x,t) \nonumber\\
&= 
\frac{1}{16~L^3}\sum\limits_{\pmb x}\sum\limits_{\mu,\nu=1}^3 \epsilon_{ijk}~\phi_i(\pmb x,t)\nonumber\\
&\quad\times\bigg(U_\mu(\pmb x,t)~\phi(\pmb x+\hat{\mu},t)-U_\mu(\pmb x-\hat{\mu},t)^\dagger~
\phi(\pmb x-\hat{\mu},t)\bigg)_j \nonumber\\
&\quad\times\bigg(U_\mu(\pmb x,t)~U_\nu(\pmb x+\hat{\mu},t)~
U_\nu(\pmb x+\hat{\mu}+\hat{\nu},t)~\phi(\pmb x+\hat{\mu}+2\hat{\nu},t)\nonumber \\
&\quad-\phantom{\bigg(}U_\mu(\pmb x-\hat{\mu},t)^\dagger~ U_\nu(\pmb x-\hat{\mu},t)~
U_\nu(\pmb x-\hat{\mu}+\hat{\nu},t)~\phi(\pmb x-\hat{\mu}+2\hat{\nu},t) \nonumber\\
&\quad+\phantom{\bigg(}U_\mu(\pmb x,t)~U_\nu(\pmb x+\hat{\mu}-\hat{\nu},t)^\dagger~
U_\nu(\pmb x+\hat{\mu}-2\hat{\nu},t)^\dagger~\phi(\pmb x+\hat{\mu}-2\hat{\nu},t)\nonumber\\
&\quad-\phantom{\bigg(}U_\mu(\pmb x-\hat{\mu},t)^\dagger ~U_\nu(\pmb x-\hat{\mu}-\hat{\nu},t)^\dagger ~
U_\nu(\pmb x-\hat{\mu}-2\hat{\nu},t)^\dagger~\phi(\pmb x-\hat{\mu}-2\hat{\nu},t) \bigg)_k\;,\nonumber\\
\label{eq:openU1_0pp}
\\
O^{1^{--}_{1}}_{\mu}(t) &= \frac{1}{L^3}\sum\limits_{\pmb x}\sum\limits_{\nu=1}^3
\epsilon_{ijk}~\Big[\phi_i~(D_{\mu}\phi)_j~(D_{\nu}D_{\nu}\phi)_k\Big](\pmb x,t) \nonumber\\
&= \frac{1}{8~L^3}\sum\limits_{\pmb x}\sum\limits_{\nu=1}^3 
\epsilon_{ijk}~\phi_i(\pmb x,t)\nonumber \\
&\quad\times\bigg(U_\mu(\pmb x,t)~\phi(\pmb x+\hat{\mu},t)-U_\mu(\pmb x-\hat{\mu},t)^\dagger~
\phi(\pmb x-\hat{\mu},t)\bigg)_j \nonumber\\
&\quad\times\bigg(U_\nu(\pmb x,t)~U_\nu(\pmb x+\hat{\nu},t)~\phi(\pmb x+2\hat{\nu},t) \nonumber\\
&\quad+\phantom{\bigg(}U_\nu(\pmb x-\hat{\nu},t)^\dagger~U_\nu(\pmb x-2\hat{\nu},t)^\dagger~\phi(\pmb x-2\hat{\nu},t) 
\bigg)_k\;. \nonumber\\
\label{eq:openU1_1mm}
\end{align}
\end{widetext}

\end{itemize}

\par
We employ a variational analysis \cite{Michael:1985ne,Luscher:1990ck,Blossier:2009kd} in order to 
get access to the energy levels of the different quantum states in the respective $J^{PC}_{\U(1)}$-channels. 
Therefore, we compute a time-sliced matrix of cross correlators for a set of basis interpolators 
$O_i$, $i=1,2,\dots,N$, defined as
\begin{align}
\begin{split}
C_{ij}(t) &= \frac{1}{L}\sum_{t^\prime=0}^{L-1}
\big\langle O_i(t^\prime)O_j^\dagger(t+t^\prime) \big\rangle_\mathrm{c} \\
&=\frac{1}{L}\sum_{t^\prime=0}^{L-1}\Big\langle\Big(O_i(t^\prime)-\big\langle O_i(t^\prime)
\big\rangle\Big)
\\
&\qquad\qquad\times\Big(O_j^\dagger(t+t^\prime)-\big\langle O_j^\dagger(t+t^\prime)
\big\rangle\Big)\Big\rangle\;,
\end{split}
\label{eq:fullcorr}
\end{align}
where we subtracted the vacuum contribution $\langle O_i(t)\rangle$ from the correlator, i.e., we only 
consider the connected contributions $\langle\cdots\rangle_\mathrm{c}$ of the correlator. 
This is necessary since states with the quantum numbers $J^{PC}=0^{++}$ mix with the vacuum, which 
has exactly these quantum numbers. 

One can show that the eigenvalues of the matrix \eqref{eq:fullcorr} behave as $\lambda_k(t)\sim\E^{-aE_k t}$, 
$k=0,1,\dots,N-1$ \cite{Seiler:1982pw}. Thus, the energy levels can be extracted as 
\begin{align}
a E_k(t+{\scriptstyle{\frac{1}{2}}}) &= \ln\frac{\lambda_k(t)}{\lambda_k(t+1)}\;.
\end{align}

Since all fields appearing in the action \eqref{eq:lattice_action} obey periodic boundary conditions, 
the propagation in $t$ and $L-t$ of all the interpolators $O_i$ is identical and thus we fit the 
eigenvalues to 
\begin{align}
\begin{split}
\lambda_k(t) &= A_k^{(1)}\cosh\big(a E_k^{(1)}(t-L/2)\big) \\
&+A_k^{(2)}\cosh\big(a E_k^{(2)}(t-L/2)\big)\;,
\end{split}
\label{eq:corrfitfunc}
\end{align}
to extract the numerical values of the energy levels. We take into account a possible excitation of the 
level $E_k$, since heavier states still can contribute for small values of $t$ to this level after the 
variational analysis. 

\par
The bound states of Table \ref{tab:interpol} are expected to have a finite extent. Approximating 
them with point-like operators can therefore create an overlap problem. Therefore, we smear all our fields. 

For the links we apply stout smearing according to the procedure in \cite{Morningstar:2003gk}.
We choose this approach due to fact that with this method a projection back 
to the gauge group is not necessary. The new link after one stout smearing step is 
\begin{align}
U^\prime_\mu(x) &= \E^{\I Q_\mu(x)}~U_\mu(x)\;,
\end{align}
where $Q_\mu(x)$ is a hermitian and traceless matrix given by
\begin{align}
\begin{split}
Q_\mu(x) &= \frac{\I}{2}\Big(\Omega_\mu(x)^\dagger - \Omega_\mu(x) - 
\frac{\mathbbm{1}}{3}~\mathrm{tr}\big[\Omega_\mu(x)^\dagger - \Omega_\mu(x)\big]\Big)\;,\\
\Omega_\mu(x) &= \Bigg(\sum_{\nu\neq\mu}\rho_{\mu\nu}~C_{\mu\nu}(x)^\dagger\Bigg)~
U_\mu(x)^\dagger\;,
\end{split}
\label{eq:stout}
\end{align}
where the so-called staples $C_{\mu\nu}(x)$ enter:
\begin{align}
\begin{split}
C_{\mu\nu}(x) &= U_\nu(x+\hat{\mu})~U_\mu(x+\hat{\nu})^\dagger~U_\nu(x)^\dagger \\
&+ U_\nu(x+\hat{\mu}-\hat{\nu})^\dagger~U_\mu(x-\hat{\nu})^\dagger~U_\nu(x-\hat{\nu})\;.
\end{split}
\label{eq:staple}
\end{align}

We set $\rho_{\mu 4}=\rho_{4\mu}=0$, $\rho_{ij}=\rho$ in Equation \eqref{eq:stout} since we want to measure 
correlations in the Euclidean time direction and thus only spatial links are allowed to be smeared. In all our 
simulations we set $\rho=0.1$, see \cite{Morningstar:2003gk}. 

Of course, this procedure can be iterated. Therefore, the new link after $(n+1)$ stout smearing steps is 
given by
\begin{align}
U^{(n+1)}_\mu(x) &= \E^{\I Q^{(n)}_\mu(x)}~U^{(n)}_\mu(x)\;.
\end{align}

The scalar field is APE smeared in our case. After $(n+1)$ APE smearing steps the 
field is then \cite{Philipsen:1996af}
\begin{align}
\begin{split}
\phi^{(n+1)}(x) &= \frac{1}{7}\Bigg(\phi^{(n)}(x) + 
\sum_{\mu=\pm 1}^{\pm 4} U^{(n)}_\mu(x)~\phi^{(n)}(x+\hat{\mu})\Bigg)\;,
\end{split}
\end{align}
where the $n$-times stout smeared links $U^{(n)}_\mu(x)$ enter in the smearing procedure of the scalar.

We usually perform $300+10L$ updates to drive the system into equilibrium. Between the 
measurements of the observables we drop $3L$ configurations for decorrelation. We also performed 
several independent runs with different random number seeds for each parameter set to further reduce 
correlations.

The integrated autocorrelation time for the plaquette is $\tau_{\mathrm{int}} \approx 1/2$, i.e., 
close to the minimal value, for the parameter sets we analyzed. Therefore, no significant correlations 
between subsequent measurements of observables are detected. We usually 
studied $V=8^4$, $12^4$, $16^4$, and $20^4$ lattices to perform a finite-size analysis of the resulting 
masses in several quantum number channels. We typically have $\mathcal{O}\big(10^5\big)$ configurations at hand to compute the correlation 
functions. E.g., in Section \ref{sec:spectrum} we used $V=8^4$ with $320000$ 
configurations, $V=12^4$ with $240000$ configurations, $V=16^4$ with $120000$ configurations, 
and $V=20^4$ with $190000$ configurations.

The errors of the correlators are computed throughout by a standard Jackknife procedure, and for 
secondary observables we use the method of error propagation unless stated otherwise.

\subsection{Techniques for gauge-variant quantities}\label{ssec:gv_tech}
In order to compute propagators of elementary fields we need to fix a gauge. Without this 
procedure the propagators would be zero \cite{Elitzur:1975im}. 
Determining these propagators is relevant, as they will be an important building block of GIPT. They will also provide additional support that we probe the theory at weak coupling and our observed results are not genuine strong-coupling effects.

\par
\begin{itemize}[leftmargin=*]
 \item {\bf Local and global gauge fixing}  
\end{itemize}

We locally fix to minimal Landau gauge as described in \cite{Ilgenfritz:2010gu,Maas:2011se} by 
the so-called stochastic overrelaxation method \cite{Cucchieri:1995pn}. Additionally we use the 
Cabibbo-Marinari trick \cite{Cabibbo:1982zn} and the method of maximal trace \cite{Suman:1993mg} for 
reunitarization of the links.

To accomplish the so-called 't Hooft-Landau gauge condition \cite{Bohm:2001yx}, which gives 
rise to a vacuum expectation value of the scalar field,  we have to fix also the 
global direction of the scalar field. We want to perform a global gauge transformation such that the 
space-time average $\bar{\phi}$ of the scalar field point into some direction $n$:
\begin{align}
g~\frac{\bar{\phi}}{\big|\bar{\phi}\big|} &= n\quad\text{with}\quad 
\bar{\phi}=\frac{1}{V}\sum_x\phi(x)\quad\text{and}\quad g\in\mathrm{SU}(3)\;,
\label{eq:globaltrafo}
\end{align}
where we set $n_i = \delta_{i,3}$ without loss of generality. We use two consecutive 
$\SU(3)$ rotations, i.e.,
\begin{align}
g\bar{\phi}&= g_2g_1\bar{\phi} = n
\;,\;\;g_1,g_2\in\mathrm{SU}(3)\;.
\end{align}
Without loss of generality we assume a normalized vector $|\bar{\phi}|=1$ in the following. 
The first transformation $g_1$ has the task to rotate the first component of $\bar{\phi}$ to zero:
\begin{align}
\begin{split}
g_1\bar{\phi} =& 
\begin{pmatrix}
g_1^{11} & g_1^{12} & 0 \\
-\big(g_1^{12}\big)^\star & \big(g_1^{11}\big)^\star & 0 \\
0 & 0 & 1
\end{pmatrix}
\begin{pmatrix}
\bar{\phi}_1\\
\bar{\phi}_2 \\
\bar{\phi}_3
\end{pmatrix}= 
\begin{pmatrix}
0\\
\bar{\phi}_2^\prime \\
\bar{\phi}_3^\prime
\end{pmatrix} = 
\bar{\phi}^\prime\;\;, \\ 
&\text{with }\big|g_1^{11}\big|^2+\big|g_1^{12}\big|^2 = 1\;.
\end{split}
\end{align}
The second transformation $g_2$ then rotates the second component of $\bar{\phi}^\prime$ to zero:
\begin{align}
\begin{split}
g_2\bar{\phi}^\prime =& 
\begin{pmatrix}
1 & 0 & 0 \\
0 & g_2^{11} & g_2^{12} \\
0 & -\big(g_2^{12}\big)^\star & \big(g_2^{11}\big)^\star
\end{pmatrix}
\begin{pmatrix}
0\\
\bar{\phi}_2^\prime \\
\bar{\phi}_3^\prime 
\end{pmatrix}= 
\begin{pmatrix}
0\\
0 \\
\bar{\phi}_3^{\prime\prime}
\end{pmatrix} = 
\bar{\phi}^{\prime\prime}\;\;,\\ 
&\text{with }\big|g_2^{11}\big|^2+\big|g_2^{12}\big|^2 = 1\;.
\end{split}
\end{align}
Solving these equations for the matrix elements $g_i^{nm}$ with the normalization constraint gives the 
desired transformation matrix $g$. To summarize, the following steps have to be performed:
\begin{enumerate}[leftmargin=*]
\item Normalize $\bar{\phi}$, $\big|\bar{\phi}\big|=1$.
\item Compute
\begin{align}
g_1^{11} &= \Bigg(1+\frac{\big|\bar{\phi}_1\big|^2}{\big|\bar{\phi}_2\big|^2}\Bigg)
^{-\frac{1}{2}} \quad,\quad
g_1^{12} = -\frac{\bar{\phi}_1\bar{\phi}_2^\star}{\big|\bar{\phi}_2\big|^2}\;.
\end{align}
\item Compute
\begin{align}
\begin{split}
\mathrm{Re}~g_2^{11} &= \frac{1}{1+\frac{\big|\bar{\phi}_2\big|^2}{\big|\bar{\phi}_3\big|^2}}~
\Bigg[\mathrm{Re}\big[\bar{\phi}_3\big]
\Bigg(1+\frac{\mathrm{Im}\big[\bar{\phi}_3\big]^2}{\mathrm{Re}\big[\bar{\phi}_3\big]^2}\Bigg)
\Bigg]^{-1}\;, \\
\mathrm{Im}~g_2^{11} &= \frac{\mathrm{Im}~\bar{\phi}_3}{\mathrm{Re}~\bar{\phi}_3}~
\mathrm{Re}~g_2^{11} \;,\\
g_2^{12} &= -\frac{\bar{\phi}_3^\star}{\big|\bar{\phi}_3\big|^2}~
\frac{\big(g_1^{11}\big)^\star}{\big|g_1^{11}\big|^2}~\bar{\phi}_2~g_2^{11}\;.
\end{split}
\end{align}
\item Construct $g=g_2g_1$ from the previous steps.
\item Apply the global gauge transformations to the scalar and gauge fields:
\begin{align}
\phi(x)\to g~\phi(x) \;,\; U_\mu(x)\to g~U_\mu(x)~g^\dagger\;,\;
\forall~x,\mu\;.
\end{align}
\end{enumerate} 
With this procedure the gauge is now completely fixed to the 
minimal 't Hooft Landau gauge. 

Note that we use less gauge-fixed configurations than for the spectrum calculations, as gauge-fixing is expensive in terms of computing time while at the same time the quantities we study are much less noisy. E.g., for the situation in Section \ref{sec:gi} we use for the $8^4$ lattice $16000$, for the $12^4$ lattice $12000$, for the 
$16^4$ lattice $4700$, and for the $20^4$ lattice $5500$ gauge-fixed configurations. 

\par
\begin{itemize}[leftmargin=*]
 \item {\bf Propagators} 
\end{itemize}

We are interested in the propagator of the gauge bosons, the scalars and the ghosts. The latter will be needed to determine the running gauge coupling.

Due to the isotropic lattice, we can take the trace over the 
Euclidean Lorentz-indices of the gauge-field propagator 
$D_{\mu\nu}^{bc}\big(p^2\big) = \big\langle A_\mu^b(p)~A_\nu^c(-p)\big\rangle$, 
with the momentum $p_\mu=\frac{2}{a}\sin\big(\frac{\pi}{L}k_\mu\big)$, $k_\mu = 0,1,\dots,L/2$. 
Further, the propagator is proportional to $\delta^{bc}$ in the minimal 't Hooft Landau gauge, and thus 
$D_{\mu\nu}^{bc} = \delta^{bc} D^c_{\mu\nu}$, with 
\begin{align}
D^c\big(p^2\big) &= \sum_{\mu=1}^4 
\big\langle A_\mu^c(p)~A_\mu^c(-p)\big\rangle \;,\;\;c=1,2,\dots,8\;.
\label{eq:gfprop}
\end{align}

For the scalar propagator we split the field $\phi$ into its real and imaginary parts and use the 
notation $\phi = \frac{1}{\sqrt{2}}\big(\phi_1+\I~\phi_2, \phi_3+\I~\phi_4, \phi_5+\I~\phi_6\big)$. 
Then, we define the propagator as 
\begin{align}
D_{ij}\big(p^2\big) &= \big\langle \phi_i(p)~\phi_j(-p)\big\rangle\;,\;\; i,j = 1,2,\dots,6\;.
\label{eq:scalarprop}
\end{align}
Again, in the minimial 't Hooft Landau gauge this propagator is diagonal, i.e., 
$D_{ij}\big(p^2\big) = D_i\big(p^2\big) \delta_{ij}$. As it is discussed in \cite{Maas:2017xzh} 
we expect, for the vev-choice $n_i=\delta_{i,3}$, that only the propagator 
$D_5(p^2\big)$ behaves like a massive propagator and the remaining ones correspond to the propagation 
of massless particles in the Landau gauge. 

\par
The ghost field propagator $G^{ab}(x,y)=\big\langle\bar{c}^a(x)c^b(y)\big\rangle$ can be computed by inverting 
the Faddeev-Popov operator $M^{ab}(x,y)$. On the lattice this operator is a linear combination of links 
mixed with the generators of the gauge group. This will be done using the methods described in \cite{Maas:2010qw}.

\par
\begin{itemize}[leftmargin=*]
 \item {\bf Running gauge coupling} 
\end{itemize}

Having computed the gauge field propagator \eqref{eq:gfprop} and the ghost propagator on the lattice, 
the running coupling $\alpha^b$ can be extracted 
for every value of $b=1,2,\dots,N^2-1$ in the miniMOM scheme \cite{vonSmekal:1997vx,vonSmekal:2009ae} as
\begin{align}
\alpha^b\big(p^2\big) &= p^6~\alpha\big(\mu^2\big)~G^b\big(p^2,\mu^2\big)^2~
D^b\big(p^2,\mu^2\big)\;,
\label{eq:runningcoupling}
\end{align}
where $\mu$ is the renormalization scale. Note that, this is a renormalization-scale invariant 
combination.

\par
\begin{itemize}[leftmargin=*]
 \item {\bf Renormalization of the scalar propagator} 
\end{itemize} 

We need to define a renormalization scheme for the 
scalar propagator $D_{ij}\big(p^2\big)$. To this end we follow \cite{Maas:2010nc,Maas:2016edk}, which assumes that the renormalization of the propagator works qualitatively as in the perturbative case 
\cite{Bohm:2001yx}. Thus, there are two renormalization constants: 
The multiplicative wave function renormalization $Z_i$ and an additive mass renormalization 
$\delta m^2_i$. This yields the renormalized scalar propagator in minimal 't Hooft Landau gauge,
\begin{align}
D_{i}^\mathrm{r}\big(p^2\big) &= \frac{1}{Z_i~\big(p^2+{m^\mathrm{r}_i}^{~2}\big)
+\Pi_i\big(p^2\big)+\delta m_i^2}\;,
\end{align}
for $i=1,2,\dots,6$ and where $m^\mathrm{r}_i$ is the renormalized mass of the $i^\text{th}$ particle and $\Pi_i\big(p^2\big)$ 
is the corresponding self energy which is obtained from the unrenormalized propagator 
\eqref{eq:scalarprop} as 
\begin{align}
\Pi_i\big(p^2\big) &= \frac{1-p^2~D_i\big(p^2\big)}{D_i\big(p^2\big)}\;.
\label{eq:selfenergy}
\end{align}
Thus, the self energy measures essentially the deviation from the tree-level propagator, i.e.,
\begin{align}
D_i\big(p^2\big) &= \frac{1}{p^2+\Pi_i\big(p^2\big)}\;.
\end{align}
Note, that the tree-level mass $m_i$ is implicitly included in the self-energy.

The scheme we use to fix the renormalization constants is:
\begin{align}
\begin{split}
D^\mathrm{r}_i\big(\mu^2\big) &= \frac{1}{\mu^2 + {m^\mathrm{r}_i}^{~2}} \;,\\
\frac{\mathrm{d}D^\mathrm{r}_i\big(p^2\big)}{\mathrm{d}p}\Bigg|_{p^2=\mu^2} &= 
-\frac{2\mu}{\big(\mu^2+{m^\mathrm{r}_i}^{~2}\big)^2}\;,
\end{split}
\end{align}
where $\mu$ is again the renormalization scale. Therefore, the renormalized propagator and its derivative 
are given by their tree-level values at $p^2=\mu^2$. From these equations the renormalization constants 
$Z$ and $\delta m$ can be derived. 
The renormalization constants are determined numerically by linear interpolation between two physical 
momenta  along the $x$-axis, with the value of $\mu$ inside the interval $(p_1,p_2)$. 
The derivative of the self-energy is obtained by analytically deriving the linear interpolation between the 
momenta points. We only choose values for $\mu$ such that $0 < p_1 < \mu < p_2<2/a$.
 
Note that both the gauge boson propagator and the ghost propagator require only a single multiplicative renormalization.
 
\par
\begin{itemize}[leftmargin=*]
 \item {\bf Position-space propagators} 
\end{itemize}

One can also compute form the momentum-space propagators the position-space correlators, also called 
Schwinger functions. The position-space correlator $\Delta(t)$ is computed by \cite{Cucchieri:2004mf}
\begin{align}
\Delta(t) &= \frac{1}{a\pi L}\sum_{p_4=0}^{L-1}\cos\Big(\frac{2\pi p_4}{L} t\Big)~
D\big(p_4^2\big)\;,
\label{eq:latschwinger}
\end{align}
for a field with propagator $D\big(p^2\big)$. Note that, the propagator $D\big(p_4^2\big)$ is evaluated 
at zero spatial momentum, as is indicated by the argument $p_4^2$, and the sum extends over the whole momentum range including the parts of the propagator 
reproduced by periodicity.

\section{The physics of an SU(3) gauge theory with a fundamental scalar}\label{sec:physics}

\subsection{Phase diagram of the theory}\label{sec:phasediagram}
Since the perturbative breaking pattern in our case is $\SU(3)\to \SU(2)$ and thus the gauge group is not fully broken, 
the Osterwalder-Seiler-Fradkin-Shenker argument \cite{Osterwalder:1977pc,Fradkin:1978dv} does not 
apply. Therefore, this theory may or may not have separated phases and a possibly rich phase 
structure. We expect (at least) two regions of the phase diagram: Due to the non-Abelian nature of our 
theory defined in Equation \eqref{eq:lattice_action}, a QCD-like region (QLR) where QCD-like physics takes place and due 
to the Higgs sector we also expect a region with BEH-like physics (HLR). 
Since we are especially interested in a situation with a perturbatively accessible BEH effect 
\cite{Frohlich:1980gj,Maas:2012tj,Maas:2013aia,Maas:2012ct} we scanned the phase diagram using the 
quantity \cite{Caudy:2007sf}
\begin{align}
\big\langle\bar{\phi}^2\big\rangle &= \left\langle\left|\frac{1}{V}\sum_x\phi(x)\right|^2\right\rangle 
=\frac{1}{V^2}\sum_{x,y}\big\langle\phi(x)^\dagger\phi(y)\big\rangle\;,
\label{eq:caudy}
\end{align} 
with $\bar{\phi}$ being the space-time average of the scalar field. This quantity is 
gauge-dependent, and thus determined after fixing to minimal 't Hooft Landau gauge gauge. If the BEH effect is active  
$\big\langle\bar{\phi}^2\big\rangle\xrightarrow{V \to \infty }\mathrm{const.}>0$, while without 
$\big\langle\bar{\phi}^2\big\rangle\sim 1/V \xrightarrow{V \to \infty } 0$ \cite{Caudy:2007sf,Maas:2016ngo}. Examples of how this quantity behaves can be found in \cite{Maas:2016ngo}.

\begin{figure}[tbh!]
\begin{center}
\includegraphics[width=0.5\textwidth]{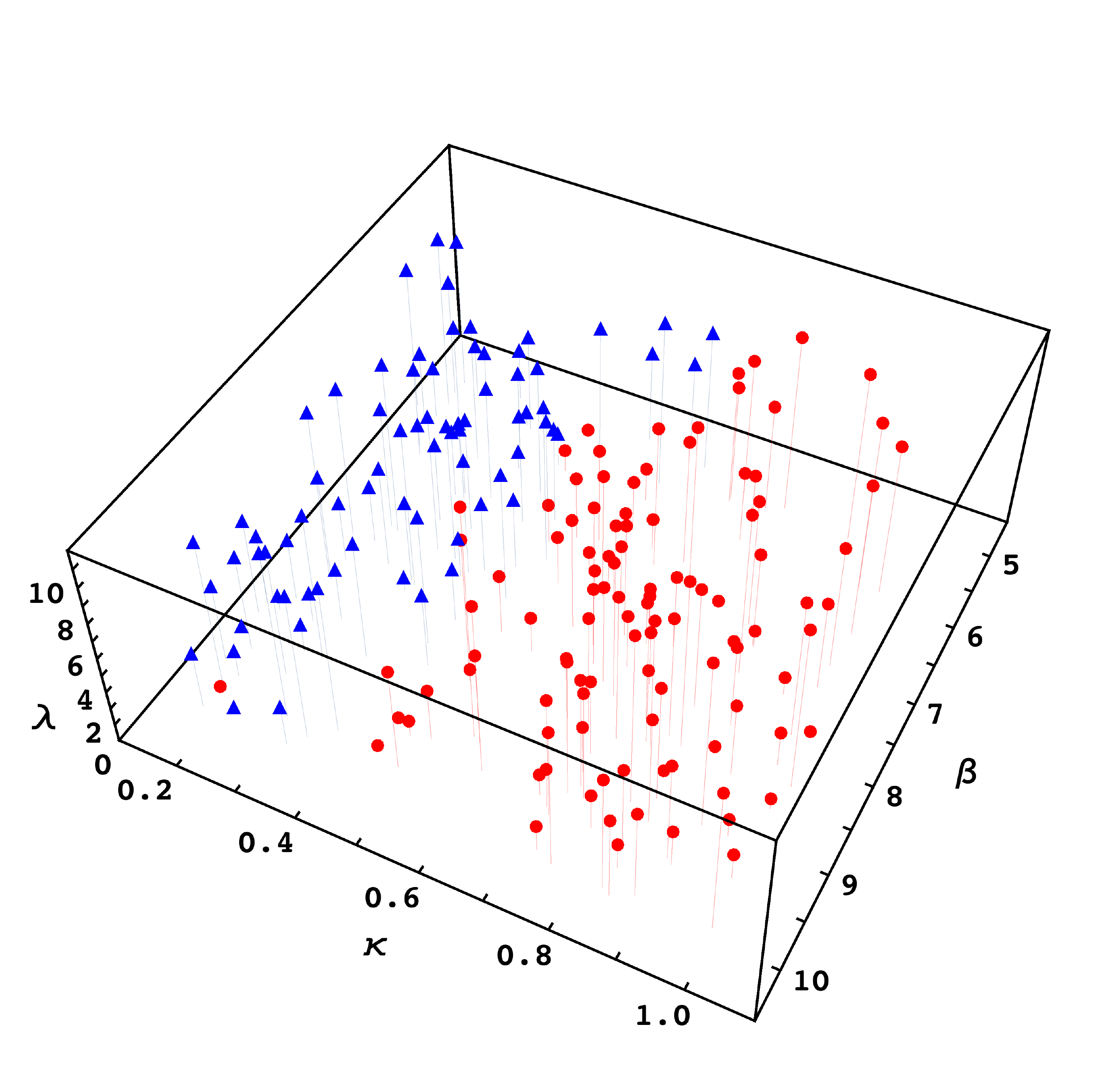}
\caption{The phase diagram of the theory according to the value of \eqref{eq:caudy}. The red dots show a BEH effect in minimal 
't Hooft Landau gauge, putting them in the HLR, while the blue triangles do not, meaning that 
they are in the QLR of the phase diagram. This is an update to the phase diagram in \cite{Maas:2016ngo}.
\label{fig:qlrhlr}}
\end{center}
\end{figure}

To scan the phase diagram quickly, we performed simulations for $V=4^4$, $6^4$, $8^4$, and $12^4$ 
lattices for randomly distributed parameters $\beta$, $\kappa$, and $\lambda$. We measured the 
quantity defined in Equation \eqref{eq:caudy} on $1000$ gauge-fixed configurations for each random 
parameter set and lattice size. Then, the volume dependence of this observable was used to decide to  
which region the parameter point belongs to. This lead to the results shown in Figure \ref{fig:qlrhlr}. 
The corresponding data can be found in Table \ref{tab:phasediagram} and Table  \ref{tab:phasediagram2}. 

\subsection{Physical spectrum}\label{sec:spectrum}

\par
In what follows, we focus on a set of parameters  
in the Higgs-like region, since our main target is to test the analytical predictions of the 
FMS mechanism in the end. 
We choose a point close to the boundary of the two regions of the phase diagram given by 
$\beta = 6.85535$,  $\kappa = 0.456074$, $\lambda = 2.3416$. This choice is motivated by the 
simulation results of the $\SU(2)$ theory, where the smallest lattice spacings, i.e., the largest cutoffs, 
have been found \cite{Maas:2014pba}.  We have also studied the spectrum for different sets of lattice parameters, which are listed in Table \ref{tab:numvals} in Appendix \ref{app:datafittables}. As far as a statistically reliable signal could be obtained we did not observe any qualitative differences. Hence, this set of parameters yields a suitable representative for the spectrum.

In the following, we investigate individually all the quantum number channels which are listed in 
Table \ref{tab:interpol}.

\par 
\begin{itemize}[leftmargin=*]
\item {\bf{$0^{++}_0$ channel}} 
\end{itemize}

The variational analysis of Section \ref{ssec:gi_tech} yielded a statistically reliable and stable result for the operator set
\begin{align}
\Big\{
O^{0^{++}_0}_{1,(10)}~,~O^{0^{++}_0}_{2,(10)}~,~O^{0^{++}_0}_{3,(10)}
~,~O^{0^{++}_0}_{4,(4)}~,~O^{0^{++}_0}_{4,(5)}
\Big\}\;,
\label{eq:0pp0set}
\end{align}
where the number in the brackets of the lower index denotes the smearing levels of the operators. Including other or more operators did not improve the result.

The operators $O^{0^{++}_0}_{1}$ and $O^{0^{++}_0}_{2}$, which contain only scalar fields, are 
smeared ten times as they are statistically very noisy due to the fact that the vacuum carries the same 
quantum numbers. For the same reason, we smear the gaugeball operator $O^{0^{++}_0}_{3}$ 
ten times as well. However, the interpolator $O^{0^{++}_0}_4$, which is a scattering state built form two 
$1^{--}_0$ operators, seems to be less noisy, and it was therefore only needed to smear it four and five times. 

\begin{figure}[tbh!]
\begin{center}
\hspace*{-1cm}
\includegraphics[width=0.5\textwidth]{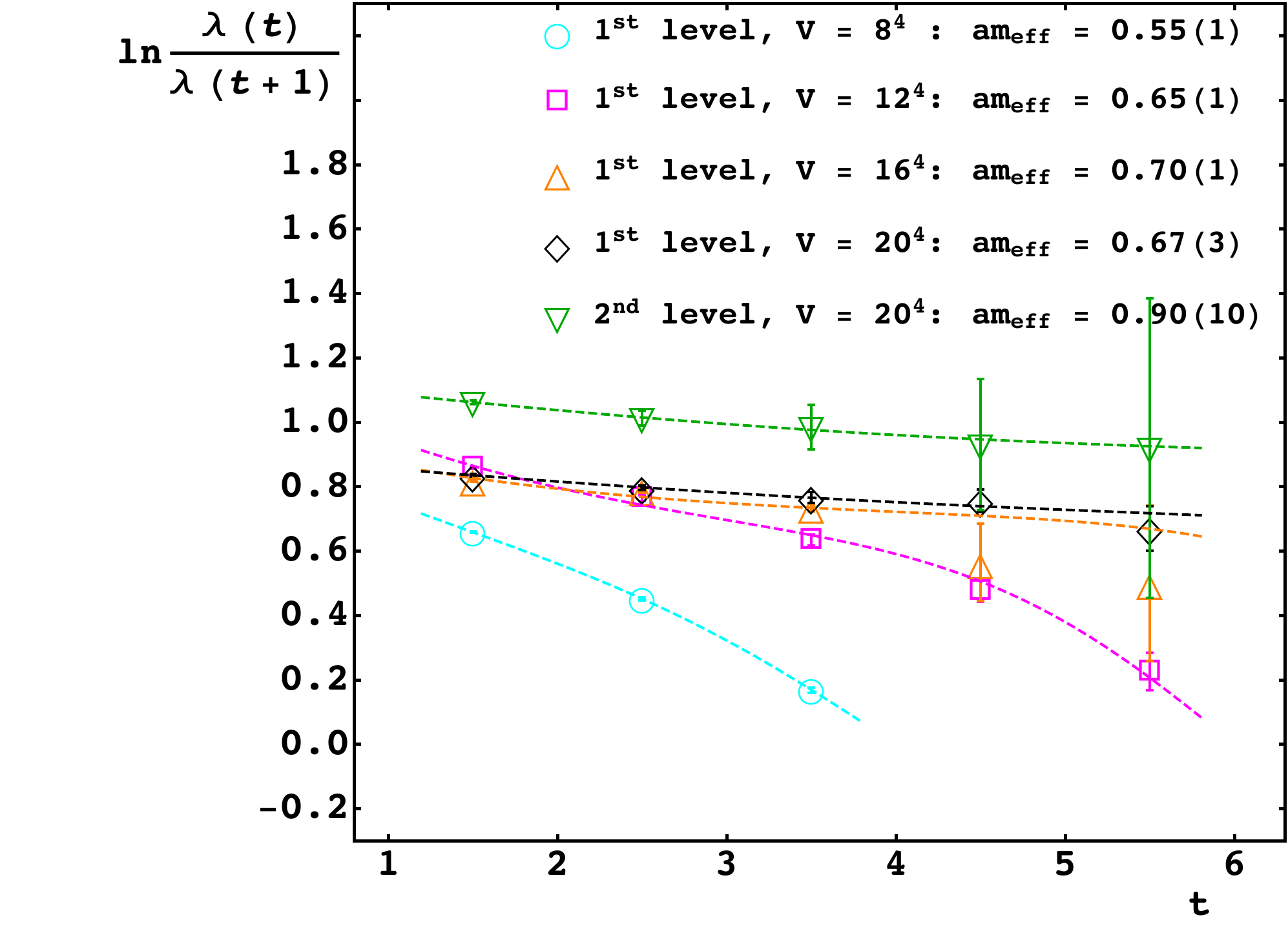}\\

\hspace*{-1cm}
\includegraphics[width=0.5\textwidth]{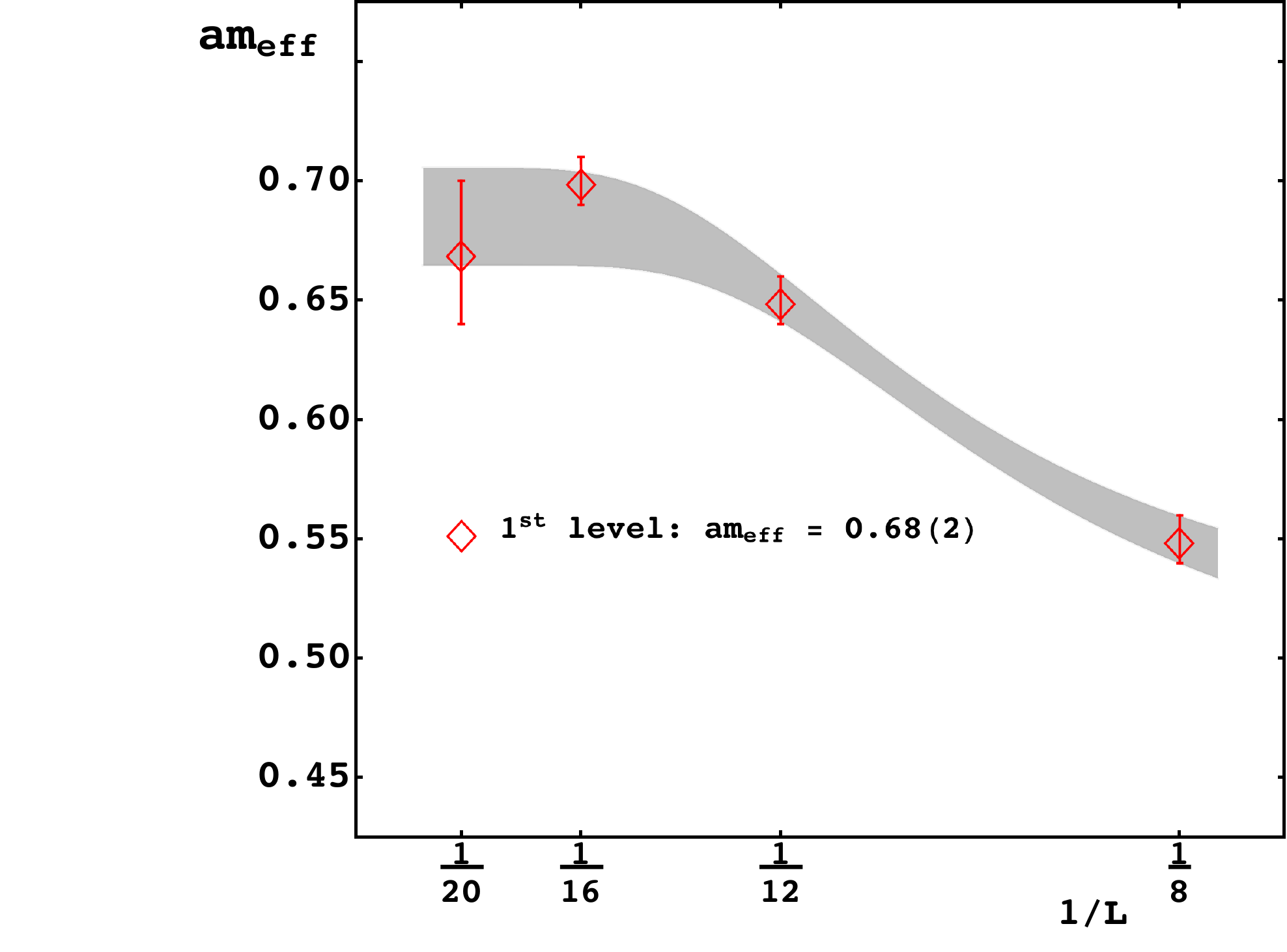}
\caption{{\bf{Top:}} Results of the variational analysis in the $0^{++}_0$ channel. The first energy levels are shown for 
the $V=8^4$, $12^4$, $16^4$, and 
$20^4$  lattices, whereas the second energy level (green triangles) is only 
shown for the largest volume for a clear display. The dashed lines are obtained by double-$\cosh$ fits of 
the eigenvalues. The lowest fitted energy values are listed in the legend as effective masses. 
{\bf{Bottom:}} First energy level of the $0^{++}_0$ channel as a 
function of the inverse lattice size. The gray bands are the error bands obtained by fits of the lower 
and upper bounds of the masses (see Table \ref{tab:fitvolumedep} in Appendix \ref{app:datafittables}). 
The extrapolated mass is $am_{0^{++}_0}=0.68(2)$.
\label{fig:varan_scalar}}
\end{center}
\end{figure}

The resulting effective mass as a function of time is plotted in Figure \ref{fig:varan_scalar}. We plot the energy of the lowest state (ground state), for each volume, and the second energy 
level (first excited state) for the largest volume. The effective masses and their errors, listed in the 
legend of Figure \ref{fig:varan_scalar}, are obtained by fitting the the mean, upper, and lower value of the 
eigenvalues by a double-$\cosh$, see Equation \eqref{eq:corrfitfunc}, for each volume. The resulting 
fit parameters are listed in Table \ref{tab:fitvalsgi} in Appendix \ref{app:datafittables}. Note that because of the large statistical noise we do not show data points for $t>6$. 

The volume dependence of the ground state mass is also plotted in Figure \ref{fig:varan_scalar}. We see that 
this state has a moderate dependence on the lattice size. Nevertheless, a fit of the lattice masses as 
a function of the volume,  $am_{0_0^{++}}(V) = am_{0_0^{++}} + \delta~\E^{-\gamma~V}$, can be 
performed and gives the gray error band (see Table \ref{tab:fitvolumedep} in Appendix 
\ref{app:datafittables} for the numerical values).
We conclude that the dimensionless ground state mass in this channel is $am_{0_0^{++}}=0.68(2)$ 
which is below the $2~am_{1_0^{--}}$ threshold, i.e., the elastic threshold,  as it can be seen in the 
discussion of the $1^{--}_0$ channel below. Since our analysis below suggests that this is the only open decay channel, this implies that the $0^{++}$ ground state is a stable particle.

The next-level state 
has an approximated mass of $am_{0_0^{++}}^\star \approx 0.9(1)$ which is compatible with the 
$2am_{1_0^{--}}= 0.78(2)$ threshold scattering state expected from the process $0_0^{++}\to 1_0^{--} + 1_0^{--} $. 
However, much more statistics for all volumes would be needed to make a definite statement.

The next expected states are the ones with mass $2am_{0_0^{++}}$ and with  
$2am_{0_0^{++}}+p^\mathrm{rel}$, where $p^\mathrm{rel}$ is a non-zero relative momentum. 
However, these states are relatively heavy and only noisy signals around this region have been found and thus 
no definite results are available.

\par
\begin{itemize}[leftmargin=*]
\item {\bf{$1^{--}_0$ channel}} 
\end{itemize}

For this channel a suitable basis of operators was found to be 
\begin{align}
\begin{split}
\Big\{
O^{1^{--}_0}_{1,\mu,(3)}~,~O^{1^{--}_0}_{1,\mu,(4)}~,~
&O^{1^{--}_0}_{2,\mu,(3)}~,~O^{1^{--}_0}_{2,\mu,(4)}~,~ \\
&\qquad O^{1^{--}_0}_{3,\mu,(3)}~,~O^{1^{--}_0}_{3,\mu,(4)}
\Big\}\;.
\label{eq:1mm0set}
\end{split}
\end{align}
The vector gaugeball interpolators $O^{1^{--}_0}_{4,\mu}$, $O^{1^{--}_0}_{5,\mu}$, 
and $O^{1^{--}_0}_{6,\mu}$ were too 
noisy even for the largest used smearing level as can be seen from the effective masses in Figure \ref{fig:vec_ten_glueballs} below. 
However, those states seem to be very high up in the spectrum 
and thus it is a justified assumption that they do not alter the infrared spectrum of the theory.

\begin{figure}[tbh!]
\begin{center}
\hspace*{-1cm}
\includegraphics[width=0.5\textwidth]{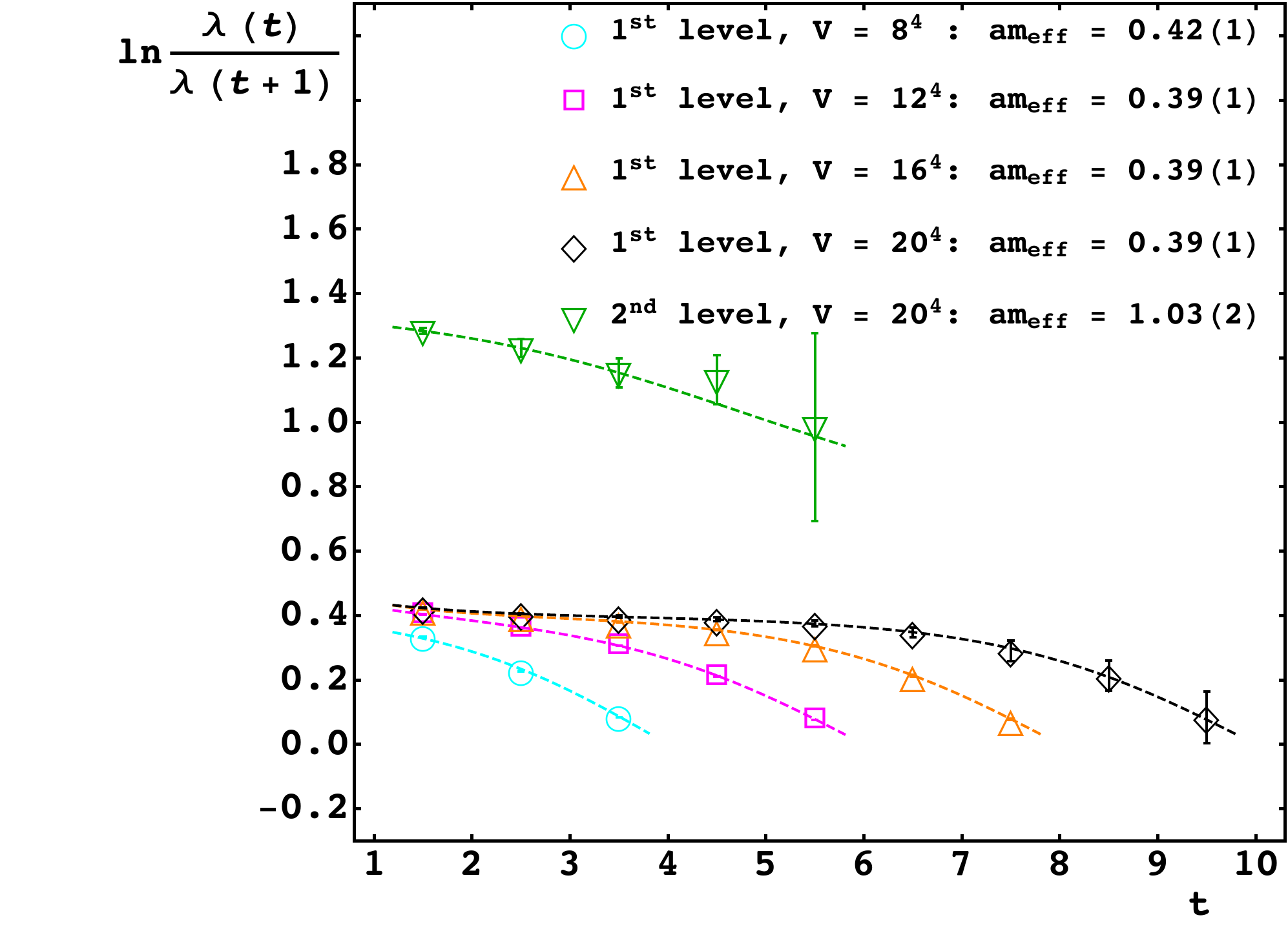}\\

\hspace*{-0.85cm}
\includegraphics[width=0.484\textwidth]{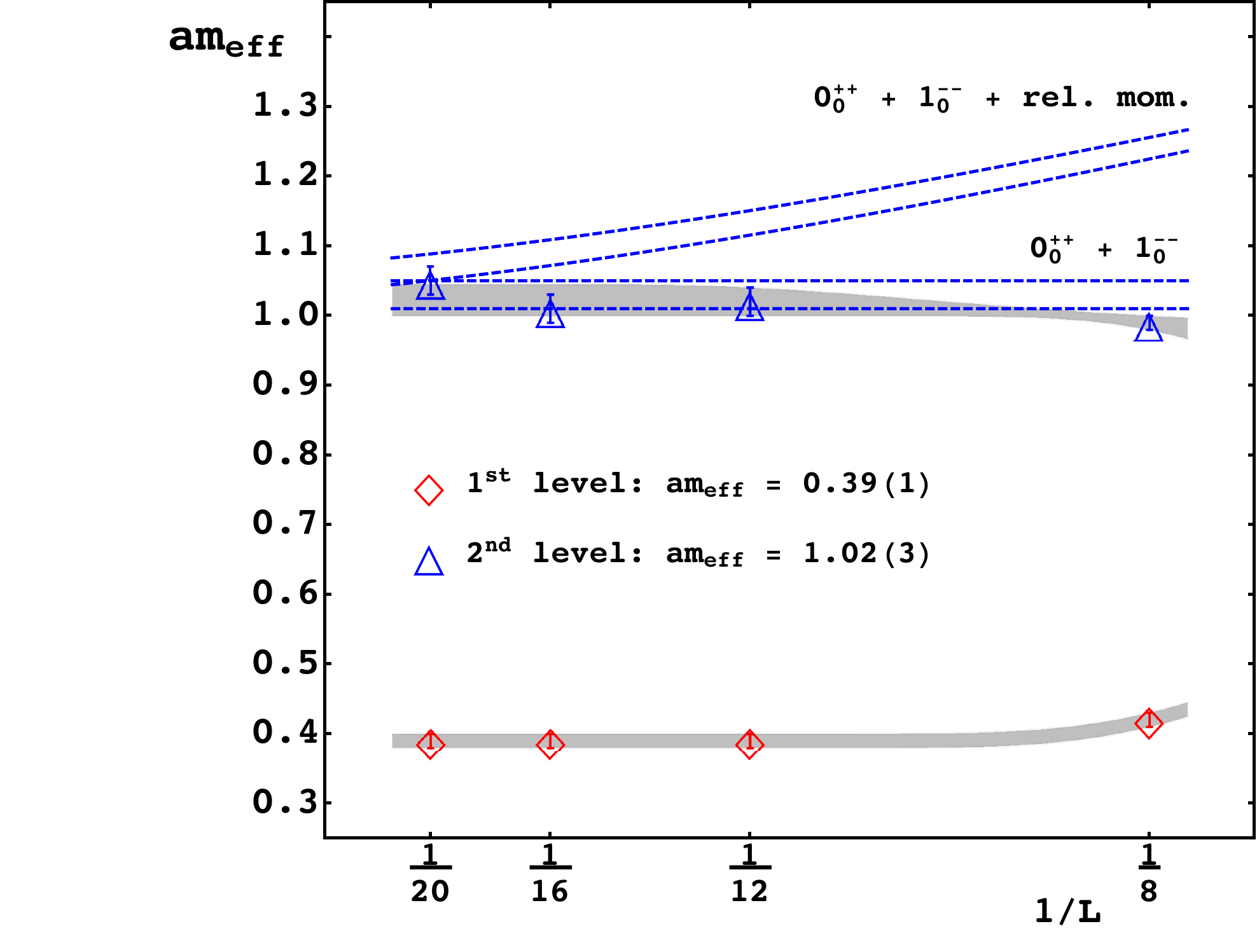}
\caption{{\bf{Top:}} Result of the variational analysis in the $1^{--}_0$ channel. The first energy levels are shown for 
the $V=8^4$, $12^4$, $16^4$, and 
$20^4$ lattices, whereas the second energy level (green triangles) is only 
shown for the largest volume for a clear display and for $t<6$. The dashed lines are obtained by 
double-$\cosh$ fits of the eigenvalues except for the smallest volume where we used a 
single-$\cosh$ fit. The lowest extracted fitted energy values are listed in the legend as effective mass. 
{\bf{Bottom:}} 
First and second energy level of the $1^{--}_0$ channel as a 
function of the inverse lattice size. The gray bands are the error bands obtained by fits of the lower 
and upper bounds of the masses (see Table \ref{tab:fitvolumedep} in Appendix \ref{app:datafittables}). 
The dashed blue lines are the expected masses of the next-level states $am_{0_0^{++}}+am_{1_0^{--}}$ 
and $p^\mathrm{rel}+am_{0_0^{++}}+am_{1_0^{--}}$, where $\pmb{p}^\mathrm{rel}=(2\pi/L,0,0)$. 
The extrapolated masses are $am_{1^{--}_0}=0.39(1)$ for the ground state, and 
$am_{1^{--}_0}^\star=1.02(3)$ for the second level.
\label{fig:varan_vector}}
\end{center}
\end{figure}

In the top of Figure \ref{fig:varan_vector}, we show the energy levels obtained from the variational analysis with 
the cross-correlation matrix built from the basis interpolators \eqref{eq:1mm0set}. The lowest energy 
level is shown for each lattice volume. As in the previous discussion the second energy level is only shown 
for the largest volume and for $t<6$. 
Again we list the effective masses in the legend of this figure, which are obtained 
by the same fit strategy as in the $0^{++}_0$ case. 
The fit parameters can be found in Table \ref{tab:fitvalsgi} in Appendix \ref{app:datafittables}.

The ground state has almost no volume dependence, hence the infinite volume extrapolated ground state 
mass is $am_{1^{--}_0}=0.39(1)$, see the bottom of Figure \ref{fig:varan_vector} and Table \ref{tab:fitvalsgi}.
Hence, the singlet vector state is lighter than the singlet scalar state, i.e., 
$m_{1^{--}_0} < m_{0^{++}_0}$ for the investigated set of bare lattice parameters. 

Next-level states are expected at a mass of $am_{0_0^{++}}+am_{1_0^{--}}$, and at 
$3am_{1_0^{--}}$ for the processes $1^{--}_0\to 0^{++}_0+1^{--}_0$ and 
$1^{--}_0\to 1^{--}_0+1^{--}_0+1^{--}_0$ respectively. Additionally, one can find states with 
relative momentum as $p^\mathrm{rel}+am_{0_0^{++}}+am_{1_0^{--}}$ and 
$p^\mathrm{rel}+3am_{1_0^{--}}$. This is possible since the operators can have overlap with such states, even if carrying zero total momentum. The energy levels $E$ can be extracted from \cite{Wurtz:2013ova}
\begin{align}
\sinh^2\left(\frac{E(L,k)}{2}\right) &= \sinh^2\left(\frac{m_\mathrm{2p}}{2}\right) + 
\sum_{i=1}^3 \sin^2\left(\frac{\pi}{L}~k_i\right)\;,
\label{eq:wurtz}
\end{align}
where $m_\mathrm{2p}$ is the mass of the two-particle state and the relative lattice momentum is 
$p^\mathrm{rel}_i = 2\pi k_i/L$, $k_i=-L/2+1,\dots,L/2$. In the continuum limit this equation turns into 
the familiar energy-momentum relation $E(\pmb p)=\sqrt{m^2+\pmb p^2}$.

The ordering of the states depends on the value of the masses of the $0_0^{++}$ and $1_0^{--}$ states.  
For the parameter set we study, the $am_{0_0^{++}}+am_{1_0^{--}}$  state should be the lightest 
next-level state, since $am_{0_0^{++}}+am_{1_0^{--}}=1.07(3)$, and 
$3am_{1_0^{--}}=1.17(3)$. Besides the ground state, we also show on the right-hand side of  
Figure \ref{fig:varan_vector} the volume dependence of the second level (blue triangles)  with its 
error band as well as the expected next-level states $am_{0_0^{++}}+am_{1_0^{--}}$ and 
$p^\mathrm{rel}+am_{0_0^{++}}+am_{1_0^{--}}$ (dashed blue lines, upper and lower bounds) with 
$\pmb{p}^\mathrm{rel}=(2\pi/L,0,0)$, i.e., the smallest possible relative momentum.
It seems that the mass of the second state is consistent with the expected $0_0^{++}+1_0^{--}$ state 
and is not in agreement with the state including relative momentum. All other energy levels are too noisy 
to comment on them.

\par
\begin{itemize}[leftmargin=*]
\item {\bf{$0^{-+}_0$, $1^{--}_0$ and $2^{++}_0$ gaugeballs}} 
\end{itemize}

Here we show the spectroscopy results of several gaugeball states. All the results shown below 
share that the signals are very noisy. This makes determinations of their masses comparatively unreliable. Still, all the masses seem to be well above 
the lowest lattice mass in the spectrum, i.e., above $am_{1^{--}_0}$.

\begin{figure}[tbh!]
\begin{center}
\hspace*{-1cm}
\includegraphics[width=0.5\textwidth]{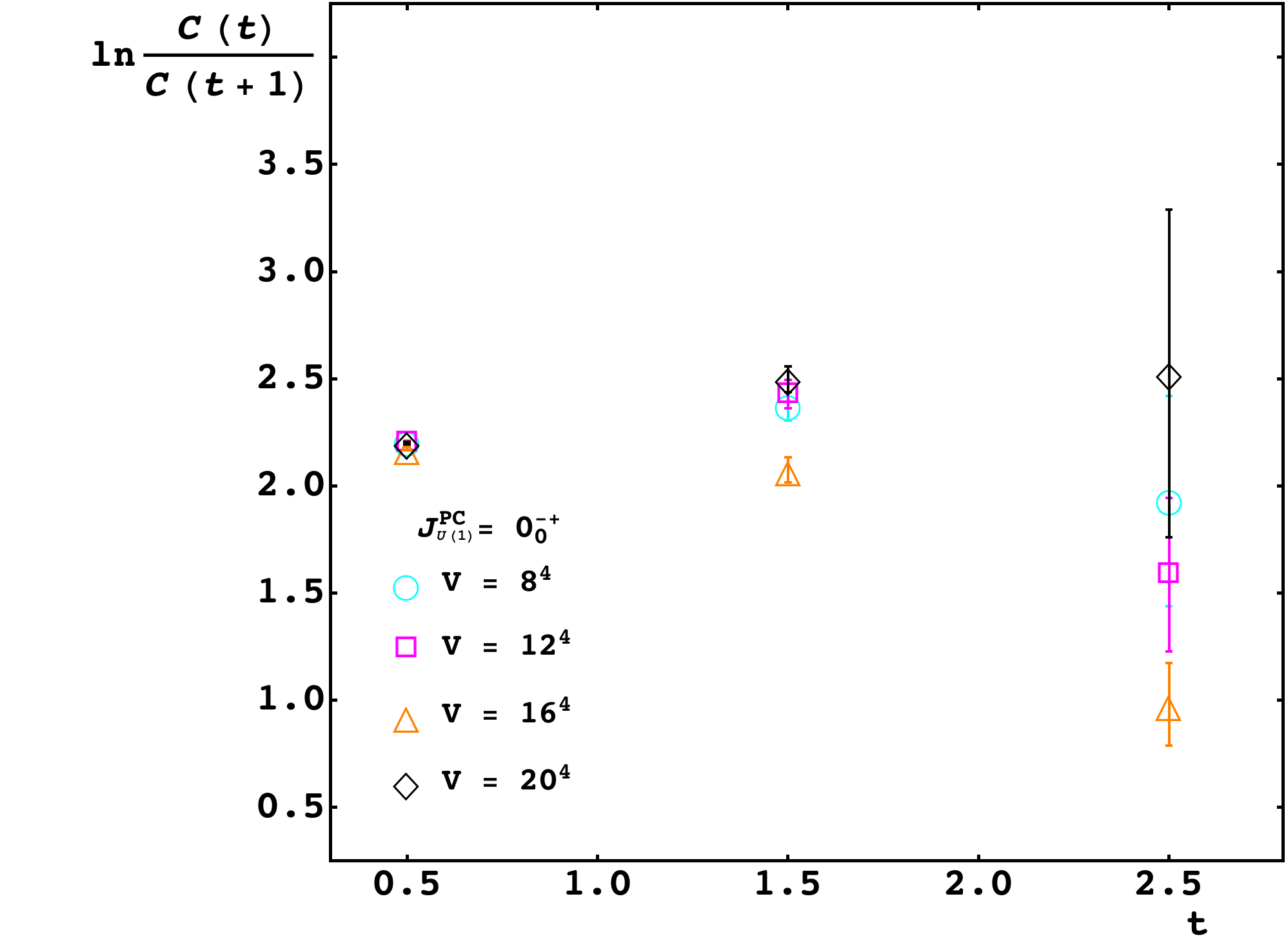}\\

\hspace*{-1cm}
\includegraphics[width=0.5\textwidth]{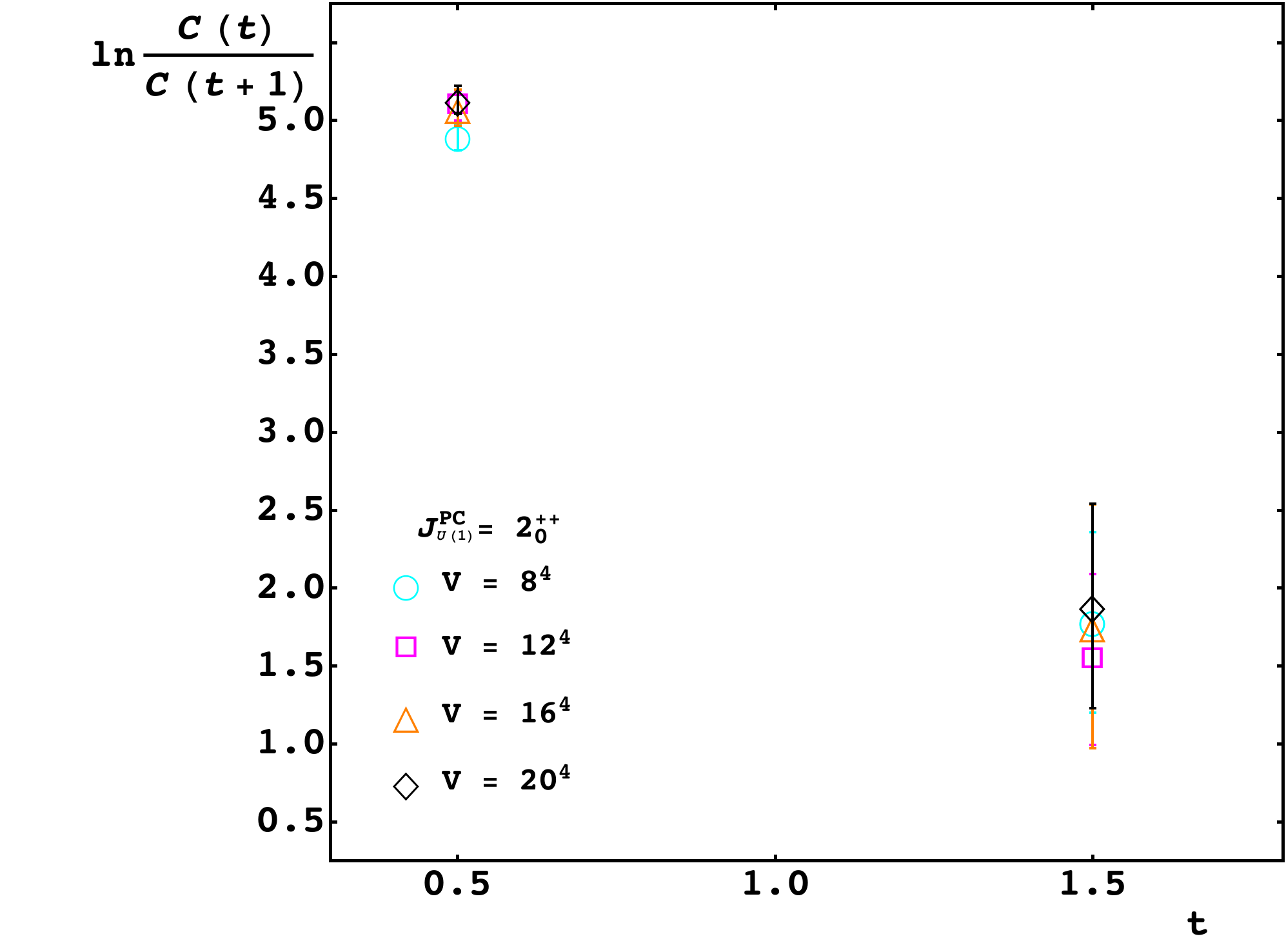}
\caption{In the top panel the effective mass of the pseudo-scalar gaugeball is shown as a 
function of Euclidean time. In the bottom panel the effective mass of the tensor gaugeball 
is plotted. 
For both, the results are shown for $V=8^4$, $12^4$, $16^4$, and $20^4$ for $10$-times smeared 
fields.
\label{fig:scalar_glueballs}}
\end{center}
\end{figure}

Figure \ref{fig:scalar_glueballs} shows the effective masses of the  $0^{-+}_0$ pseudo-scalar gaugeball in the top panel and the $2^{++}_0$ gaugeball in the bottom panel as a function of Euclidean time for 
several lattice volumes. We do not show data points for $t>3$ and $t>2$ respectively, since these regions 
are dominated by noise even though we used $10$-times smeared operators.
 
The effective masses in both channels are around 
$am_{0^{-+}_0}\approx am_{2^{++}_0}\approx 2.0$, i.e., above the lattice cutoff. These approximate 
masses  are of course just crude estimates. 

We also performed a variational analysis with sets of different smeared operators in these channels. 
However, this procedure did not improve the signal substantially and therefore we do not show the 
results here. 

Some, but not all, possible decay channels for the two states with the available channels are:
\begin{itemize}[label=-]
\item $0^{-+}_0$ channel: two $1^{--}_0$ in a $p$-wave

\item $2^{++}_0$ channel: two $0^{++}_0$ in a $d$-wave

\hspace*{2.05cm} two $1^{--}_0$ in a $s$-wave

\hspace*{2.05cm} $1^{--}_1$ and $1^{--}_{-1}$ in a $s$-wave
\end{itemize}
The masses in both channels are compatible with the last option at both points, i.e., a decay in 
$1^{--}_1$ and $1^{--}_{-1}$ in an $s$-wave (see below). Nonetheless, this is very speculative since 
more statistics and more operators including better overlap with the decay channels would be needed 
to make precise statements. Of course, another option is that those signals are just too noise dominated.

\begin{figure*}[tbh!]
\begin{center}
\includegraphics[width=0.48\textwidth]{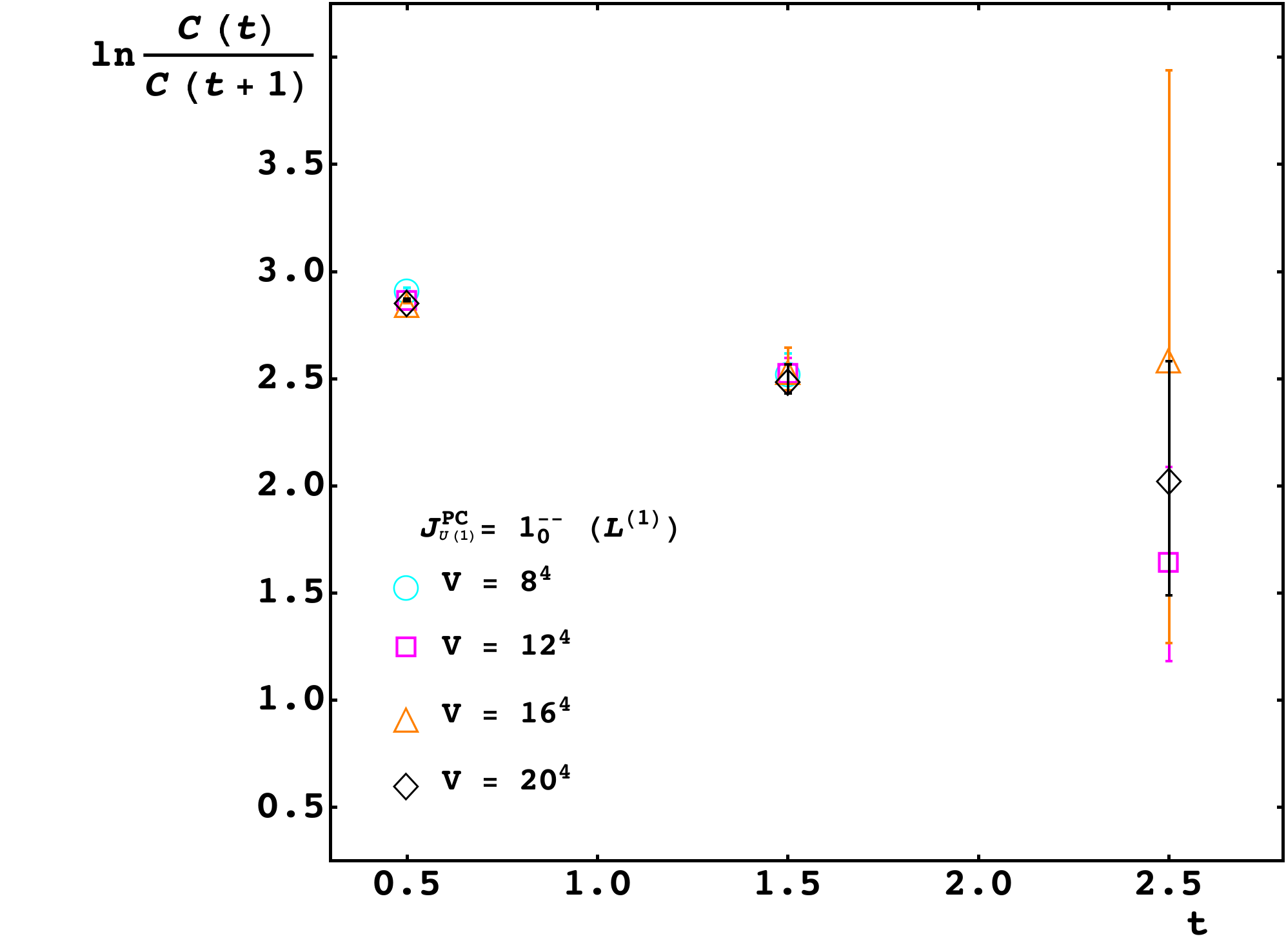}
\hspace*{0.15cm}
\includegraphics[width=0.48\textwidth]{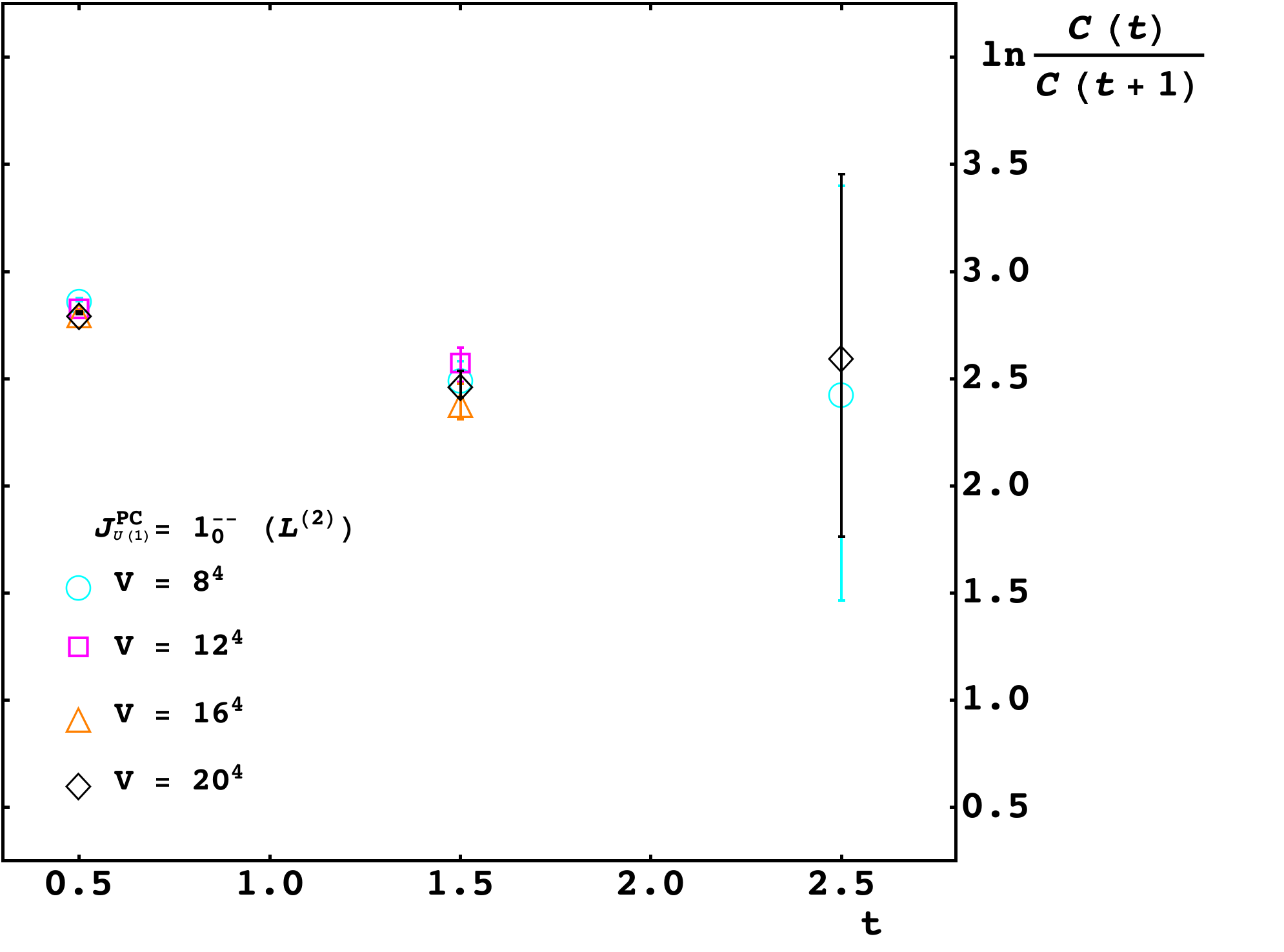}

\vspace*{0.15cm}
\hspace*{-9.2cm}
\includegraphics[width=0.48\textwidth]{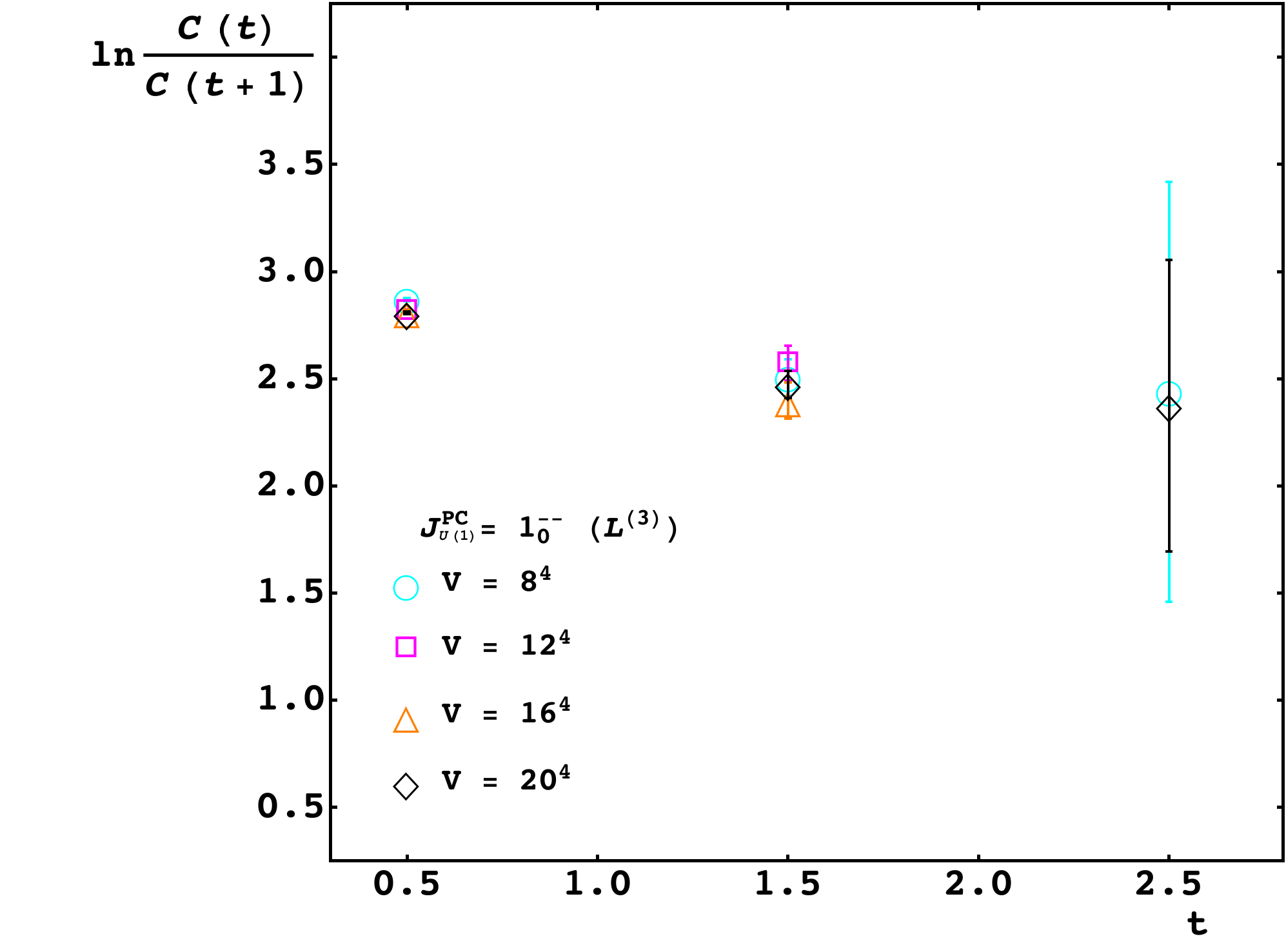}
\caption{The three panels show the effective masses of the $1^{--}_0$ gaugeballs $L^{(1)}$, 
$L^{(2)}$ and $L^{(3)}$ as a function of Euclidean time $t<3$. 
The results are shown for $V=8^4$ , $12^4$, $16^4$ , and $20^4$ for $10$-times smeared fields. 
\label{fig:vec_ten_glueballs}}
\end{center}
\end{figure*}

In Figure \ref{fig:vec_ten_glueballs} we show the effective masses of the three $1^{--}_0$ gaugeballs, 
$L^{(1)}$, $L^{(2)}$, and $L^{(3)}$, for $V=8^4$, $12^4$, $16^4$ and $20^4$ lattices. 
As before, we do not plot the whole time region in all the plots due to the large fluctuations of 
the correlators and thus the effective masses. The results are shown for $10$-times smeared 
operators as before.

Even though the signals are again noisy we deduce that the effective masses of the three $1^{--}_0$ 
gaugeballs are approximately $7$-times larger than the extracted ground state mass in this channel, and 
thus well above the lattice cutoff. 
As already argued in the discussion of the $1^{--}_0$ channel, they do not alter the ground state 
and thus the infrared spectrum, since they are too high up in the spectrum to generate any significant 
contribution. 

We are well aware that the effective mass plateau of three points which are still inclined, are probably still contaminated by excited state contributions, and higher statistics would be needed.

\par
\begin{itemize}[leftmargin=*]
\item {\bf{$0^{++}_{\pm 1}$ and $1^{--}_{\pm 1}$ open $\U(1)$ channels}} 
\end{itemize}

Finally, we study quantum number channels with an open $\U(1)$ quantum number, i.e.,
the $0^{++}_{\pm 1}$ and $1^{--}_{\pm 1}$ states. At least the lightest state with non-vanishing $\U(1)$ charge is necessarily stable, as this custodial charge is conserved in the theory.

\begin{figure*}[tbh!]
\begin{center}
\includegraphics[width=0.4883\textwidth]{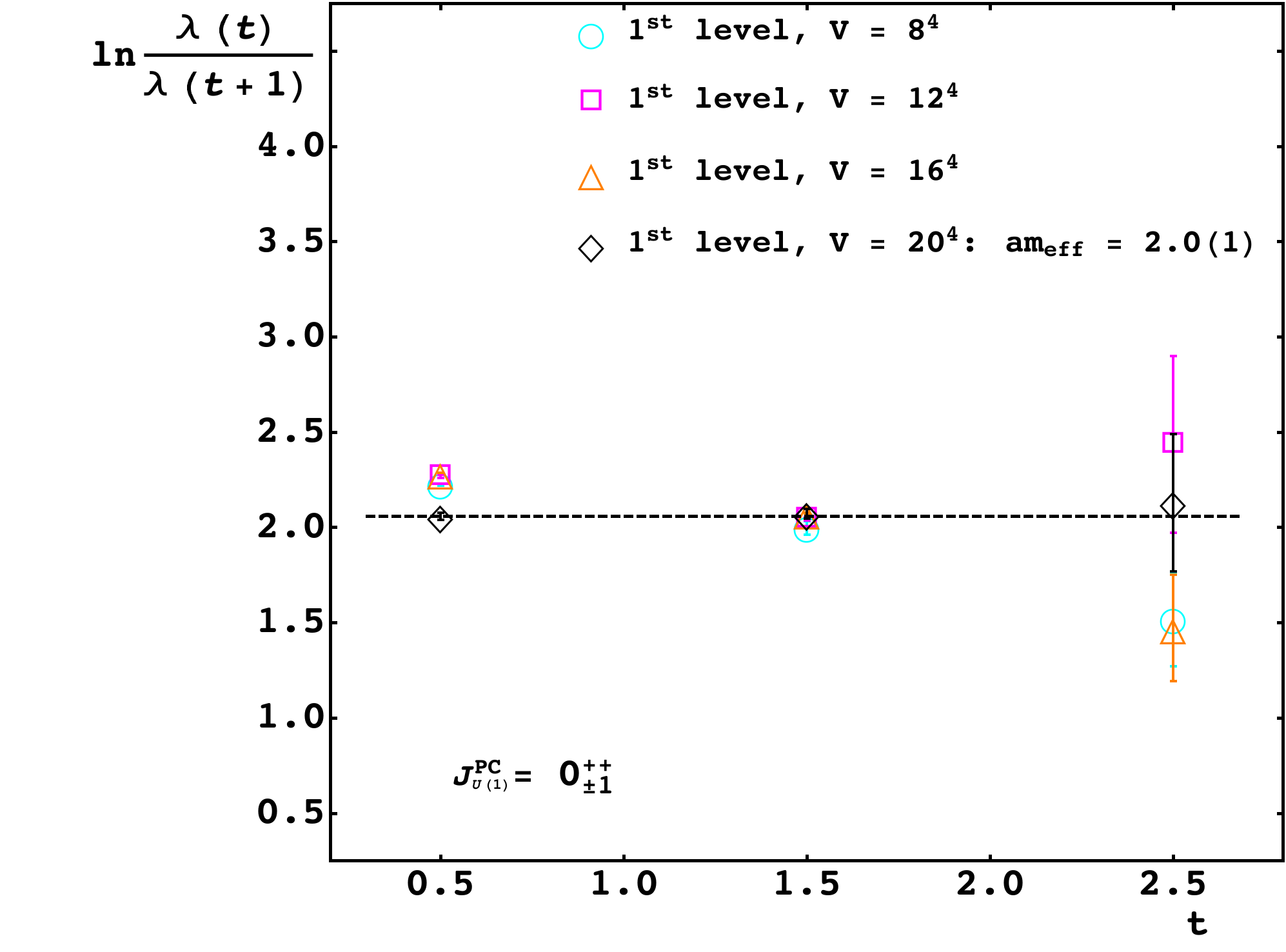}
\hspace*{0.05cm}
\includegraphics[width=0.498\textwidth]{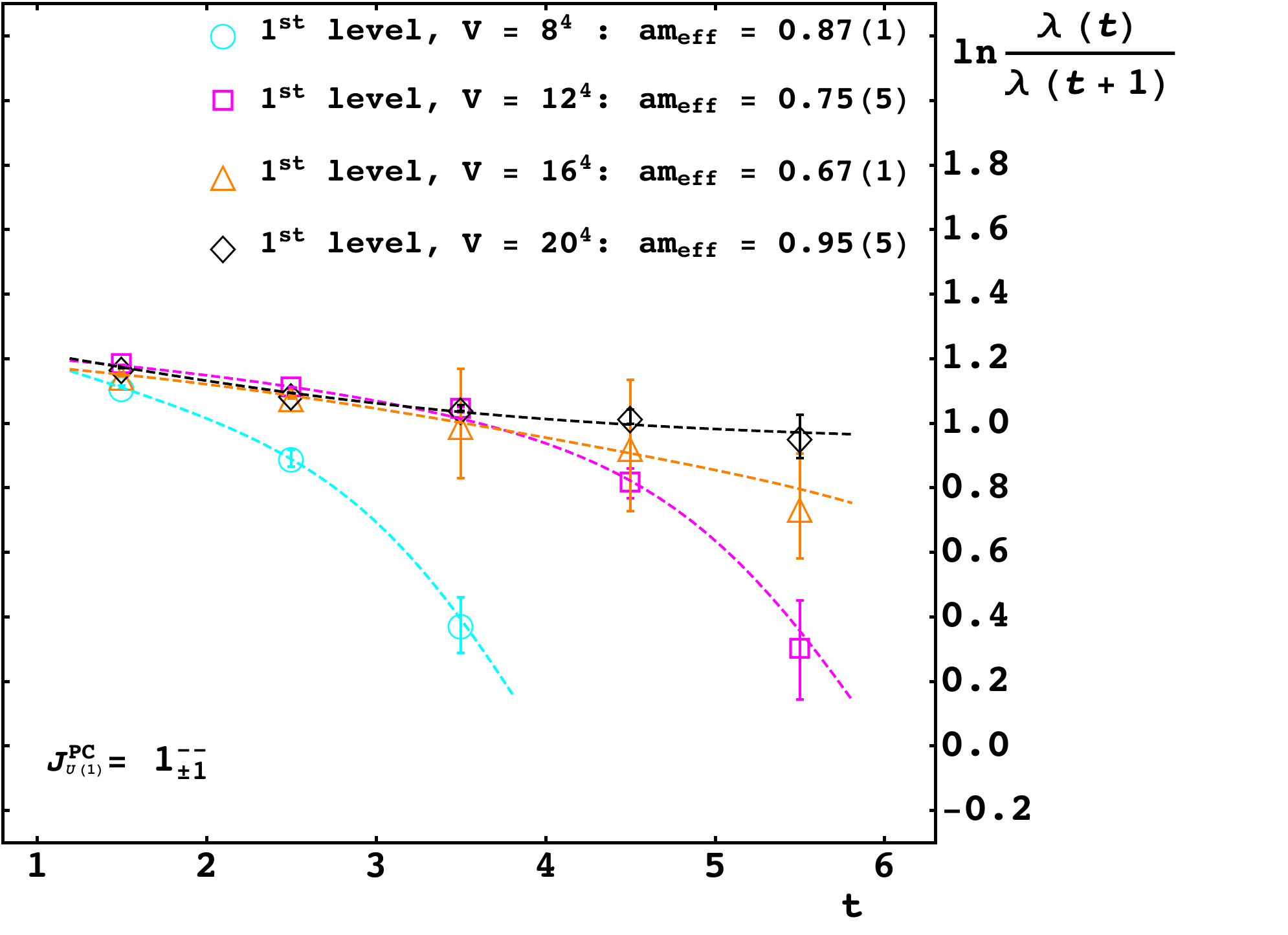}
\caption{On the left-hand side the effective mass of the $0^{++}_{\pm 1}$ state is shown as a 
function of Euclidean time $t<3$. On the right-hand side the effective mass of the 
$1^{--}_{\pm 1}$ state is plotted for $t<6$. 
For both, the results are shown for $V=8^4$, $12^4$, $16^4$ , and $20^4$ lattices. 
The dashed lines in the left and right panels are results of single- and double-$\cosh$ fits, respectively.
\label{fig:open_u1}}
\end{center}
\end{figure*}

In Figure \ref{fig:open_u1} we present results for the effective masses of the $0^{++}_{\pm 1}$ 
(left plot) and $1^{--}_{\pm 1}$ (right plot) channels for different lattice volumes. 
In both channels we performed a variational analysis with different smearing levels of the corresponding 
operators: In the scalar sector the basis consists of $6$- to $10$-times smeared interpolators, whereas in 
the vector sector we included $8$- to $10$-times smeared interpolators in the basis. 
The effective masses
\footnote{Note that, $m_{0^{++}_{+1}}=m_{0^{++}_{-1}}$ , i.e., particle and anti-particle states. Thus 
the effective mass given in the left-hand side of Figure \ref{fig:open_u1} is the mass of the particle and 
anti-particle. Certainly, the same is true for the  $1^{--}_{\pm 1}$ channel.}
of both states are listed in the legends of the figure and the corresponding values from the fit are given in 
Table \ref{tab:fitvolumedep} in Appendix \ref{app:datafittables}.

The scalar sector is dominated by noise and only the points for $t<3$ are reliable. The mass in this channel 
is roughly $am_{0^{++}_{\pm 1}}\approx 2$. Of course this is only a coarse estimation and larger 
statistics as well as larger lattices can alter the result. 

The vector channel is not so much dominated by noise and thus more time slices can be 
used for the fit ($t<6$). However, from $V=8^4$ to $V=16^4$ the effective mass seems to drop but 
for the largest lattice slightly rises again. 
Again, more statistics could still change this behavior. Nonetheless, we estimate a mass of
$am_{1^{--}_{\pm 1}}\approx 0.8(2)$. 

Higher levels were unaccessible due to the amount of statistical noise. Besides increasing the statistics
also including more operators could improve the result in 
both cases. 

\par
\begin{itemize}[leftmargin=*]
\item {\bf{Summary of the spectrum}} 
\end{itemize}
\begin{figure}[tbh!]
\begin{center}
\includegraphics[width=0.49\textwidth]{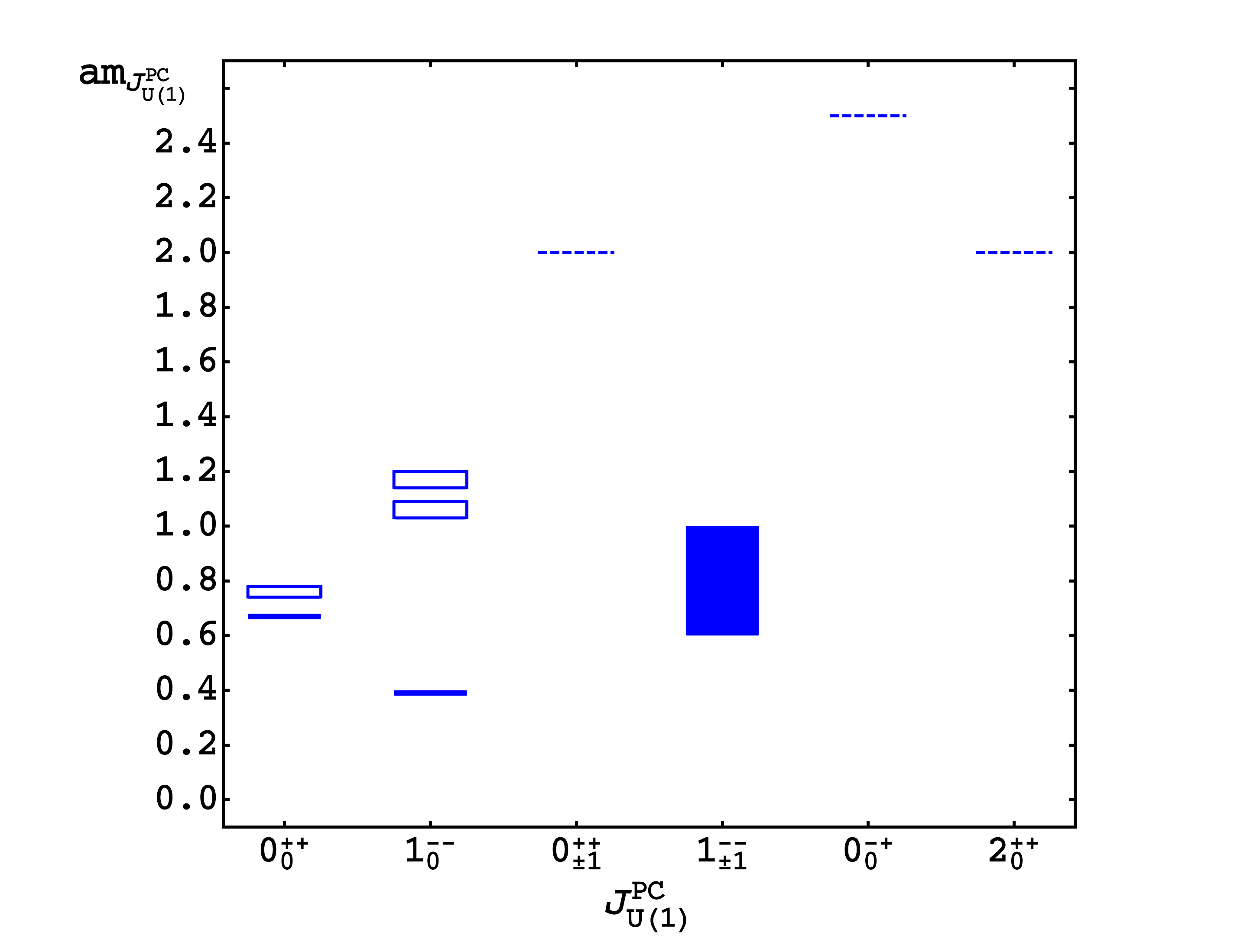}
\caption{Spectrum of the theory for the lattice 
parameter set $\beta = 6.85535$,  $\kappa = 0.456074$, $\lambda = 2.3416$. 
The description is given in the main text. Dashed levels are only estimates. 
\label{fig:spectrum_gi}}
\end{center}
\end{figure}

We summarize the computed spectrum of states in Figure \ref{fig:spectrum_gi}. The filled boxes 
correspond to the ground states, the empty boxes are the elastic thresholds  for the 
scalar and vector singlet channels as discussed above, and the dashed 
lines are the estimated ground state masses of the $0_{\pm 1}^{++}$, $0_{0}^{-+}$, $2_{0}^{++}$ 
channels. Where available, results for other lattice parameters can be found in the appendix.

Of course, it would be important to track the development of the spectrum along lines of constant physics, even though we find qualitatively the same situation everywhere in the HLR. Due to the fine-tuning character of the theory, the required numerical resources for this purpose, also given that at least three states have to be determined reliably for this, unfortunately exceed our resources by far.

\section{Gauge-variant observables and running gauge coupling}\label{sec:gi}

\subsection{Spectrum from tree-level perturbation theory}\label{ssec:treelevel}

For future reference, we briefly rehearse the spectrum of the theory at tree-level perturbation theory, see \cite{Maas:2017xzh} for details. For this we use a continuum setup, and employ 't Hooft-Landau gauge. 

We split the scalar field into its vev and a fluctuation part $\varphi$ around the vev
\begin{align}
 \phi (x) 
 = \frac{v}{\sqrt{2}} n + \varphi(x)\;.
 \label{eq:SplitHiggsF}
\end{align}
\noindent The spectrum then contains one real-valued massive scalar degree of freedom and $8$ would-be Goldstone modes. 
The non-Goldstone Higgs boson and the would-be 
Goldstones can be described in a gauge-covariant manner without specifying $n$ 
by $h \equiv \sqrt{2}\, \re(n^{\dagger}\phi)$ and 
$\breve{\varphi} \equiv \phi - \re(n^{\dagger}\phi)n = \varphi - \re(n^{\dagger}\varphi)n$, respectively. 
However, without loss of generality, in the following we will usually make the explicit choice $n_i = \delta_{i,3}$.

Rewriting the scalar kinetic term of the Lagrangian by splitting the Higgs field into the vev and the 
fluctuation part, we obtain
\begin{align}
\begin{split}
 (D_{\m} \phi)^{\dagger} (D^{\m} \phi) &= \pdm \varphi^{\dagger} \pd^{\m} \varphi + \frac{g^{2}v^{2}}{2}\, n^{\dagger} T^{a}T^{b} n \,  A^{\mu\,a}A_{\mu}^b \\ 
 &\quad + \sqrt{2}gv\, \im(n^{\dagger} T^{a}\pd^{\mu} \varphi)A_{\mu}^{a}  + \dots\;,
\end{split}
\label{eq:CovDev}
\end{align}
where we have the usual \cite{Bohm:2001yx} mass matrix for the gauge bosons in the first line and the mixing 
between the longitudinal parts of the gauge bosons and the Goldstone bosons in the second line.
Note that only those gauge bosons mix with the Goldstone bosons which acquire a mass, i.e., which correspond 
to the broken generators of the gauge group. These mixing terms are removed by the 't Hooft Landau gauge fixing 
condition \cite{Bohm:2001yx}. 

The mass matrix $(M^{2}_{A})^{{ab}}$ of the gauge bosons is already diagonal for our choice of $n$, and is given by,
\begin{align}
\begin{split}
 (\MA^{2})^{ab} &= \frac{g^{2}v^{2}}{2}\, n^{\dagger} \{T^{a},T^{b}\} n  \\
 &=  \frac{g^{2}v^{2}}{4} \diag\Big(0,0,0,1,1,1,\frac{4}{3}\Big)^{ab}\;.
 \end{split}
 \label{eq:gbmassmatrix}
\end{align}
Thus, we obtain $3$ massless gauge bosons, $4$ degenerated massive gauge bosons with mass $\mA = \frac{1}{2}gv$ 
and one with mass $\MA = \sqrt{4/3}~\mA$. 
Moreover, the elementary Higgs field has a mass $\mh^{2} = \lambda_\mathrm{c}^{2}v^{2}$, where $\lambda_\mathrm{c}$ is the 
four-Higgs coupling, i.e., the term $\frac{\lambda_\mathrm{c}}{2} (\phi^{\dagger}\phi)^{2}$ in the continuum setup. 

The situation is now that which is, in an abuse of language, usually called 'spontaneously broken' in case the system experiences the BEH effect.
The breaking pattern reads $\SU(3) \to \SU(2)$. 
With respect to the subgroup $\SU(2)$ the gauge bosons are in the adjoint representation (massless), 
a fundamental and an anti-fundamental representation (mass $\mA$) and a singlet representation 
(mass $\MA$), explaining their degeneracy pattern.

\subsection{Spectrum from the lattice}\label{ssec:lattice_pt}

Here we again study the same set of lattice parameters as before, as again the behavior is representative for all other cases.

\par
\begin{itemize}[leftmargin=*]
 \item {\bf Gauge-field propagators} 
\end{itemize}

Let us now focus on the propagator of the gauge bosons 
$D^c\big(p^2\big)$, $c=1,2,\dots,8$,  
defined in Equation \eqref{eq:gfprop}. 
The lattice momenta $p_\mu=2\pi k_\mu/L$ are along the links, and along all possible diagonals of the 
lattice, i.e., $(k,0,0,0)$, $(k,k,0,0)$, $(k,k,k,0)$, and $(k,k,k,k)$, 
$k=0,1,\dots,L/2$. 

\begin{figure}[tbh!]
\begin{center}
\hspace*{-0.5cm}
\includegraphics[width=0.5\textwidth]{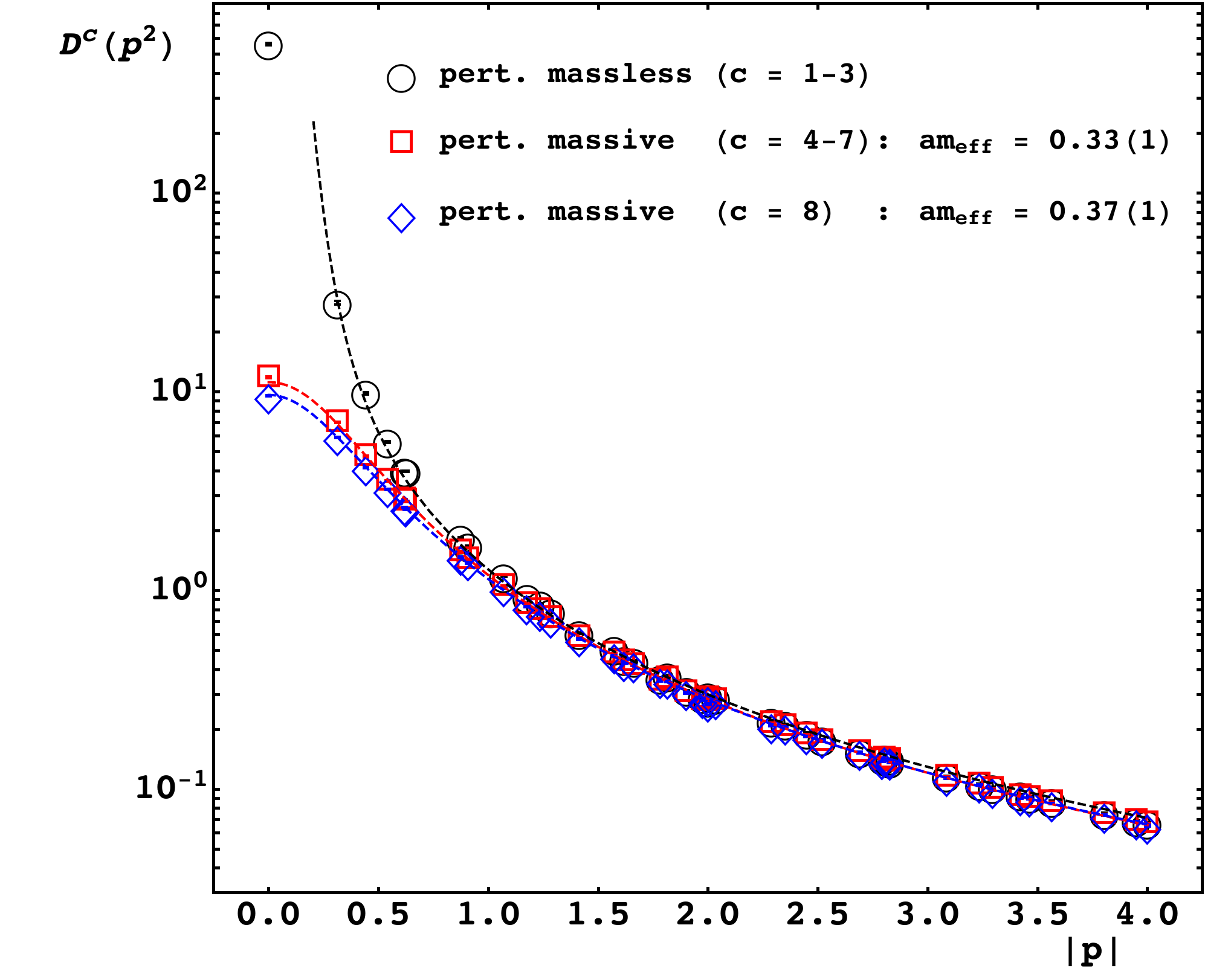}\\

\hspace*{-0.5cm}
\includegraphics[width=0.5\textwidth]{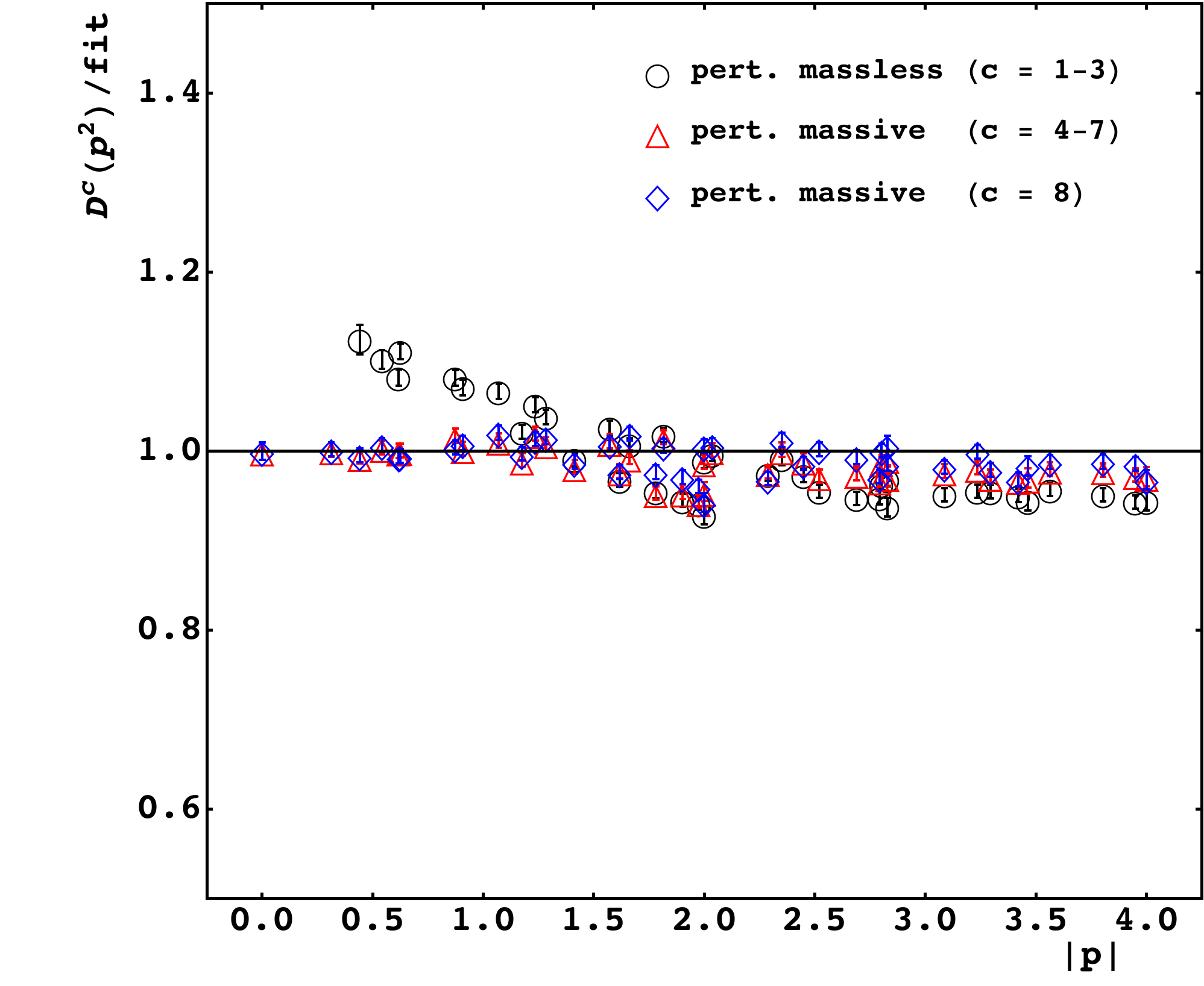}
\caption{{\textbf{Top:}} Plot of the gauge-boson propagators on a $20^4$ lattice for the $3$ perturbatively massless 
modes (black circles), the $4$ degenerate perturbatively massive modes (red squares), and the heaviest 
mode (blue diamonds) as a function of the absolute value of the physical momentum $|p|$. The dashed lines are results of the fits described in the main text.
{\textbf{Bottom:}} Here, the data points are divided by the corresponding fitted values as a function of 
$|p|$ and thus shows the qualities of the fits.
\label{fig:propV20}}
\end{center}
\end{figure}

In the top panel of Figure \ref{fig:propV20} we show the propagators, evaluated on a $20^4$  
lattice, of the perturbatively $3$ massless modes  ($c=1,2,3$, black circles), of the perturbatively $4$ 
degenerate massive modes ($c=4,5,6,7$, red squares), and the perturbatively heaviest mode 
($c=8$, blue diamonds). Those are plotted as a function of the absolute value of the physical momentum 
$|p| \equiv \sqrt{p_\mu~p_\mu}$. The degenerate 
modes are averaged over to improve the statistics.

The dashed lines are fits according to one-loop-inspired fit formulas
\begin{widetext}
\begin{align}
\begin{split}
D^c\big(p^2\big) &=\frac{Z}{p^2}\Bigg(\frac{A}{Vp^4} + \frac{p^2}
{\big(am^c\big)^2+~b^2~p^2~
\big(1+d^2\ln\frac{p^2+\Lambda^2}{\Lambda^2}\big)^\gamma}\Bigg)\;,\;\; c= 1,2,3\;, \\
D^c\big(p^2\big) &= \frac{Z}{p^2+\ln\big(p^2+b^2\big)^d+\big(am^c\big)^2}\;,\;\; c=4,5,6,7,8\;,
\end{split}
\label{eq:gbpropfitfunc}
\end{align}
\end{widetext}
where $Z$ are wave function renormalization constants and $am^c$ is the effective mass in lattice units. 
The first term in the first line is a pure finite-volume effect. The logarithmic corrections of leading loop-corrections are taken into account for both cases. A list of the fit parameters 
can be found in Table \ref{tab:fitvalsgv} in Appendix \ref{app:datafittables}. Also fits with the 
tree-level propagators have been performed but those fit functions did not resolve the UV-behavior well. 
Only for coarser lattices, i.e., larger $am_{1^{--}_0}$ masses, the tree-level form is a good fit ansatz 
at least for the massive modes. Those larger masses dominate, such that the logarithmic corrections 
only play a minor role, see \cite{Maas:2016ngo}.

The effective masses extracted for the different sectors are listed in the legend of the figure for the 
$20^4$ lattice. 
The fitted effective masses for the perturbatively massless modes are indeed very small and comparable to 
zero. This suggests a Coulomb-like behavior,  although corrections deep in the infrared may still alter this.
 
In the bottom panel of Figure \ref{fig:propV20} the ratio of data to the fit is shown.
For the massive modes the fit according to \eqref{eq:gbpropfitfunc} shows only small deviations from 
the data for the whole momentum range, whereas larger deviations for the massless modes are visible towards the infrared. The latter is accounted for in the fit as a finite-volume effect, which is to be expected for a massless mode.

\begin{figure}[tbh!]
\begin{center}
\hspace*{-2cm}
\includegraphics[width=0.51\textwidth]{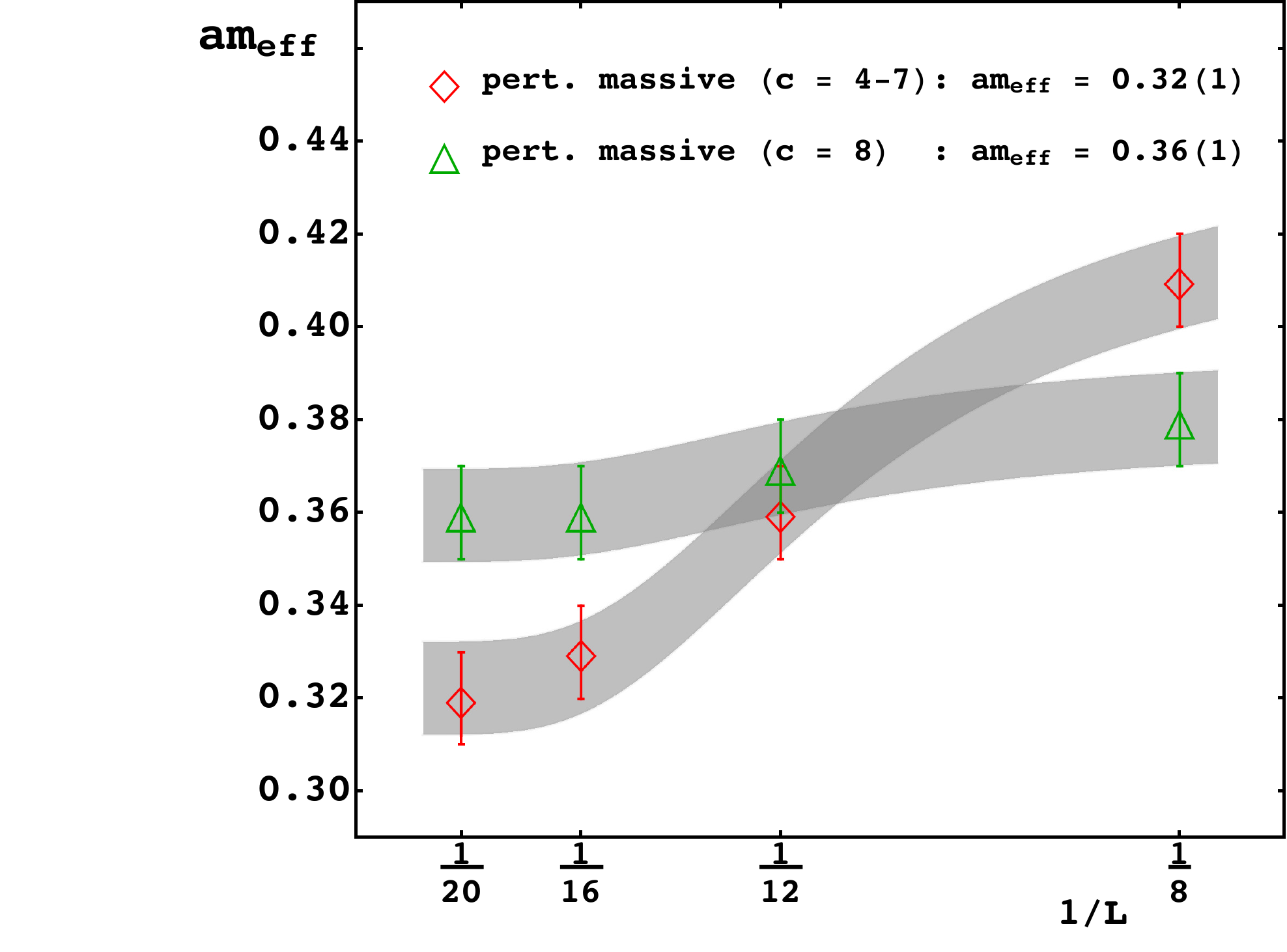}\\

\hspace*{-1.7cm}
\includegraphics[width=0.492\textwidth]{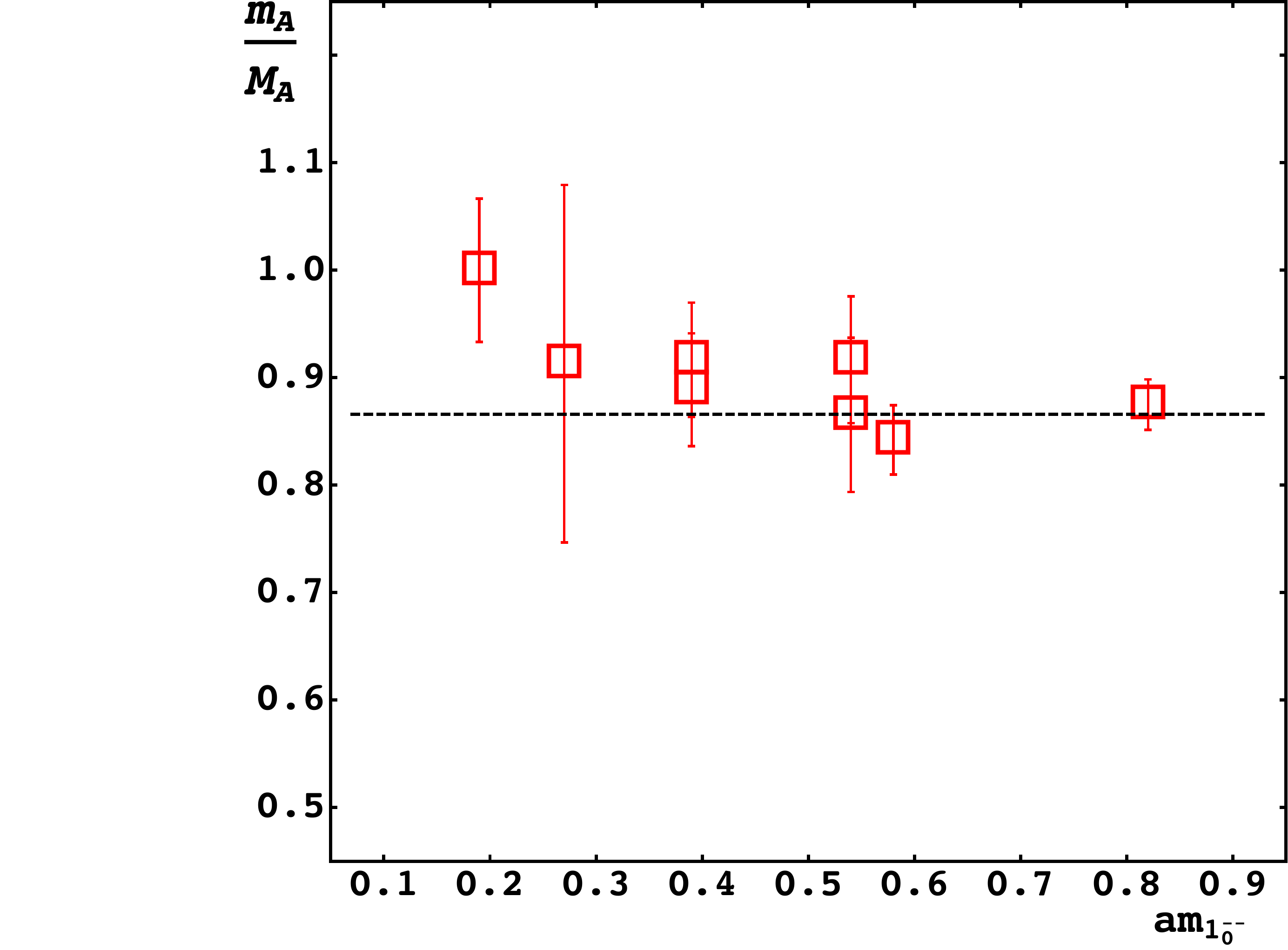}
\caption{{\bf{Top:}} Masses of the $4$ degenerate massive (red diamonds) and the heaviest (green triangles) 
gauge bosons as a function of the inverse lattice size $L$. The gray areas are the corresponding  
error bands obtained from a fit to $am + \alpha~\E^{-\gamma~ V}$, see Table \ref{tab:fitvolumedep}. 
{{\bf{Bottom:}} Ratio of the masses of the 4 degenerate lightest gauge bosons $\mA$ to the mass of the 
heaviest gauge boson $\MA$ as a function of all the lattice parameter sets we studied, see Table 
\ref{tab:numvals}. The dashed line is the prediction from tree-level perturbation theory, i.e., 
$\mA/\MA=\sqrt{3/4}$, see Equation \eqref{eq:gbmassmatrix}.
} 
\label{fig:propmassvol}}
\end{center}
\end{figure}

The extracted masses from the fits for $c=4,5,6,7$ (red diamonds) and $c=8$ (green triangles) are 
shown in the top panel of Figure \ref{fig:propmassvol} as a function of the inverse lattice size 
$L$. 
The extrapolated infinite volume values are $\mA = 0.32(1)$ for the $4$ degenerate massive and 
$\MA = 0.36(1)$ for the heaviest gauge boson, see the legend in the figure and 
Table \ref{tab:fitvolumedep}. The ratio of the lighter and heavier mass is $\mA/\MA = 0.89(5)$ which is in 
good agreement with the tree-level ratio of $\sqrt{3/4}\approx 0.87$, see Equation \eqref{eq:gbmassmatrix}. 
Together with the (almost) masslessness of the propagator in the unbroken sector this implies 
that the spectrum of the elementary fields coincides with the one expected from perturbation theory, 
especially of three massless and five massive states. 

In the bottom panel of Figure \ref{fig:propmassvol} we show the ratio of the masses of the  
$4$ degenerate gauge bosons to the mass of the heaviest gauge boson, i.e., $\mA/\MA$, as 
a function of the lattice mass of the singlet vector state $am_{1_0^{--}}$ obtained for different 
lattice parameter sets, see Table \ref{tab:numvals}. The dashed line is the prediction of 
tree-level perturbation theory, i.e., $\mA/\MA=\sqrt{3/4}$, see Equation \eqref{eq:gbmassmatrix}. 
Almost all the data points we studied are in good agreement with this prediction signaling that 
next-to-leading-order effects should only play a minor role.

\begin{figure*}[tbh!]
\begin{center}
\includegraphics[width=0.49\textwidth]{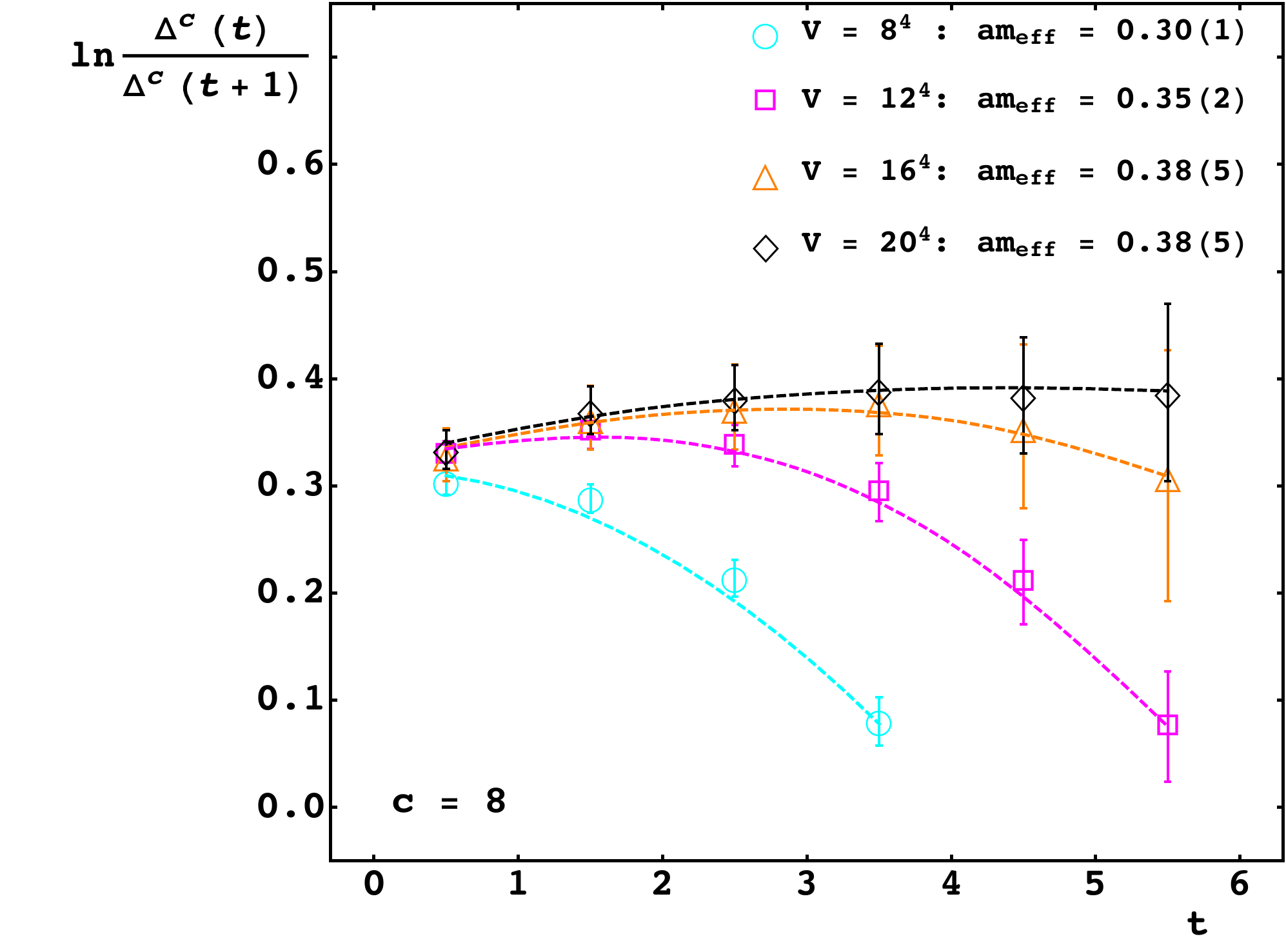}
\hspace*{0.15cm}
\includegraphics[width=0.49\textwidth]{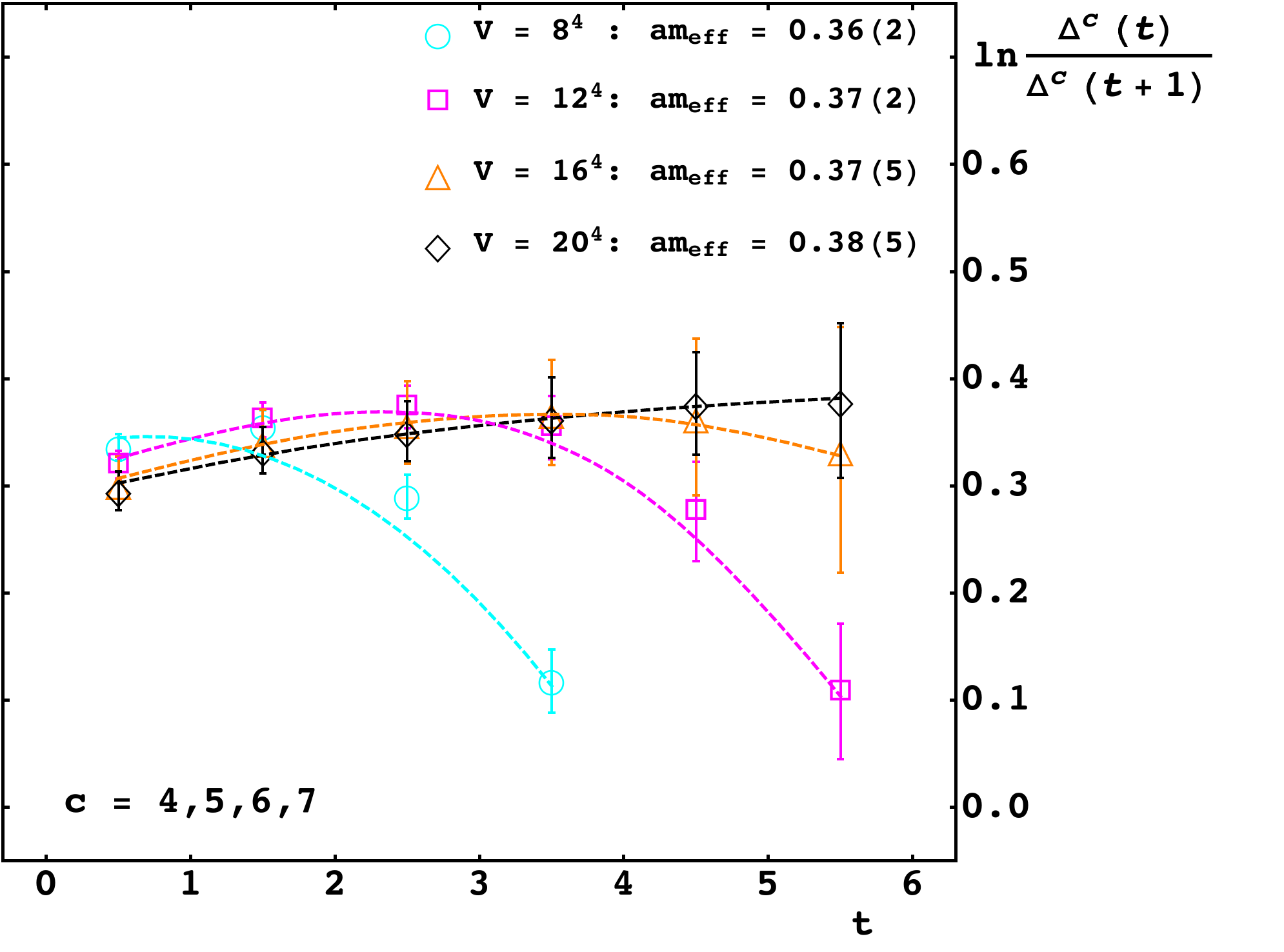}

\hspace*{-9.7cm}
\includegraphics[width=0.515\textwidth]{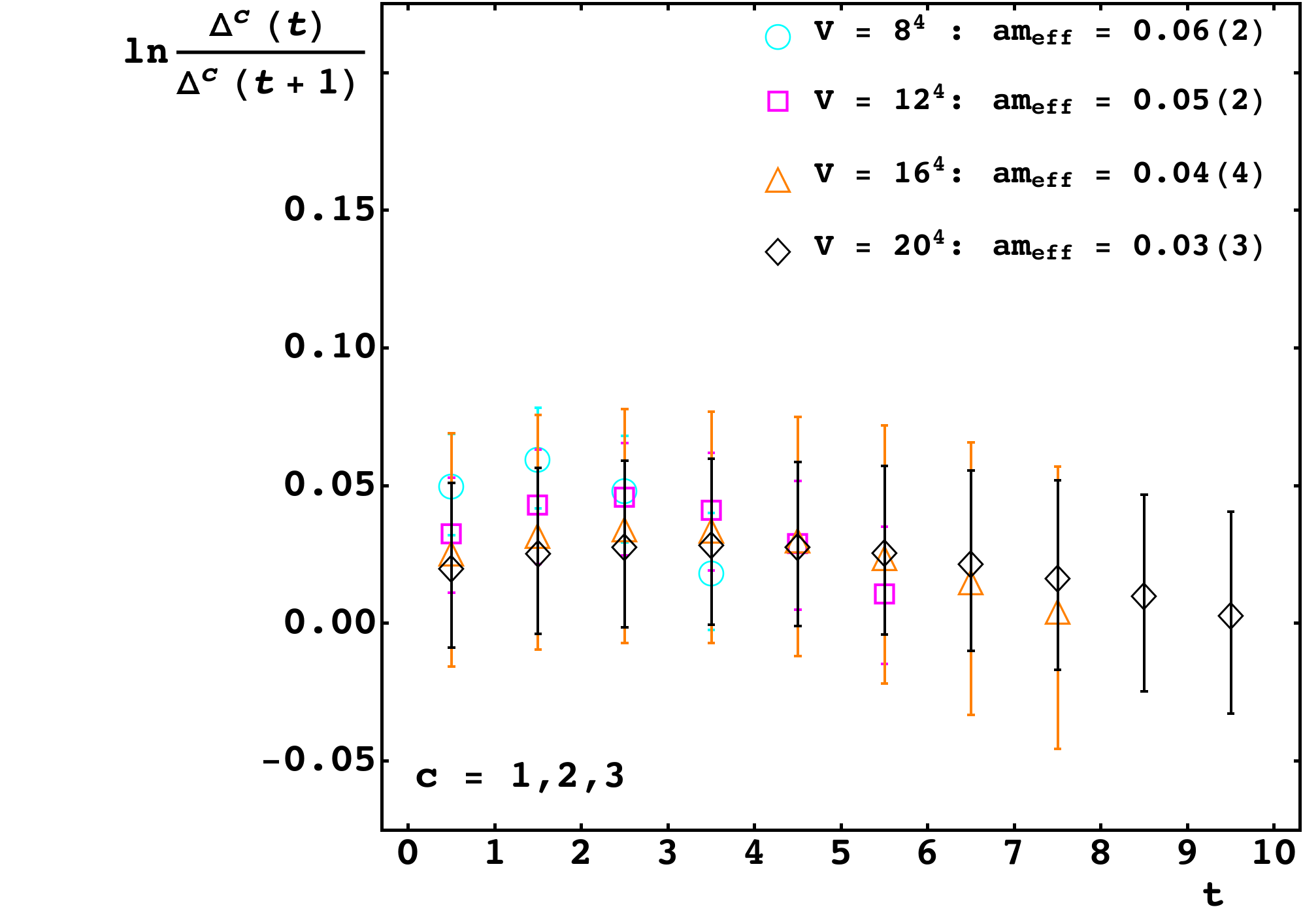}
\caption{Effective masses from the space-time correlators for the heaviest mode (top left panel), the $4$ 
degenerate modes (top right panel), and the $3$ massless modes (bottom left panel), for $8^4$, 
$12^4$, $16^4$, and $20^4$ lattices. The masses in the legends in each panel are obtained by 
taking the maximum value of the functions for each volume. The dashed lines in the top panels 
correspond to the space-time-transformed fit functions \eqref{eq:gbpropfitfunc}. 
\label{fig:schwingerV20}}
\end{center}
\end{figure*}

\par
\begin{itemize}[leftmargin=*]
 \item {\bf Space-time correlators / Schwinger functions} 
\end{itemize}
We also computed the space-time correlators or Schwinger functions $\Delta^c(t)$, $c=1,2,\dots,8$, as described in Section  
\ref{ssec:gv_tech}, along the lines of \cite{Maas:2014pba}. 
Again, Schwinger functions where degeneracies are expected, i.e., $c=1,2,3$, and $c=4,5,6,7$, 
are averaged over.
\footnote{These expected degenerate states overlap within error bars.} 
The resulting effective masses obtained from $\ln \Delta^c(t)/\Delta^c(t+1)$ for different lattice 
volumes are given in Figure \ref{fig:schwingerV20}. The errors are computed  from the propagators 
by the method of error propagation.
The top left panel shows the effective mass as a function of Euclidean time for the heaviest mode, the top 
right panel the effective mass of the $4$ degenerate massive modes, and the remaining panel shows 
a plot 
of the effective mass for the $3$ degenerate massless modes. Due to the relatively large error bars  
for the massive modes for $t>6$, we do not show those points here. 

The dashed lines in the top panel correspond to the space-time correlator obtained by inserting 
the corresponding fit functions \eqref{eq:gbpropfitfunc}
into the definition of the lattice space-time correlator \eqref{eq:latschwinger}. 

From the maximum values of the effective mass curves one can deduce the masses for 
each volume. The results are given in the legend of each plot. Of course, the errors are still too large and 
more statistics is needed to make a final statement. But the trend is clear and the obtained masses are 
in agreement within the large error bars with the ones obtained from the fits of the propagators 
with the functions defined in Equation \eqref{eq:gbpropfitfunc}.
Furthermore, the effective masses of the particles in the unbroken subsector ($c=1,2,3$) tend 
to zero for $V\to\infty$.

\begin{figure}[tbh!]
\begin{center}
\hspace*{-1cm}
\includegraphics[width=0.47\textwidth]{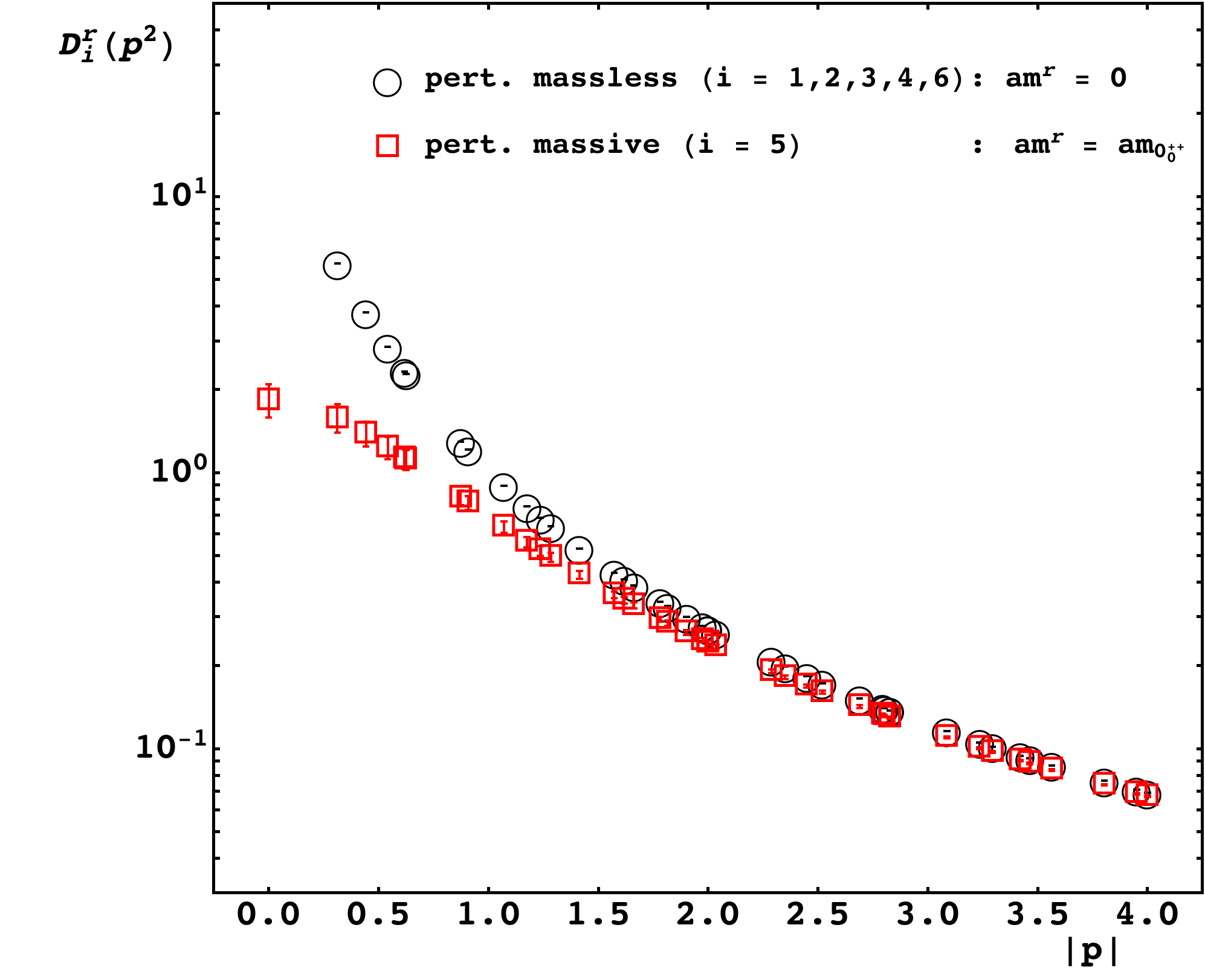}
\caption{Perturbatively massless and massive scalar propagators for a $20^4$ lattice and for 
$a\mu = 0.85$. The renormalized masses are $am^\mathrm{r} = 0$ for the (averaged) 
massless mode and $am^\mathrm{r} = am_{0^{++}_0}$ for the massive mode. 
\label{fig:scalarprop}}
\end{center}
\end{figure} 

\par
\begin{itemize}[leftmargin=*]
 \item {\bf Scalar-field propagator} 
\end{itemize}
In the scalar sector we computed the renormalized propagators of the real components of the scalar field 
$D_i^\mathrm{r}\big(p^2\big)$, $i=1,2,3,4,6$, as described in Section \ref{ssec:gv_tech}. 
We choose the arbitrary dimensionless renormalization scale to be 
$a\mu = 0.85$ for each propagator. Under the assumption that the pole scheme works 
\cite{Maas:2016edk,Maas:2017wzi} we set the renormalized masses to 
$am^\mathrm{r} = am_{0^{++}_0}$ for the perturbatively massive propagator ($i=5$) and to 
$am^\mathrm{r} = 0$ for the perturbatively massless propagators ($i=1,2,3,4,6$). 
The degenerate massless renormalized propagators are averaged over to increase the statistics.

Having determined both renormalization constants, the renormalized propagator $D_i^\mathrm{r}$ can 
be computed. The result is shown in Figure \ref{fig:scalarprop}, where again both, the 
perturbatively massless (black circles) and massive (red squares) modes are shown. 
Both propagators show the expected behaviors, namely the ones of a massless and a massive propagator.

In order to extract the effective masses, the space-time correlators need to be computed. Unfortunately, 
the statistics is too low at this point and thus the error bars too large to extract the effective mass from 
the Schwinger functions. Therefore, no results on this are presented here. However, in \cite{Maas:2013aia} 
\footnote{An $\SU(2)$ gauge theory with a fundamental scalar was used there.} 
it was found that under similar circumstances the effective mass agreed reasonably with the renormalized mass, supporting that the employed scheme acts like a pole scheme. Still, this will require further scrutiny, as this is not necessarily always the case \cite{Maas:2016edk}.

\begin{figure}[tbh!]
\begin{center}
\hspace*{-1cm}
\includegraphics[width=0.465\textwidth]{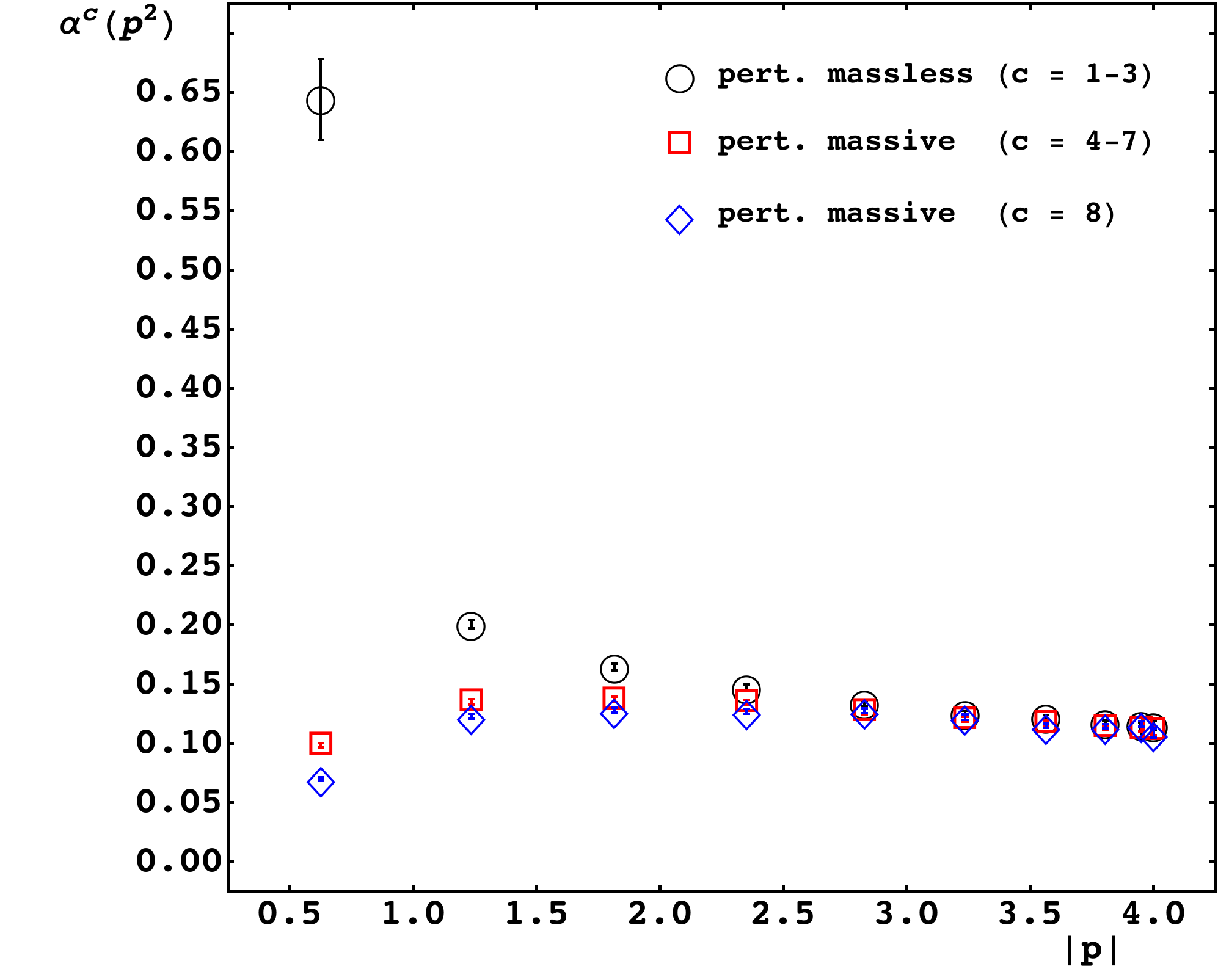}
\caption{Renormalized running coupling for the three different sectors. The renormalization 
has been performed such that the couplings agree with the perturbative one for large momenta, see 
\cite{Maas:2013aia} for details. The lattice couplings are in this case 
$\beta = 6.85535$, 
$\kappa = 0.456074$, and 
$\lambda = 2.3416$.
\label{fig:running}}
\end{center}
\end{figure} 

\par
\begin{itemize}[leftmargin=*]
 \item {\bf Running gauge coupling and the ghosts} 
\end{itemize}
The ghost propagators are all very close to the one of a massless particle, and thus close to perturbation theory. There is only a little deviation towards the infrared, which is larger the smaller the associated gauge boson mass is. As a consequence, the running coupling is mainly dominated by the gauge-boson propagator.

Thus, we show here only the latter, the renormalized running gauge coupling $\alpha^c\big(p^2\big)$ 
for the different sectors in Figure \ref{fig:running}: 
The perturbatively massless sector (black circles), the sector with the 
$4$ degenerate massive modes (red squares), and the sector with the heaviest mode (blue diamonds). The coupling to the massive modes show the typical behavior already seen for the 
$\SU(2)$ case \cite{Maas:2013aia}. The coupling of the massless modes is infrared (mildly) enhanced, and does not (yet) saturate. Still, because all propagators are rather close to the perturbative ones, so are the gauge couplings. In particular, all unify, implementing the simple picture of a grand-unified theory, in the ultraviolet. Only at small momenta the BEH effect induces the differences. The suppression of the massive couplings can be interpreted as a decoupling of the massive modes from the massless dynamics. However, this statement is only true for the gauge sector, as the gauge-invariant physics of Section \ref{sec:spectrum} does not show any sign of this separation.

The couplings stay small throughout the whole momentum range, signaling that leading order GIPT should already be quite reliable. This is also supported by the fact that the propagators can be fitted well with one-loop expressions.

\section{Test of the Fr\"ohlich-Morchio-Strocchi mechanism}\label{sec:testfms}
Thus, this section is dedicated to test GIPT and the underlying FMS mechanism 
\cite{Frohlich:1980gj,Frohlich:1981yi}. To this end, we first recapitulate the predictions of GIPT for this theory. The generalization of these predictions to general $\SU(N)$ gauge groups 
can be found in \cite{Maas:2017xzh}.

\subsection{Predictions from gauge-invariant perturbation theory}\label{subsec:predictions} 

The gauge-invariant, and thus experimentally observable, spectrum consists of states which are 
either singlets or non-singlets with respect to the custodial $\U(1)$ group, see Section 
\ref{ssec:gi_tech}.

\par
\begin{itemize}[leftmargin=*]
 \item {\bf $\U(1)$-singlet states} 
\end{itemize}
Let us start the discussion with the $\U(1)$-singlet states:  A gauge-invariant composite operator  describing a scalar, 
positive (charge-) parity boson, i.e., $J_{\U(1)}^{PC}=0^{++}_0$, is 
$O^{0_0^{++}}(x)=\big(\phi^\dagger\phi\big)(x)$. We apply the FMS mechanism and expand 
the correlator first in Higgs fluctuations, using Equation \eqref{eq:SplitHiggsF}, and then the 
resulting propagators to leading order in standard perturbation theory, yielding \cite{Maas:2017xzh}
\begin{align}
\begin{split}
\big\langle O^{0_0^{++}}(x)O^{0_0^{++}}(y)^\dagger\big\rangle &= \frac{v^4}{2} 
+v^2 \big\langle h(x)h(y)\big\rangle_\mathrm{tl} \\
&+\big\langle h(x)h(y)\big\rangle_\mathrm{tl}^2 +\cdots \;,
\end{split}
\label{eq:fmsscalar}
\end{align}
where 'tl' means 'tree level'. Here, the Higgs field $h$ is identified with $\phi_5$, see Equation 
\eqref{eq:scalarprop}. The second term on the right-hand side of Equation \eqref{eq:fmsscalar} 
describes the propagation of a single elementary Higgs boson and the third term describes 
two non-interacting Higgs bosons propagating both from $x$ to $y$. 
Comparing poles on both sides of Equation \eqref{eq:fmsscalar} predicts the mass of the 
left-hand side, and thus of the observable particle. This scalar boson should therefore have a 
mass equal to the mass of the elementary Higgs $\mh$. Also, a next state should 
exist in this quantum number channel which is a scattering state of twice this mass.

Next, consider a singlet vector operator 
$O_\mu^{1^{--}_0}(x)=\I\big(\phi^\dagger D_\mu\phi\big)(x)$. The same expansion yields 
\cite{Maas:2017xzh}
\begin{align}
\begin{split}
\big\langle O_\mu^{1^{--}_0}(x)O_\nu^{1^{--}_0}(y)^\dagger\big\rangle &= 
\frac{v^4g^2}{12}\big\langle A_\mu(x)^8 A_\nu(y)^8\big\rangle_\mathrm{tl} + \cdots\;.
\end{split}
\label{eq:fmsvector}
\end{align}
The poles of the right-hand side are at the mass $\MA$ of the heaviest gauge boson $A^8_\mu$. 

All the remaining states which can be constructed from the fields expand to trivial scattering states, e.g., gaugeball states, and thus do not give rise to stable particles, see also Table \ref{tab:interpol}. 

\par
\begin{itemize}[leftmargin=*]
 \item {\bf $\U(1)$-non-singlet states} 
\end{itemize}
Let us now focus on states with open $\U(1)$ quantum numbers. Since the corresponding charge 
is conserved, the lightest such state is absolutely stable. The scalar  as well as the vector non-singlet 
states, $O^{0^{++}_{1}}(x)$ and $O^{1^{--}_{1}}_\mu(x)$, are defined at the end 
of Table \ref{tab:interpol}. The lattice versions are given in Equations \eqref{eq:openU1_0pp} and 
\eqref{eq:openU1_1mm}. Applying the FMS mechanism and employing a tree-level analysis to the 
bound state correlators of both operators yields, after a cumbersome calculation \cite{Maas:2017xzh}, a ground state 
mass of $2\mA$ for both quantum number channels. This arises
as in leading order a product of propagators of one of the massless elementary gauge boson as well as two gauge bosons 
with mass $\mA$ contribute to this state. We expect also a higher order excitation with
mass $2\mA+\MA$ in both channels. There exists, of course, an
anti-particle of the same mass but opposite $\U(1)$ charge for both channels as well.

\begin{figure*}[tbh!]
\begin{center}
\begin{minipage}{0.49\textwidth}
\includegraphics[width=0.93\textwidth]{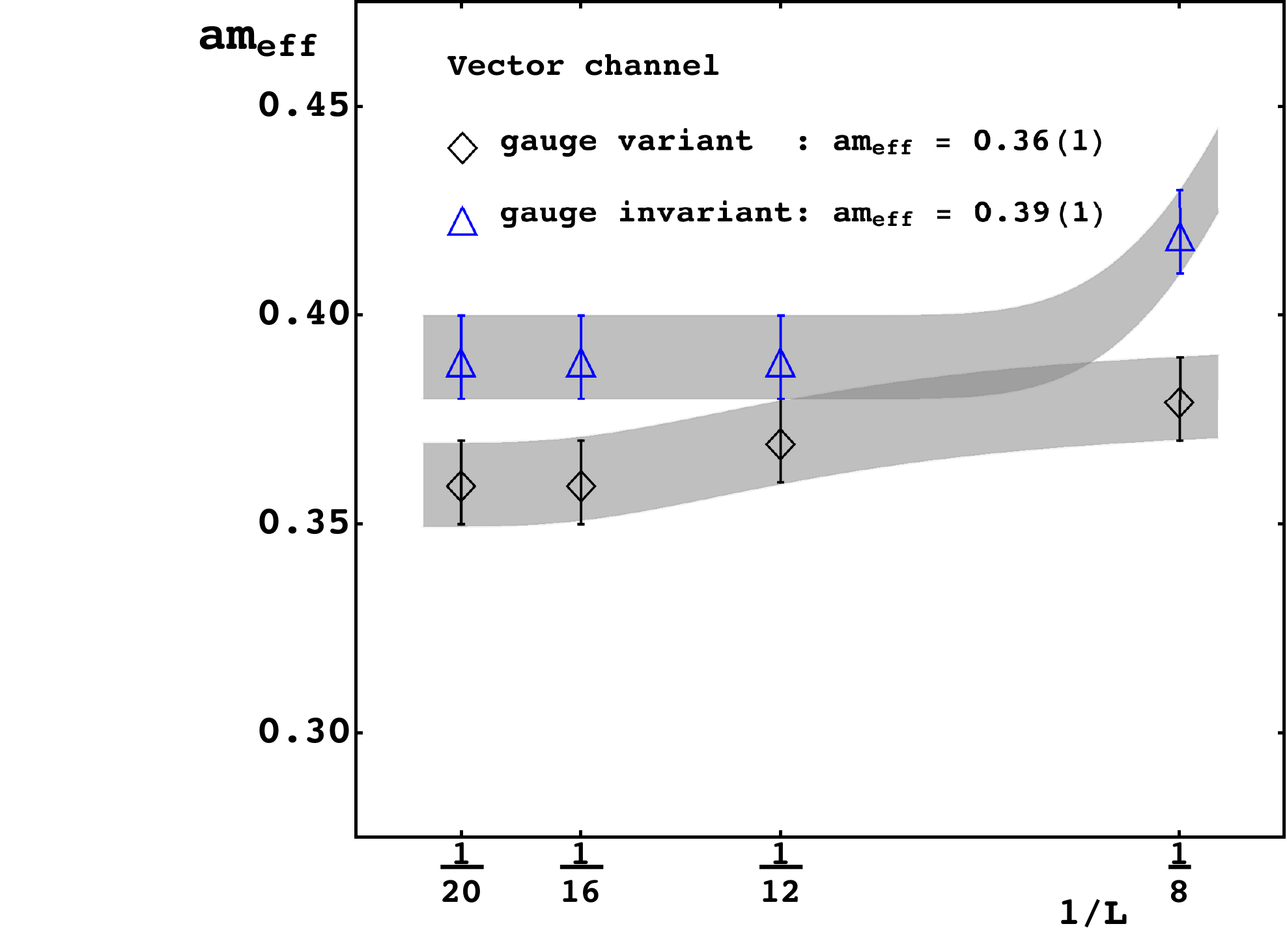}
\end{minipage}
\begin{minipage}{0.49\textwidth}

\vspace*{-0.21cm}
\includegraphics[width=0.897\textwidth]{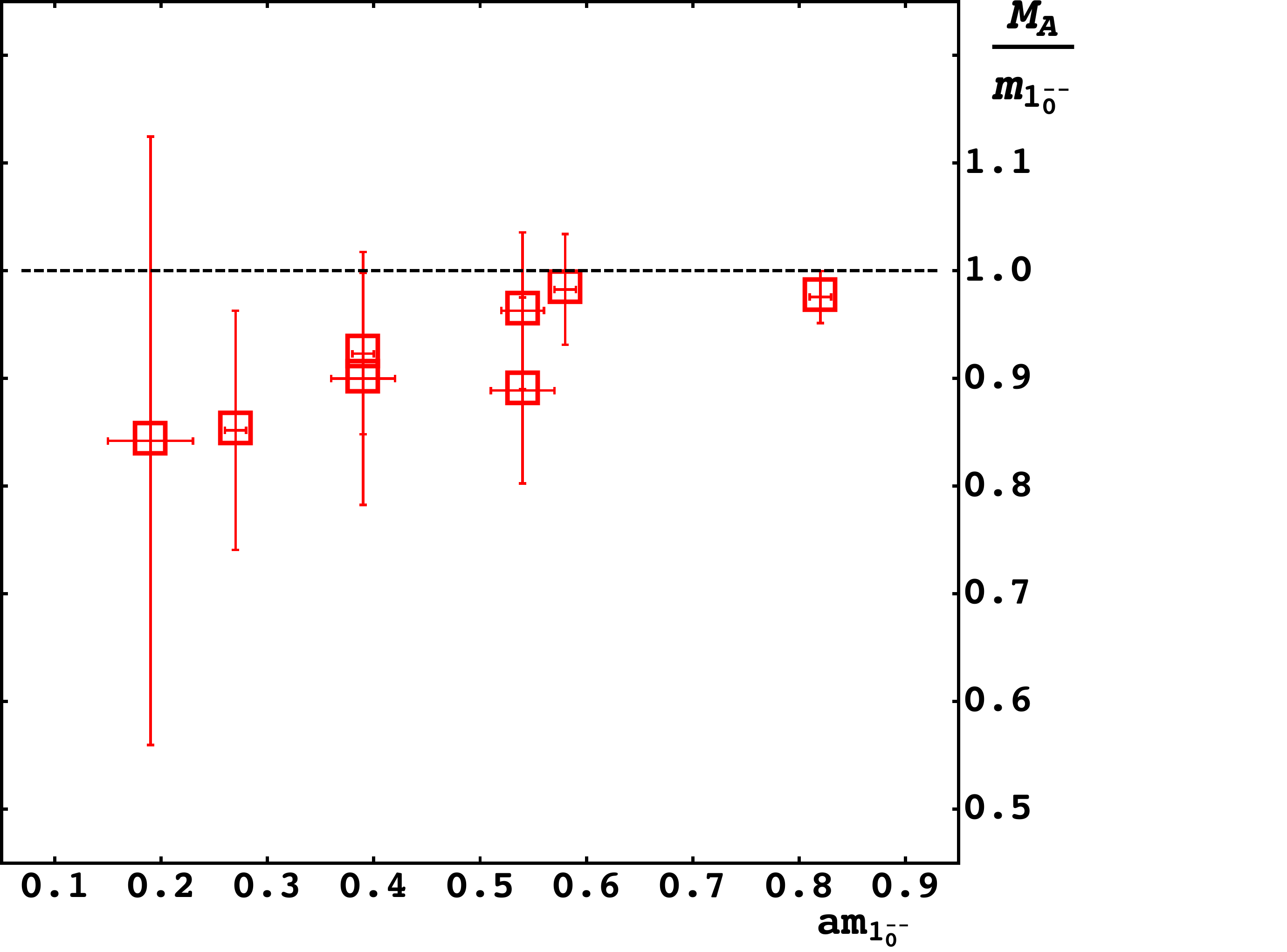}
\end{minipage}
\caption{{\bf{Left:}} Effective masses of the singlet vector channel obtained from the gauge-variant propagator 
and from a variational analysis of gauge-invariant operators as a function of the inverse lattice size. 
The discrepancy of the infinite volume extrapolated values is discussed in the main text. 
{\bf{Right:}} Ratio of the heaviest vector boson mass $\MA$ and the ground state mass in the 
$1^{--}_0$ channel, $m_{1^{--}_0}$, as a function of the lattice mass $am_{1^{--}_0}$. 
The points are obtained from simulations at points in the phase diagram in the 
HLR, see Table \ref{tab:numvals}. The dashed line is the GIPT prediction.
\label{fig:comparevol}}
\end{center}
\end{figure*}

\par
\begin{itemize}[leftmargin=*]
 \item {\bf Summary} 
\end{itemize}

The prediction for the physical (gauge-invariant) spectrum obtained from leading-order gauge-invariant perturbation as 
well as the gauge-variant spectrum from standard perturbation theory are summarized in Table \ref{tab:su3}.

\begin{table}[tbh!]
\begin{center}
\caption{
{\textbf{Left}}: Gauge-variant spectrum of an $\SU(3)$ gauge theory with a single scalar field in 
the fundamental representation. We set the direction $n$ of the vev to $n_i=\delta_{i,3}$. 
{\textbf{Right}}: Prediction of the gauge-invariant (physical) spectrum of the theory using leading-order gauge-invariant perturbation theory. Here $\mh$ denotes the mass of the 
elementary Higgs field, $\MA$ is the mass of the heaviest elementary gauge boson and $\mA$ the mass 
of the degenerated lighter massive gauge bosons. We assign a custodial $\U(1)$ charge of $1/3$ to the 
scalar field $\phi$. The column 'N.l. state' (next-level state) lists the masses of possible additional bound states or 
resonances. Whether these states are 
indeed bound states or resonances or only nontrivial scattering states can not be decided here. 
Trivial scattering states are ignored.}
{
\begin{tabular}{@{}c|ccc|ccccc@{}}
\multicolumn{4}{r}{\quad elementary spectrum} & \multicolumn{4}{c}{\qquad gauge-invariant spectrum} \cr

\hline

\hline

$J^P$ &  Field  &  Mass  &  Deg.  &  $\U(1)$  &  Op. & Mass & N.-l. state &  Deg.  \cr
\hline

$0^+$ & $h$ & $\mh$ & $1$ & $0$ & $O^{0^{++}_0}$ & $\mh$ & - &$1$\cr
$\quad$ & $\quad$ & $\quad$ & $\quad$ & $\pm 1$ & $O^{0^{++}_{\pm 1}}$ & 
$2\mA$ & $2\mA + \MA$ & $1/\bar{1}$\cr
\hline
$1^-$ & $A_\mu^{1,2,3}$ & $0$ & $3$ & $0$ & $O_\mu^{1^{--}_{0}}$ & 
$\MA$ & - & $1$\cr 
 & $A_\mu^{4,\dots,7}$ & $\mA$ & $4$ & $\pm 1$ & $O_\mu^{1^{--}_{\pm 1}}$ & $2\mA$ 
& $2\mA + \MA$ & $1/\bar{1}$ \cr 
& $A_\mu^{8}$ & $\MA$ & $1$  &  &  & &  & \cr

\hline

\hline

\end{tabular}
}
\label{tab:su3}
\end{center}
\end{table}

\subsection{Comparison between the spectra}\label{subsec:compare}
We focus again on the lattice parameter set $\beta = 6.85535$,  $\kappa = 0.456074$, and 
$\lambda = 2.3416$ for our investigations. 

From the findings of the previous subsection and the predictions of the gauge-invariant physical 
spectrum, we are able to check the predictions of gauge-invariant perturbation theory utilizing the FMS mechanism explicitly.

In the $0^{++}_0$ channel we found one stable state with a mass of $am_{0^{++}_0}=0.68(2)$ which 
is the one that we expect to find from the Schwinger function of the renormalized scalar propagator. 
While consistent, the results are still too strongly affected by statistical errors to provide an unambiguous conclusion. 
The remaining states in this channel are high up in the spectrum and consistent with trivial scattering states.

In the $1^{--}_0$ channel we extracted a ground state lattice mass of $am_{1^{--}_0}=0.39(1)$. 
The mass extracted from the heaviest gauge boson is $a\MA = 0.36(1)$. The volume dependence 
of these masses is shown on the left-hand side of Figure \ref{fig:comparevol}. They are already in pretty good agreement.

However, at leading order, those masses should be equal, but there remains a bit more than a $1\sigma$ discrepancy between 
them. There are several explanations for that: First, the prediction
relies on the smallness of the Higgs fluctuations and on the applicability of standard perturbation theory. Higher-order effects or genuine 
 nonperturbative effects could explain this deviation.

Another possibility is that this discrepancy of the masses could stem from finite volume and 
discretization effects. We observe that the larger the mass of the lightest state is\footnote{Which is the singlet vector state mass in our case.}, 
i.e., the larger the lattice spacing $a$ is and thus the larger the physical volume is, 
the better is the agreement with the vector boson mass, see right-hand side of 
Figure \ref{fig:comparevol} and \cite{Maas:2016ngo}.

Lastly, also the infinite volume extrapolation we used, see Table \ref{tab:fitvolumedep}, 
does not take into account that the broken sector of the theory still interacts weakly with 
the unbroken sector, i.e., the  sector  of  massless particles.
The extrapolation we used does not take this into account and more sophisticated fitting 
procedures could change the results slightly.

In view of these additional systematic uncertainties, the results is already quite well in agreement with the prediction.

At the same time, the right-hand side of Figure \ref{fig:comparevol} shows that the result is not a coincidence, and that the agreement is generic. It contains all our results in the HLR, see Table \ref{tab:numvals}. In this plot the dimensionless ratio of the vector boson mass $\MA$ to 
the singlet vector mass $m_{1^{--}_0}$ as a function of the lattice singlet vector mass $am_{1^{--}_0}$  
is shown. The dashed line at $\MA/m_{1^{--}_0} = 1$ reflects  agreement with the
prediction in the vector channel. All in all, good agreement to the GIPT prediction is found, in particular for larger 
physical volumes, corresponding due to the fixed lattice volumes to larger lattice masses of the singlet vector state and thus larger $a$. 

The open $\U(1)$ quantum number channels, i.e., the $0^{++}_{\pm 1}$ and $1^{--}_{\pm 1}$ 
channels, still suffer from low statistics. The extracted ground state of the vector state would be 
consistent with both, the ground state, $2a\mA = 0.64(2)$, and the predicted next-level state, 
$a\mA + a\MA = 1.00(3)$ since 
$am_{1^{--}_{\pm 1}} \approx 0.8(2)$. The scalar state has a mass of 
$am_{0^{++}_{\pm 1}}\approx 2.0(1)$ and 
is relatively high up in the spectrum. Thus, we neither can confirm nor disprove the FMS prediction in these  
channels firmly, even though the results are consistent.

Nevertheless, the mass for the vector is already significantly smaller than one would naively expect from a simple constituent model or from ordinary perturbation theory. These would place the mass at at least $3m_h\approx 2.0(1)$. This is especially important, as the lightest such particle is stable and in a realistic theory could serve as a dark matter candidate.

\begin{figure}[tbh!]
\begin{center}
\includegraphics[width=0.49\textwidth]{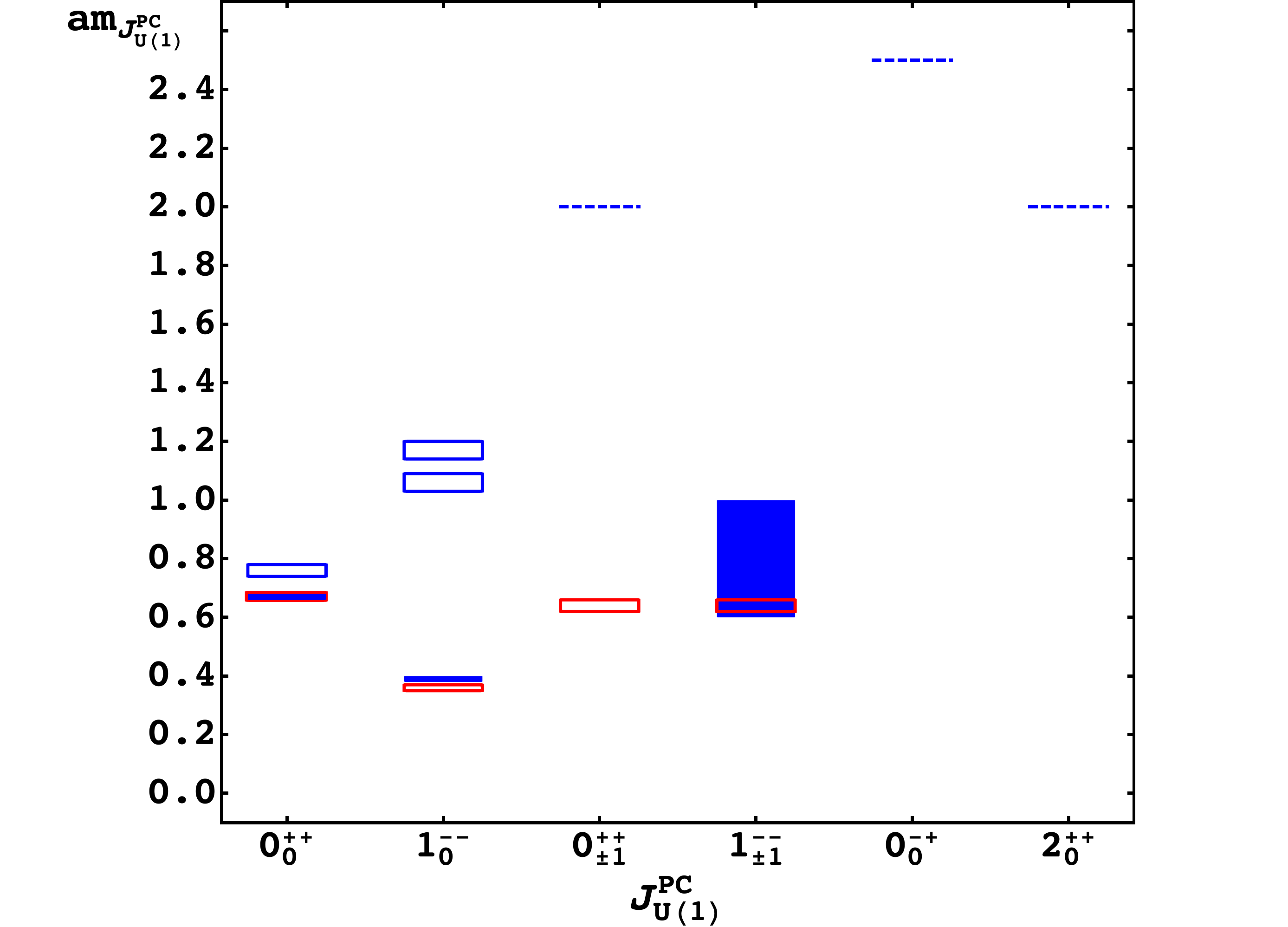}
\caption{Physical (gauge-invariant) spectrum of the theory (blue boxes) compared to the 
predictions from the FMS mechanism to leading order (red boxes) for the lattice parameter 
set $\beta=6.85535$, $\kappa=0.456074$, $\lambda=2.3416$. 
\label{fig:spectrum_gv}}
\end{center}
\end{figure}

Summarizing, in Figure \ref{fig:spectrum_gv} we show the physical spectrum of the theory for different 
quantum number channels and compare it to the predictions from GIPT. 
The filled, blue boxes correspond to the ground states, the empty, blue boxes are the 
elastic thresholds for the scalar and vector singlet channels as discussed above, and the dashed, blue  
lines are the approximate ground state masses of the $0_{\pm 1}^{++}$, $0_{0}^{-+}$, and
$2_{0}^{++}$. The red boxes are the predictions of leading-order gauge-invariant perturbation theory for the ground states. The overall agreement shows that the spectrum is, even at leading order, well predicted.

\section{Summary and conclusions}\label{sec:conc}

Summarizing, we have presented a detailed study of the spectroscopy of an 
$\SU(3)$ gauge theory with a fundamental scalar, a toy model for grand-unified theories. 

To this end, we determined the spectrum using leading-order standard perturbation theory, leading-order gauge-invariant perturbation theory utilizing the FMS mechanism \cite{Maas:2017wzi}, and in a full non-perturbative lattice investigation. As in the general case \cite{Maas:2017xzh} the predictions form standard perturbation theory and gauge-invariant perturbation theory disagree qualitatively. As was already seen in the exploratory study \cite{Maas:2016ngo}, it is found that the predictions of gauge-invariant perturbation theory describe the spectrum obtained from the lattice not only qualitatively, but within some $10$\% quantitatively, wherever the lattice results were statistically sufficiently significant. Given the remaining systematic uncertainties from the lattice and the fact that the analytical computations were done at leading order, this is quite an impressive agreement. In particular, this worked even in cases where the lowest order in the analytical calculation vanished surprisingly good.

At the same time the results of standard perturbation theory are even qualitatively off, especially the theory shows strong evidence for a mass gap of the order the GUT scale. This included also the pure gauge sector, which could be argued to have light gaugeballs in standard approaches \cite{Bohm:2001yx,Langacker:1980js}.

This strongly suggest that only gauge-invariant perturbation theory is adequate in describing the theory, and its dynamics, analytically. This is in agreement with all other comparative studies, see \cite{Maas:2017wzi} for a review of those. However, this work is the first systematic study over a range of parameters and in several channels simultaneously for a theory which was expected to show qualitative disagreement to standard methods.

This strongly suggest that the physics picture behind the FMS mechanism is the correct one to describe theories with a BEH effect. Moreover, this implies that gauge-invariant perturbation theory is the analytical toolkit to work with for these theories, and that BSM predictions should be performed using it. Of course, as always, this is evidence, and further investigations will be necessary for a firm establishing of these conclusions. But given the additional effort need for gauge-invariant perturbation theory in comparison to standard perturbation theory, there is little reason not to use it to make BSM predictions.

\section*{Acknowledgments}
PT has been supported by the FWF doctoral school W1203-N16. 
The computational results presented have been obtained using the Vienna Scientific Cluster 
(VSC), the HPC center at the University of Graz and the Graz University of Technology.

\appendix
\section{Lattice parameter sets and fit tables}\label{app:datafittables}
In this appendix, we collect all the parameter sets of the phase diagram points were we performed 
simulations for different lattice sizes in order to obtain data for spectroscopy and for propagators 
of gauge-variant fields.
Additionally, we list the fit parameters which we obtained and which are used in the figures shown in 
Section \ref{sec:spectrum} and \ref{sec:testfms} .

\subsection{Lattice parameters sets and some observables}\label{app:numvals}

Here, we provide the numerical values for the ground state energy levels in the vector singlet channel in 
lattice units $am_{1^{--}_0}$, the masses of the $4$ degenerate gauge bosons $\mA$ and the 
heaviest gauge boson mass $a\MA$ also in lattice units, the average plaquette defined for gauge group 
$\SU(3)$ as
\begin{align}
U_\mathrm{P} &= \frac{1}{18~V}\sum_x\sum_{\mu<\nu}\mathrm{Re}~\tr\big[U_{\mu\nu}(x)\big]\;,
\end{align}
where $U_{\mu\nu}(x)$ is defined in Equation \eqref{eq:plaq}, 
for several values of the lattice couplings $\beta$, $\kappa$ and $\lambda$. We also provide the 
expectation values of $\bar{\phi}^2$ defined in Equation \eqref{eq:caudy} as well as for the length of 
the scalar field $|\phi|$ given by
\begin{align}
|\phi| &= \frac{1}{V}\sum_x\sqrt{\phi(x)^\dagger\phi(x)}\;.
\end{align}
Only such values are shown for which a 
BEH effect was found.

\begin{table*}[tbhp!]
\begin{center}
\caption{Numerical values of the ground state energy level in the $1^{--}_0$ channel, the masses of the 
gauge-variant vector states  $a\MA$ and $a\mA$, the plaquette expectation 
value $\langle U_\mathrm{P}\rangle$, $\langle \bar{\phi}^2\rangle$, and $\langle |\phi|\rangle$, for various values of $\beta$, $\kappa$, and $\lambda$ in the 
Higgs-like region of the phase diagram. We also list all the lattice volumes we studied for these values and how 
many (gauge-fixed) configurations we used.}
\begin{tabular}{@{}ccclcccccc@{}}

\hline

\hline

$\beta$ & $\kappa$ & $\lambda$ & $(V,\text{\# configs},\text{\# fixed configs})$ & $am_{1^{--}_0}$ & $a\MA$ & $a\mA$ & $\langle U_\mathrm{P}\rangle$ & $\langle\bar{\phi}^2\rangle$ & $\langle|\phi|\rangle$ \cr 

\hline

\multicolumn{ 1}{c}{$5.798500$} & \multicolumn{ 1}{c}{$0.419035$} & \multicolumn{ 1}{c}{$1.259900$} & $(\phantom{0}8^4,11200\phantom{0},\phantom{0}112)$, $(12^4,19500\phantom{0},1170)$,  & \multicolumn{ 1}{c}{$0.54(3)$} & \multicolumn{ 1}{c}{$0.48(2)$} &  \multicolumn{ 1}{c}{$0.44(1)$} & \multicolumn{ 1}{c}{$0.5832(1)$} & \multicolumn{ 1}{c}{$0.3085(3)$} & \multicolumn{ 1}{c}{$1.6274(3)$} \cr 
\multicolumn{ 1}{c}{} & \multicolumn{ 1}{c}{} & \multicolumn{ 1}{c}{} & $(16^4,7700\phantom{00},\phantom{0}462)$ & \multicolumn{ 1}{c}{} & \multicolumn{ 1}{c}{} & \multicolumn{ 1}{c}{} & \multicolumn{ 1}{c}{} & \multicolumn{ 1}{c}{} \cr 
\multicolumn{ 1}{c}{$6.855350$} & \multicolumn{ 1}{c}{$0.456074$} & \multicolumn{ 1}{c}{$2.341600$} & $(\phantom{0}8^4,320000,\phantom{0}250)$, $(12^4,240000,1000)$,  & \multicolumn{ 1}{c}{$0.39(1)$} & \multicolumn{ 1}{c}{$0.36(1)$} &  \multicolumn{ 1}{c}{$0.32(1)$} &\multicolumn{ 1}{c}{$0.6674(1)$} & \multicolumn{ 1}{c}{$0.2761(5)$} & \multicolumn{ 1}{c}{$1.39780(7)$} \cr 
\multicolumn{ 1}{c}{} & \multicolumn{ 1}{c}{} & \multicolumn{ 1}{c}{} & $(16^4,120000,3000)$, $(20^4,190000,5473)$ & \multicolumn{ 1}{c}{} & \multicolumn{ 1}{c}{} & \multicolumn{ 1}{c}{} & \multicolumn{ 1}{c}{} & \multicolumn{ 1}{c}{} \cr 
\multicolumn{ 1}{c}{$7.912200$} & \multicolumn{ 1}{c}{$0.493113$} & \multicolumn{ 1}{c}{$3.423300$} & $(\phantom{0}8^4,11200\phantom{0},\phantom{0}112)$, $(12^4,24950\phantom{0},1479)$,  & \multicolumn{ 1}{c}{$0.39(1)$} & \multicolumn{ 1}{c}{$0.36(1)$} & \multicolumn{ 1}{c}{$0.33(1)$} & \multicolumn{ 1}{c}{$0.7204(1)$} & \multicolumn{ 1}{c}{$0.3291(1)$} & \multicolumn{ 1}{c}{$1.3166(1)$} \cr
\multicolumn{ 1}{c}{} & \multicolumn{ 1}{c}{} & \multicolumn{ 1}{c}{} & $(16^4,25000\phantom{0},1500)$ & \multicolumn{ 1}{c}{} & \multicolumn{ 1}{c}{} & \multicolumn{ 1}{c}{} & \multicolumn{ 1}{c}{} & \multicolumn{ 1}{c}{} \cr
\multicolumn{ 1}{c}{$8.172900$} & \multicolumn{ 1}{c}{$0.490558$} & \multicolumn{ 1}{c}{$6.483650$} & $(\phantom{0}8^4,25000\phantom{0},\phantom{0}250)$, $(12^4,10000\phantom{0},\phantom{0}250)$,  & \multicolumn{ 1}{c}{$0.27(1)$} & \multicolumn{ 1}{c}{$0.23(2)$} & \multicolumn{ 1}{c}{$0.21(2)$}  & \multicolumn{ 1}{c}{$0.7291(1)$} & \multicolumn{ 1}{c}{$0.1389(2)$} & \multicolumn{ 1}{c}{$1.1528(1)$} \cr
\multicolumn{ 1}{c}{} & \multicolumn{ 1}{c}{} & \multicolumn{ 1}{c}{} & $(16^4,10000\phantom{0},\phantom{0}250)$, $(20^4,10000\phantom{0},\phantom{0}400)$ & \multicolumn{ 1}{c}{} & \multicolumn{ 1}{c}{} & \multicolumn{ 1}{c}{} & \multicolumn{ 1}{c}{} & \multicolumn{ 1}{c}{} \cr 
\multicolumn{ 1}{c}{$8.433600$} & \multicolumn{ 1}{c}{$0.488003$} & \multicolumn{ 1}{c}{$9.544000$} & $(\phantom{0}8^4,10000\phantom{0},\phantom{0}100)$, $(12^4,10000\phantom{0},1000)$,  & \multicolumn{ 1}{c}{$0.19(4)$} & \multicolumn{ 1}{c}{$0.16(2)$} & \multicolumn{ 1}{c}{$0.16(2)$}  & \multicolumn{ 1}{c}{$0.7382(1)$} & \multicolumn{ 1}{c}{$0.0683(2)$} & \multicolumn{ 1}{c}{$1.0984(1)$} \cr 
\multicolumn{ 1}{c}{} & \multicolumn{ 1}{c}{} & \multicolumn{ 1}{c}{} & $(16^4,5775\phantom{00},\phantom{0}579)$, $(20^4,1981\phantom{00},\phantom{0}261)$ & \multicolumn{ 1}{c}{} & \multicolumn{ 1}{c}{} & \multicolumn{ 1}{c}{} & \multicolumn{ 1}{c}{} & \multicolumn{ 1}{c}{} \cr
\multicolumn{ 1}{c}{$9.590550$} & \multicolumn{ 1}{c}{$0.444462$} & \multicolumn{ 1}{c}{$0.411800$} & $(\phantom{0}8^4,100000,1000)$, $(12^4,102000,3400)$,  & \multicolumn{ 1}{c}{$0.82(1)$} & \multicolumn{ 1}{c}{$0.80(1)$} & \multicolumn{ 1}{c}{$0.70(1)$}  & \multicolumn{ 1}{c}{$0.7844(1)$} & \multicolumn{ 1}{c}{$2.7405(7)$} & \multicolumn{ 1}{c}{$3.75918(7)$} \\ 
\multicolumn{ 1}{c}{} & \multicolumn{ 1}{c}{} & \multicolumn{ 1}{c}{} & $(16^4,102000,3400)$, $(20^4,130848,2802)$ & \multicolumn{ 1}{c}{} & \multicolumn{ 1}{c}{} & \multicolumn{ 1}{c}{} & \multicolumn{ 1}{c}{} & \multicolumn{ 1}{c}{} \cr 
\multicolumn{ 1}{c}{$9.607400$} & \multicolumn{ 1}{c}{$0.174193$} & \multicolumn{ 1}{c}{$0.030100$} & $(\phantom{0}8^4,11200\phantom{0},\phantom{0}112)$, $(12^4,25000\phantom{0},1500)$,  & \multicolumn{ 1}{c}{$0.54(2)$} & \multicolumn{ 1}{c}{$0.52(2)$} & \multicolumn{ 1}{c}{$0.45(2)$}  & \multicolumn{ 1}{c}{$0.7786(1)$} & \multicolumn{ 1}{c}{$2.802(1)$} & \multicolumn{ 1}{c}{$5.6397(3)$} \cr
\multicolumn{ 1}{c}{} & \multicolumn{ 1}{c}{} & \multicolumn{ 1}{c}{} & $(16^4,25000\phantom{0},1500)$, $(20^4,1543\phantom{0},\phantom{0}172)$ & \multicolumn{ 1}{c}{} & \multicolumn{ 1}{c}{} & \multicolumn{ 1}{c}{} & \multicolumn{ 1}{c}{} & \multicolumn{ 1}{c}{} \cr
\multicolumn{ 1}{c}{$10.05222$} & \multicolumn{ 1}{c}{$0.420352$} & \multicolumn{ 1}{c}{$0.717362$} & $(\phantom{0}8^4,20000\phantom{0},\phantom{0}200)$, $(12^4,100000,1000)$,  & \multicolumn{ 1}{c}{$0.58(1)$} & \multicolumn{ 1}{c}{$0.57(1)$} & \multicolumn{ 1}{c}{$0.48(1)$}  & \multicolumn{ 1}{c}{$0.7896(1)$} & \multicolumn{ 1}{c}{$1.3101(1)$} & \multicolumn{ 1}{c}{$2.4002(2)$} \cr 
\multicolumn{ 1}{c}{} & \multicolumn{ 1}{c}{} & \multicolumn{ 1}{c}{} & $(16^4,42494\phantom{0},1000)$, $(20^4,9195\phantom{00},\phantom{0}749)$ & \multicolumn{ 1}{c}{} & \multicolumn{ 1}{c}{} & \multicolumn{ 1}{c}{} & \multicolumn{ 1}{c}{} & \multicolumn{ 1}{c}{} \cr

\hline

\hline

\end{tabular}
\label{tab:numvals}
\end{center}
\end{table*}

We do not list higher energy levels in this channel as well as the lattice masses in the scalar singlet channel, 
since only for the main simulation point defined in Section \ref{sec:spectrum} enough 
statistics was gained. There, the mass $am_{0^{++}_0}$ was below the elastic threshold and also the 
higher levels were not to noisy to draw conclusions. Also, points have not been included where no 
BEH effect was found and/or where the singlet vector mass was above $1$ in lattice units.

The errors listed in Table \ref{tab:numvals} are obtained by fitting the lower and upper bounds of the 
eigenvalues for the gauge-invariant case and of the propagators in the gauge-variant case. 
Subsequently, the method of error propagation is used. Systematic errors are not included.

\subsection{Tables of fit parameters}
In this section we show the fit parameters used in the figures shown in Section \ref{sec:spectrum} and 
\ref{sec:testfms} for the parameter values $\beta=6.855350$, $\kappa=0.456074$, and 
$\lambda= 2.341600$. All the errors are obtained as described previously. We use fit 
routines provided by {\tt Mathematica} \cite{Mathematica} throughout.

\begin{table*}[tbh!]
\begin{center}
\caption{Fit parameters from a double-$\cosh$ fit of the eigenvalues, 
$\lambda(t) = A~\cosh\big(am_{\text{eff}}^{(1)}(t-L/2)\big)
+B~\cosh\big(am_{\text{eff}}^{(2)}(t-L/2)\big),$ 
obtained from a variational analysis 
for several lattice volumes $V=L^4$ in the scalar and vector channels. 
The dash indicates that only a single-$\cosh$ fit has been used.}
\begin{tabular}{@{}c|cccccrr@{}}

\hline

\hline
																													
$J^{PC}_{\U1}$ & $V$ & Level & Fit-range $[t_{\text{min}},t_{\text{max}}]$   & 
$am_{\text{eff}}^{(1)}$	&  $am_{\text{eff}}^{(2)}$ & $A$ & $B$\cr
\hline																							
$0^{++}_0$&$8^4$ & $1^\text{st}$ & $[1,4]$ & $0.55(1)$ & $1.44(1)$ & $0.1156(3)$ & 
$0.0031(1)$ \cr
&$12^4$ & $1^\text{st}$ & $[1,6]$ & $0.65(1)$ & $1.55(1)$ & $0.0180(4)$ & $0.0009(1)$ \cr
&$16^4$ & $1^\text{st}$ & $[1,6]$ & $0.70(1)$ & $1.45(3)$ & $0.0044(1)$ & $0.000007(2)$ \cr
&$20^4$ & $1^\text{st}$ & $[2,7]$ & $0.67(3)$ & $1.08(5)$ & $0.0010(1)$ & $0.00003(2)$ \cr
&$8^4$ & $2^\text{nd}$ & $[1,4]$ & $0.95(4)$ & $1.4(2)$ & $0.0282(2)$ & $0.002(1)$ \cr
&$12^4$ & $2^\text{nd}$ & $[1,6]$ & $0.80(1)$ & $1.4(2)$ & $0.0045(5)$ & $0.00030(15)$ \cr
&$20^4$ & $2^\text{nd}$& $[2,6]$ & $0.90(10)$ & $1.3(1)$ & $0.00013(6)$ 
& $0.000003(2)$ \cr
\hline
$1^{--}_0$&$8^4$ & $1^\text{st}$ & $[2,4]$ & $0.42(1)$ & $-$ & $0.337(1)$ 
& $-$ \cr
&$12^4$ & $1^\text{st}$ & $[2,6]$ & $0.39(1)$ & $1.50(4)$ & $0.159(1)$ & $0.0003(1)$ \cr
&$16^4$ & $1^\text{st}$ & $[2,8]$ & $0.39(1)$ & $1.5(2)$ & $0.072(1)$ & $0.000003(2)$ \cr
&$20^4$ & $1^\text{st}$ & $[2,9]$ & $0.39(1)$ & $1.4(1)$ & $0.033(1)$ & $0.0000005(5)$ \cr
&$8^4$ & $2^\text{nd}$ & $[2,4]$ & $0.99(1)$ & $-$ & $0.026(1)$ 
& $-$ \cr
&$12^4$ & $2^\text{nd}$ & $[2,4]$ & $1.02(2)$ & $1.7(2)$ & $0.018(1)$ & $0.00026(1)$ \cr
&$16^4$ & $2^\text{nd}$ & $[2,6]$ & $1.01(1)$ & $1.9(2)$ & $0.0010(1)$ & $0.0000003(2)$ \cr
&$20^4$ & $2^\text{nd}$ & $[2,6]$ & $1.03(2)$ & $1.4(1)$ & $0.0003(1)$ & $0.000002(1)$ \cr
\hline
$0^{++}_{\pm 1}$ & $20^4$ & $1^\text{st}$ & $[1,3]$ & $2.0(1)$ & $-$ 
& $2.3(1)\cdot 10^{-9}$ 
& $-$\cr
\hline
$1^{--}_{\pm 1}$ & $8^4$ & $1^\text{st}$ & $[2,6]$ & $0.87(1)$ & $1.65(1)$ & $0.0236(1)$ 
& $0.0017(1)$\cr
& $12^4$ & $1^\text{st}$ & $[2,6]$ & $0.75(5)$ & $1.35(5)$ & $0.0025(1)$ 
& $0.0005(2)$\cr
& $16^4$ & $1^\text{st}$ & $[2,6]$ & $0.67(1)$ & $1.27(2)$ & $0.0007(1)$ 
& $0.00006(1)$\cr
& $20^4$ & $1^\text{st}$ & $[2,6]$ & $0.95(5)$ & $1.60(5)$ & $0.00006(6)$ 
& $0.00000002(1)$\cr

\hline

\hline																							

\end{tabular}
\label{tab:fitvalsgi}
\end{center}
\end{table*}


\begin{table}[t!]
\begin{center}
\caption{Fit parameters from a fit of the perturbatively massless and massive propagators 
for several lattice volumes $V$. The fit functions are defined in Equation \eqref{eq:gbpropfitfunc}. 
}
\begin{tabular}{@{}c|ccccccc@{}}
\multicolumn{8}{c}{Perturbatively massless ($c=1,2,3$)}\cr 

\hline

\hline
																													
$V$ & $Z$  & $A/V$ & $b$ & $d^2$ & $\Lambda^2$ & $\gamma$ & $am^c$\cr
\hline																							
$8^4$   & $1$ & $0.445(2)$ & $0.8$ & $0.4$ & $0.18$ & $0.25$ & $0.26(2)$\cr  
$12^4$ & $1$ & $0.120(4)$ & $0.8$ & $0.4$ & $0.18$ & $0.25$ & $\sim 10^{-10}$\cr
$16^4$ & $1$ & $0.030(3)$ & $0.8$ & $0.4$ & $0.18$ & $0.25$ & $\sim10^{-10}$\cr
$20^4$ & $1$ & $0.013(1)$ & $0.8$ & $0.4$ & $0.18$ & $0.25$ & $\sim10^{-10}$\cr

\hline

\hline	
\end{tabular}

\begin{tabular}{@{}c|ccccc@{}}
\multicolumn{6}{c}{Perturbatively massive ($c=4,5,6,7$)}\cr 																		

\hline

\hline
																													
$V$ & $Z$  & $b$ & $c$ & $\gamma$  & $am^c$  \cr
\hline	
$8^4$  & $1.33(1)$ & $1.044(1)$ & $1$ & $0.249(1)$ & $0.41(1)$ \cr
$12^4$ & $1.28(1)$ & $1.013(4)$ & $1$ & $0.199(1)$ & $0.36(1)$ \cr
$16^4$ & $1.27(1)$ & $1.027(3)$ & $1$ & $0.166(1)$ & $0.34(1)$ \cr
$20^4$ & $1.26(1)$ & $0.997(1)$ & $1$ & $0.249(1)$ & $0.33(1)$ \cr
\hline

\hline	
\end{tabular}

\begin{tabular}{@{}c|ccccc@{}}
\multicolumn{6}{c}{Perturbatively massive ($c=8$)}\cr 																		

\hline

\hline
																													
$V$ & $Z$  & $b$ & $c$ & $\gamma$  & $am^c$  \cr
\hline	
$8^4$  & $1.33(1)$ & $1.419(4)$ & $1$ & $0.244(2)$ & $0.38(1)$  \cr
$12^4$ & $1.34(1)$ & $1.449(3)$ & $1$ & $0.249(1)$ & $0.37(1)$  \cr
$16^4$ & $1.29(4)$ & $1.134(8)$ & $1$ & $0.152(1)$ & $0.36(1)$  \cr
$20^4$ & $1.25(1)$ & $1.040(4)$ & $1$ & $0.135(1)$ & $0.36(1)$  \cr
\hline

\hline

\end{tabular}
\label{tab:fitvalsgv}
\end{center}
\end{table}

\begin{table}[t!]
\begin{center}
\caption{Infinite volume extrapolations of the gauge-invariant singlet scalar ($am_{0^{++}_0}$) and 
vector ($am_{1^{--}_0}$) lattice masses, as well as the extrapolation of the gauge-variant lattice masses 
of the gauge-boson propagator $D^c\big(p^2\big)$. In all those cases the fit function is 
$am_{\mathrm{eff}}(V) = am+\alpha~\E^{-\gamma~V}$.}
\begin{tabular}{@{}c|crr@{}}

\hline

\hline
																									
State & $am$ & $\delta$ & $\gamma$ \cr

\hline																							

$0^{++}_0$ & $0.68(2)$ & $-0.140(5)$ & $0.00007(5)$ \cr
$1^{--}_0$ & $0.39(1)$ & $0.8701(1)$ & $0.000822(1)$\cr
\hline
$c=4,5,6,7$ & $0.33(1)$ & $0.106(1)$ & $0.000048(1)$ \cr
$c=8$ & $0.36(1)$ & $0.248(1)$ & $0.000043(1)$\cr

\hline

\hline																							

\end{tabular}
\label{tab:fitvolumedep}
\end{center}
\end{table}


\clearpage
\twocolumngrid

\begin{table}[htbp!]
\renewcommand{\thetable}{\Roman{table}a}
\caption{Data points used in Figure \ref{fig:qlrhlr}. Parameters on the left are in the  QLR and 
on the right in the HLR.}
{\footnotesize{
\begin{tabular}{@{}ccc|ccc@{}}

\hline

\hline

$\beta$ & $\kappa$ & $\lambda$ & $\beta$ & $\kappa$ & $\lambda$ \cr

\hline

$5.001350$ & $0.582573$ & $5.432500$ & $5.197100$ & $0.409979$ & $0.081700$ \cr 
$5.059800$ & $0.512301$ & $5.486200$ & $5.272050$ & $0.699105$ & $7.728100$ \cr 
$5.102750$ & $0.207329$ & $2.328300$ & $5.300650$ & $0.772447$ & $9.909500$ \cr 
$5.107700$ & $0.274441$ & $3.234200$ & $5.379650$ & $0.744762$ & $6.064800$ \cr 
$5.160350$ & $0.171077$ & $8.581500$ & $5.401400$ & $0.887851$ & $9.573700$ \cr 
$5.167050$ & $0.620915$ & $8.179000$ & $5.421050$ & $0.686636$ & $6.646600$ \cr  
$5.250850$ & $0.204756$ & $3.255700$ & $5.451750$ & $0.686295$ & $7.562000$ \cr 
$5.337550$ & $0.168969$ & $6.854000$ & $5.669450$ & $0.741210$ & $3.126300$ \cr 
$5.342250$ & $0.443841$ & $7.157200$ & $5.798500$ & $0.419035$ & $1.259900$ \cr 
$5.426000$ & $0.315120$ & $1.118300$ & $5.807200$ & $0.565423$ & $1.937700$ \cr 
$5.458500$ & $0.324675$ & $5.803900$ & $5.866600$ & $0.735811$ & $4.016500$ \cr 
$5.509550$ & $0.172845$ & $9.857800$ & $5.895350$ & $0.972639$ & $8.525100$ \cr 
$5.574550$ & $0.570725$ & $9.613500$ & $5.920550$ & $0.940701$ & $9.520400$ \cr 
$5.605750$ & $0.201746$ & $5.547600$ & $6.084750$ & $0.552647$ & $0.163700$ \cr 
$5.670550$ & $0.127774$ & $8.257900$ & $6.167900$ & $0.572055$ & $2.671000$ \cr 
$5.721350$ & $0.169362$ & $3.214100$ & $6.360750$ & $0.621580$ & $1.835700$ \cr 
$5.755400$ & $0.331019$ & $2.264100$ & $6.486150$ & $0.538411$ & $5.369000$ \cr 
$5.790750$ & $0.405788$ & $2.250800$ & $6.380150$ & $0.677711$ & $7.140400$ \cr 
$5.793600$ & $0.153236$ & $4.163900$ & $6.497800$ & $0.966706$ & $8.734100$ \cr 
$6.008900$ & $0.351835$ & $1.344400$ & $6.533800$ & $0.638196$ & $8.017500$ \cr 
$6.068550$ & $0.280759$ & $8.057500$ & $6.553100$ & $0.504496$ & $3.610100$ \cr 
$6.167150$ & $0.180624$ & $4.885100$ & $6.615000$ & $0.705641$ & $9.185900$ \cr 
$6.207200$ & $0.262217$ & $2.780300$ & $6.713150$ & $0.469278$ & $2.234300$ \cr 
$6.215350$ & $0.409988$ & $7.701600$ & $6.726000$ & $0.763173$ & $0.284600$ \cr 
$6.273150$ & $0.190975$ & $2.091600$ & $6.812250$ & $0.820240$ & $7.319600$ \cr 
$6.372100$ & $0.437174$ & $5.576000$ & $6.855350$ & $0.456074$ & $2.341600$ \cr 
$6.472300$ & $0.459049$ & $5.750400$ & $6.869350$ & $0.553558$ & $1.101700$ \cr 
$6.487400$ & $0.344371$ & $6.804900$ & $6.926950$ & $0.960870$ & $7.011400$ \cr 
$6.545900$ & $0.301934$ & $2.308100$ & $7.012700$ & $0.619996$ & $2.964100$ \cr 
$6.605300$ & $0.289701$ & $4.131000$ & $7.023000$ & $0.842342$ & $5.491900$ \cr  
$6.628550$ & $0.414380$ & $8.934200$ & $7.031000$ & $0.600326$ & $2.198400$ \cr 
$6.931900$ & $0.349044$ & $6.758700$ & $7.043650$ & $0.845501$ & $1.041900$ \cr 
$6.970950$ & $0.296308$ & $4.862800$ & $7.105250$ & $0.615691$ & $2.320600$ \cr 
$6.977200$ & $0.487910$ & $9.244100$ & $7.105850$ & $0.596599$ & $7.122900$ \cr  
$7.173700$ & $0.451051$ & $4.375700$ & $7.117050$ & $0.817956$ & $0.342010$ \cr 
$7.234100$ & $0.232861$ & $7.159300$ & $7.148850$ & $0.523361$ & $2.828200$ \cr 
$7.268050$ & $0.247745$ & $9.878900$ & $7.174550$ & $0.588776$ & $1.782800$ \cr 
$7.305150$ & $0.323949$ & $8.068000$ & $7.204800$ & $0.683967$ & $1.665400$ \cr 
$7.339400$ & $0.358870$ & $8.085100$ & $7.208950$ & $0.960266$ & $5.082000$ \cr 
$7.346350$ & $0.411554$ & $4.349400$ & $7.360400$ & $0.736564$ & $1.736900$ \cr 
$7.418900$ & $0.920576$ & $1.052100$ & $7.374350$ & $0.845641$ & $9.252600$ \cr 
$7.432050$ & $0.391114$ & $8.650700$ & $7.437300$ & $0.697871$ & $2.504100$ \cr 
$7.483150$ & $0.226824$ & $3.598700$ & $7.441350$ & $0.661384$ & $6.554300$ \cr 
$7.506900$ & $0.189312$ & $8.790900$ & $7.564250$ & $0.584331$ & $6.390000$ \cr 
$7.510100$ & $0.384604$ & $2.640300$ & $7.672450$ & $0.628484$ & $3.437600$ \cr 
$7.568650$ & $0.469619$ & $7.639800$ & $7.735250$ & $0.984959$ & $0.237900$ \cr 
$7.678250$ & $0.271300$ & $5.989000$ & $7.766750$ & $0.692516$ & $3.139500$ \cr 
$7.762950$ & $0.321140$ & $4.994100$ & $7.808100$ & $0.960520$ & $7.907700$ \cr 
$7.796300$ & $0.329934$ & $9.307400$ & $7.858100$ & $0.736625$ & $3.223500$ \cr 
$7.813900$ & $0.207180$ & $3.728300$ & $7.866600$ & $0.645056$ & $9.138700$ \cr 
$7.899400$ & $0.322706$ & $1.125500$ & $7.901600$ & $0.813695$ & $7.029700$ \cr 
$8.005100$ & $0.422098$ & $3.445900$ & $7.912200$ & $0.493113$ & $3.423300$ \cr 
$8.068750$ & $0.203724$ & $9.813100$ & $7.950100$ & $0.875409$ & $4.801300$ \cr 
$8.087350$ & $0.209009$ & $6.371400$ & $7.987850$ & $0.551466$ & $2.057400$ \cr

\hline

\hline

\end{tabular}}}
\label{tab:phasediagram}
\end{table}

\begin{table}[h!]
\addtocounter{table}{-1}
\renewcommand{\thetable}{\Roman{table}b}
\caption{Continuation of Table \ref{tab:phasediagram}. Parameters on the left are in the  QLR and 
on the right in the HLR.}
{\footnotesize{
\begin{tabular}{@{}ccc|ccc@{}}

\hline

\hline

$\beta$ & $\kappa$ & $\lambda$ & $\beta$ & $\kappa$ & $\lambda$ \cr

\hline

$8.353950$ & $0.407782$ & $7.748300$ & $8.012300$ & $0.673406$ & $5.704600$ \cr 
$8.436150$ & $0.182435$ & $3.852300$ & $8.113800$ & $0.698703$ & $9.904300$ \cr 
$8.472200$ & $0.267625$ & $3.397300$ & $8.160300$ & $0.964134$ & $2.041000$ \cr
$8.582100$ & $0.429369$ & $4.599900$ & $8.172900$ & $0.490558$ & $6.483650$ \cr 
$8.593950$ & $0.146744$ & $4.287900$ & $8.220800$ & $0.760451$ & $7.768400$ \cr 
$8.602650$ & $0.230192$ & $2.043700$ & $8.285350$ & $0.875041$ & $0.405600$ \cr 
$8.623950$ & $0.172898$ & $1.173800$ & $8.293750$ & $0.803808$ & $9.327600$ \cr 
$8.743200$ & $0.147487$ & $5.834700$ & $8.320100$ & $0.996596$ & $9.316800$ \cr 
$8.744950$ & $0.338999$ & $7.049800$ & $8.352050$ & $0.827021$ & $9.685900$ \cr 
$8.916800$ & $0.279954$ & $8.572500$ & $8.433600$ & $0.488003$ & $9.544000$ \cr 
$8.926350$ & $0.136314$ & $0.164600$ & $8.435450$ & $0.911922$ & $4.289900$ \cr 
$8.938250$ & $0.250790$ & $1.945800$ & $8.444900$ & $0.778887$ & $4.404500$ \cr 
$9.134200$ & $0.177666$ & $5.825000$ & $9.022300$ & $0.911371$ & $9.650900$ \cr 
$9.245700$ & $0.250974$ & $7.596700$ & $9.023450$ & $0.951263$ & $1.072200$ \cr 
$9.342200$ & $0.154872$ & $4.764900$ & $9.070300$ & $0.553505$ & $4.155700$ \cr 
$9.448100$ & $0.189995$ & $1.870400$ & $9.077950$ & $0.805575$ & $9.828100$ \cr 
$9.576000$ & $0.375749$ & $8.755900$ & $9.101500$ & $0.565895$ & $8.109800$ \cr 
$9.580600$ & $0.176039$ & $8.591700$ & $9.126750$ & $0.731445$ & $6.487600$ \cr 
$9.641300$ & $0.295739$ & $9.858000$ & $9.446000$ & $0.705353$ & $2.410400$ \cr 
$9.751650$ & $0.341738$ & $8.630300$ & $9.188600$ & $0.842640$ & $6.243700$ \cr 
$9.817400$ & $0.225222$ & $0.721100$ & $9.221400$ & $0.743818$ & $6.391400$ \cr 
$9.882900$ & $0.176520$ & $3.603400$ & $9.235300$ & $0.443518$ & $0.315800$ \cr 
$9.897950$ & $0.322277$ & $2.636400$ & $9.236950$ & $0.720000$ & $8.094800$ \cr 
           &            &            & $8.477500$ & $0.688544$ & $6.273400$ \cr 
           &            &            & $9.269100$ & $0.501241$ & $3.053700$ \cr 
           &            &            & $9.317000$ & $0.470739$ & $0.950400$ \cr 
           &            &            & $8.486750$ & $0.788898$ & $7.609300$ \cr 
           &            &            & $9.360650$ & $0.592932$ & $6.956300$ \cr 
           &            &            & $9.360900$ & $0.686409$ & $1.289500$ \cr 
           &            &            & $9.386500$ & $0.984136$ & $1.510600$ \cr 
           &            &            & $8.523750$ & $0.658269$ & $3.503700$ \cr 
           &            &            & $9.451700$ & $0.892261$ & $5.746600$ \cr 
           &            &            & $9.466250$ & $0.776157$ & $2.063100$ \cr 
           &            &            & $9.495750$ & $0.901869$ & $2.136300$ \cr 
           &            &            & $9.498000$ & $0.827010$ & $4.656400$ \cr 
           &            &            & $9.511350$ & $0.713814$ & $6.990200$ \cr 
           &            &            & $8.532100$ & $0.738112$ & $8.422200$ \cr 
           &            &            & $8.543450$ & $0.497295$ & $8.338300$ \cr 
           &            &            & $9.590550$ & $0.444462$ & $0.411800$ \cr 
           &            &            & $9.607400$ & $0.174196$ & $0.030100$ \cr 
           &            &            & $8.609250$ & $0.984941$ & $0.645700$ \cr 
           &            &            & $9.652350$ & $0.483514$ & $5.815200$ \cr 
           &            &            & $9.724100$ & $0.838204$ & $1.513900$ \cr 
           &            &            & $8.686450$ & $0.826199$ & $0.206800$ \cr 
           &            &            & $8.779400$ & $0.717174$ & $4.457700$ \cr 
           &            &            & $9.868400$ & $0.726492$ & $1.530000$ \cr 
           &            &            & $8.948050$ & $0.835929$ & $6.733200$ \cr 
           &            &            & $9.891100$ & $0.993516$ & $8.291200$ \cr 
           &            &            & $8.984150$ & $0.722012$ & $2.819000$ \cr 
           &            &            & $9.901600$ & $0.879329$ & $5.160900$ \cr 
           &            &            & $9.935000$ & $0.754458$ & $7.992600$ \cr 
           &            &            & $9.948950$ & $0.833908$ & $6.667000$ \cr 
           &            &            & $9.992500$ & $0.847794$ & $4.726900$ \cr 
           &            &            & $10.05222$ & $0.420352$ & $0.717362$ \cr

\hline

\hline

\end{tabular}}}
\label{tab:phasediagram2}
\end{table}

\clearpage

\bibliography{bib}

\begin{thebibliography}{46}%
\makeatletter
\providecommand \@ifxundefined [1]{%
 \@ifx{#1\undefined}
}%
\providecommand \@ifnum [1]{%
 \ifnum #1\expandafter \@firstoftwo
 \else \expandafter \@secondoftwo
 \fi
}%
\providecommand \@ifx [1]{%
 \ifx #1\expandafter \@firstoftwo
 \else \expandafter \@secondoftwo
 \fi
}%
\providecommand \natexlab [1]{#1}%
\providecommand \enquote  [1]{``#1''}%
\providecommand \bibnamefont  [1]{#1}%
\providecommand \bibfnamefont [1]{#1}%
\providecommand \citenamefont [1]{#1}%
\providecommand \href@noop [0]{\@secondoftwo}%
\providecommand \href [0]{\begingroup \@sanitize@url \@href}%
\providecommand \@href[1]{\@@startlink{#1}\@@href}%
\providecommand \@@href[1]{\endgroup#1\@@endlink}%
\providecommand \@sanitize@url [0]{\catcode `\\12\catcode `\$12\catcode
  `\&12\catcode `\#12\catcode `\^12\catcode `\_12\catcode `\%12\relax}%
\providecommand \@@startlink[1]{}%
\providecommand \@@endlink[0]{}%
\providecommand \url  [0]{\begingroup\@sanitize@url \@url }%
\providecommand \@url [1]{\endgroup\@href {#1}{\urlprefix }}%
\providecommand \urlprefix  [0]{URL }%
\providecommand \Eprint [0]{\href }%
\providecommand \doibase [0]{http://dx.doi.org/}%
\providecommand \selectlanguage [0]{\@gobble}%
\providecommand \bibinfo  [0]{\@secondoftwo}%
\providecommand \bibfield  [0]{\@secondoftwo}%
\providecommand \translation [1]{[#1]}%
\providecommand \BibitemOpen [0]{}%
\providecommand \bibitemStop [0]{}%
\providecommand \bibitemNoStop [0]{.\EOS\space}%
\providecommand \EOS [0]{\spacefactor3000\relax}%
\providecommand \BibitemShut  [1]{\csname bibitem#1\endcsname}%
\let\auto@bib@innerbib\@empty
\bibitem [{\citenamefont {Weinberg}(1996)}]{Weinberg:1996kr}%
  \BibitemOpen
  \bibfield  {author} {\bibinfo {author} {\bibfnamefont {S.}~\bibnamefont
  {Weinberg}},\ }\href@noop {} {\emph {\bibinfo {title} {{The quantum theory of
  fields. Vol. 2: Modern applications}}}}\ (\bibinfo  {publisher} {Cambridge
  University Press},\ \bibinfo {address} {Cambridge},\ \bibinfo {year} {1996})\
  \bibinfo {note} {cambridge, UK: Univ. Pr. (1996) 489 p}\BibitemShut {NoStop}%
\bibitem [{\citenamefont {Haag}(1992)}]{Haag:1992hx}%
  \BibitemOpen
  \bibfield  {author} {\bibinfo {author} {\bibfnamefont {R.}~\bibnamefont
  {Haag}},\ }\href@noop {} {\emph {\bibinfo {title} {{Local quantum physics:
  Fields, particles, algebras}}}}\ (\bibinfo  {publisher} {Springer},\ \bibinfo
  {address} {Berlin},\ \bibinfo {year} {1992})\ p.\ \bibinfo {pages} {356},\
  \bibinfo {note} {berlin, Germany: Springer (1992) 356 p. (Texts and
  monographs in physics)}\BibitemShut {NoStop}%
\bibitem [{\citenamefont {'t~Hooft}(1980)}]{'tHooft:1979bj}%
  \BibitemOpen
  \bibfield  {author} {\bibinfo {author} {\bibfnamefont {G.}~\bibnamefont
  {'t~Hooft}},\ }\href@noop {} {\bibfield  {journal} {\bibinfo  {journal} {NATO
  Adv.Study Inst.Ser.B Phys.}\ }\textbf {\bibinfo {volume} {59}},\ \bibinfo
  {pages} {101} (\bibinfo {year} {1980})}\BibitemShut {NoStop}%
\bibitem [{\citenamefont {Osterwalder}\ and\ \citenamefont
  {Seiler}(1978)}]{Osterwalder:1977pc}%
  \BibitemOpen
  \bibfield  {author} {\bibinfo {author} {\bibfnamefont {K.}~\bibnamefont
  {Osterwalder}}\ and\ \bibinfo {author} {\bibfnamefont {E.}~\bibnamefont
  {Seiler}},\ }\href {\doibase 10.1016/0003-4916(78)90039-8} {\bibfield
  {journal} {\bibinfo  {journal} {Annals Phys.}\ }\textbf {\bibinfo {volume}
  {110}},\ \bibinfo {pages} {440} (\bibinfo {year} {1978})}\BibitemShut
  {NoStop}%
\bibitem [{\citenamefont {Banks}\ and\ \citenamefont
  {Rabinovici}(1979)}]{Banks:1979fi}%
  \BibitemOpen
  \bibfield  {author} {\bibinfo {author} {\bibfnamefont {T.}~\bibnamefont
  {Banks}}\ and\ \bibinfo {author} {\bibfnamefont {E.}~\bibnamefont
  {Rabinovici}},\ }\href {\doibase 10.1016/0550-3213(79)90064-6} {\bibfield
  {journal} {\bibinfo  {journal} {Nucl.Phys.}\ }\textbf {\bibinfo {volume}
  {B160}},\ \bibinfo {pages} {349} (\bibinfo {year} {1979})}\BibitemShut
  {NoStop}%
\bibitem [{\citenamefont {Fr\"ohlich}\ \emph {et~al.}(1980)\citenamefont
  {Fr\"ohlich}, \citenamefont {Morchio},\ and\ \citenamefont
  {Strocchi}}]{Frohlich:1980gj}%
  \BibitemOpen
  \bibfield  {author} {\bibinfo {author} {\bibfnamefont {J.}~\bibnamefont
  {Fr\"ohlich}}, \bibinfo {author} {\bibfnamefont {G.}~\bibnamefont {Morchio}},
  \ and\ \bibinfo {author} {\bibfnamefont {F.}~\bibnamefont {Strocchi}},\
  }\href {\doibase 10.1016/0370-2693(80)90594-8} {\bibfield  {journal}
  {\bibinfo  {journal} {Phys.Lett.}\ }\textbf {\bibinfo {volume} {B97}},\
  \bibinfo {pages} {249} (\bibinfo {year} {1980})}\BibitemShut {NoStop}%
\bibitem [{\citenamefont {Fr\"ohlich}\ \emph {et~al.}(1981)\citenamefont
  {Fr\"ohlich}, \citenamefont {Morchio},\ and\ \citenamefont
  {Strocchi}}]{Frohlich:1981yi}%
  \BibitemOpen
  \bibfield  {author} {\bibinfo {author} {\bibfnamefont {J.}~\bibnamefont
  {Fr\"ohlich}}, \bibinfo {author} {\bibfnamefont {G.}~\bibnamefont {Morchio}},
  \ and\ \bibinfo {author} {\bibfnamefont {F.}~\bibnamefont {Strocchi}},\
  }\href {\doibase 10.1016/0550-3213(81)90448-X} {\bibfield  {journal}
  {\bibinfo  {journal} {Nucl.Phys.}\ }\textbf {\bibinfo {volume} {B190}},\
  \bibinfo {pages} {553} (\bibinfo {year} {1981})}\BibitemShut {NoStop}%
\bibitem [{\citenamefont {Patrignani}\ \emph {et~al.}(2016)\citenamefont
  {Patrignani} \emph {et~al.}}]{pdg}%
  \BibitemOpen
  \bibfield  {author} {\bibinfo {author} {\bibfnamefont {C.}~\bibnamefont
  {Patrignani}} \emph {et~al.} (\bibinfo {collaboration} {Particle Data
  Group}),\ }\href {\doibase 10.1088/1674-1137/40/10/100001} {\bibfield
  {journal} {\bibinfo  {journal} {Chin. Phys.}\ }\textbf {\bibinfo {volume}
  {C40}},\ \bibinfo {pages} {100001} (\bibinfo {year} {2016})}\BibitemShut
  {NoStop}%
\bibitem [{\citenamefont {Maas}(2013{\natexlab{a}})}]{Maas:2012tj}%
  \BibitemOpen
  \bibfield  {author} {\bibinfo {author} {\bibfnamefont {A.}~\bibnamefont
  {Maas}},\ }\href {\doibase 10.1142/S0217732313501034} {\bibfield  {journal}
  {\bibinfo  {journal} {Mod.Phys.Lett.}\ }\textbf {\bibinfo {volume} {A28}},\
  \bibinfo {pages} {1350103} (\bibinfo {year} {2013}{\natexlab{a}})},\ \Eprint
  {http://arxiv.org/abs/1205.6625} {arXiv:1205.6625 [hep-lat]} \BibitemShut
  {NoStop}%
\bibitem [{\citenamefont {Maas}\ and\ \citenamefont
  {Mufti}(2014)}]{Maas:2013aia}%
  \BibitemOpen
  \bibfield  {author} {\bibinfo {author} {\bibfnamefont {A.}~\bibnamefont
  {Maas}}\ and\ \bibinfo {author} {\bibfnamefont {T.}~\bibnamefont {Mufti}},\
  }\href {\doibase 10.1007/JHEP04(2014)006} {\bibfield  {journal} {\bibinfo
  {journal} {JHEP}\ }\textbf {\bibinfo {volume} {1404}},\ \bibinfo {pages}
  {006} (\bibinfo {year} {2014})},\ \Eprint {http://arxiv.org/abs/1312.4873}
  {arXiv:1312.4873 [hep-lat]} \BibitemShut {NoStop}%
\bibitem [{\citenamefont {Maas}(2017)}]{Maas:2017wzi}%
  \BibitemOpen
  \bibfield  {author} {\bibinfo {author} {\bibfnamefont {A.}~\bibnamefont
  {Maas}},\ }\href@noop {} {\  (\bibinfo {year} {2017})},\ \Eprint
  {http://arxiv.org/abs/1712.04721} {arXiv:1712.04721 [hep-ph]} \BibitemShut
  {NoStop}%
\bibitem [{\citenamefont {Maas}(2015)}]{Maas:2015gma}%
  \BibitemOpen
  \bibfield  {author} {\bibinfo {author} {\bibfnamefont {A.}~\bibnamefont
  {Maas}},\ }\href {\doibase 10.1142/S0217732315501357} {\bibfield  {journal}
  {\bibinfo  {journal} {Mod. Phys. Lett.}\ }\textbf {\bibinfo {volume} {A30}},\
  \bibinfo {pages} {1550135} (\bibinfo {year} {2015})},\ \Eprint
  {http://arxiv.org/abs/1502.02421} {arXiv:1502.02421 [hep-ph]} \BibitemShut
  {NoStop}%
\bibitem [{\citenamefont {Maas}\ and\ \citenamefont
  {Pedro}(2016)}]{Maas:2016qpu}%
  \BibitemOpen
  \bibfield  {author} {\bibinfo {author} {\bibfnamefont {A.}~\bibnamefont
  {Maas}}\ and\ \bibinfo {author} {\bibfnamefont {L.}~\bibnamefont {Pedro}},\
  }\href {\doibase 10.1103/PhysRevD.93.056005} {\bibfield  {journal} {\bibinfo
  {journal} {Phys. Rev.}\ }\textbf {\bibinfo {volume} {D93}},\ \bibinfo {pages}
  {056005} (\bibinfo {year} {2016})},\ \Eprint
  {http://arxiv.org/abs/1601.02006} {arXiv:1601.02006 [hep-ph]} \BibitemShut
  {NoStop}%
\bibitem [{\citenamefont {Maas}\ \emph
  {et~al.}(2017{\natexlab{a}})\citenamefont {Maas}, \citenamefont
  {Sondenheimer},\ and\ \citenamefont {Törek}}]{Maas:2017xzh}%
  \BibitemOpen
  \bibfield  {author} {\bibinfo {author} {\bibfnamefont {A.}~\bibnamefont
  {Maas}}, \bibinfo {author} {\bibfnamefont {R.}~\bibnamefont {Sondenheimer}},
  \ and\ \bibinfo {author} {\bibfnamefont {P.}~\bibnamefont {Törek}},\
  }\href@noop {} {\  (\bibinfo {year} {2017}{\natexlab{a}})},\ \Eprint
  {http://arxiv.org/abs/1709.07477} {arXiv:1709.07477 [hep-ph]} \BibitemShut
  {NoStop}%
\bibitem [{\citenamefont {Maas}\ \emph
  {et~al.}(2017{\natexlab{b}})\citenamefont {Maas}, \citenamefont
  {Sondenheimer},\ and\ \citenamefont {T\"orek}}]{Maas:2017pcw}%
  \BibitemOpen
  \bibfield  {author} {\bibinfo {author} {\bibfnamefont {A.}~\bibnamefont
  {Maas}}, \bibinfo {author} {\bibfnamefont {R.}~\bibnamefont {Sondenheimer}},
  \ and\ \bibinfo {author} {\bibfnamefont {P.}~\bibnamefont {T\"orek}}\
  }(\bibinfo {year} {2017})\ \Eprint {http://arxiv.org/abs/1710.01941}
  {arXiv:1710.01941 [hep-lat]} \BibitemShut {NoStop}%
\bibitem [{\citenamefont {T\"orek}\ and\ \citenamefont
  {Maas}(2016)}]{Torek:2016ede}%
  \BibitemOpen
  \bibfield  {author} {\bibinfo {author} {\bibfnamefont {P.}~\bibnamefont
  {T\"orek}}\ and\ \bibinfo {author} {\bibfnamefont {A.}~\bibnamefont {Maas}},\
  }\bibfield  {booktitle} {\emph {\bibinfo {booktitle} {{Proceedings, 34th
  International Symposium on Lattice Field Theory (Lattice 2016): Southampton,
  UK, July 24-30, 2016}}},\ }\href@noop {} {\bibfield  {journal} {\bibinfo
  {journal} {PoS}\ }\textbf {\bibinfo {volume} {LATTICE2016}},\ \bibinfo
  {pages} {203} (\bibinfo {year} {2016})},\ \Eprint
  {http://arxiv.org/abs/1610.04188} {arXiv:1610.04188 [hep-lat]} \BibitemShut
  {NoStop}%
\bibitem [{\citenamefont {Maas}\ and\ \citenamefont
  {T\"orek}(2017)}]{Maas:2016ngo}%
  \BibitemOpen
  \bibfield  {author} {\bibinfo {author} {\bibfnamefont {A.}~\bibnamefont
  {Maas}}\ and\ \bibinfo {author} {\bibfnamefont {P.}~\bibnamefont {T\"orek}},\
  }\href {\doibase 10.1103/PhysRevD.95.014501} {\bibfield  {journal} {\bibinfo
  {journal} {Phys. Rev.}\ }\textbf {\bibinfo {volume} {D95}},\ \bibinfo {pages}
  {014501} (\bibinfo {year} {2017})},\ \Eprint
  {http://arxiv.org/abs/1607.05860} {arXiv:1607.05860 [hep-lat]} \BibitemShut
  {NoStop}%
\bibitem [{\citenamefont {Montvay}\ and\ \citenamefont
  {M\"unster}(1994)}]{Montvay:1994cy}%
  \BibitemOpen
  \bibfield  {author} {\bibinfo {author} {\bibfnamefont {I.}~\bibnamefont
  {Montvay}}\ and\ \bibinfo {author} {\bibfnamefont {G.}~\bibnamefont
  {M\"unster}},\ }\href@noop {} {\emph {\bibinfo {title} {{Quantum fields on a
  lattice}}}}\ (\bibinfo  {publisher} {Cambridge University Press},\ \bibinfo
  {address} {Cambridge},\ \bibinfo {year} {1994})\ p.\ \bibinfo {pages} {491},\
  \bibinfo {note} {cambridge, UK: Univ. Pr. (1994) 491 p. (Cambridge monographs
  on mathematical physics)}\BibitemShut {NoStop}%
\bibitem [{\citenamefont {Gattringer}\ and\ \citenamefont
  {Lang}(2010)}]{Gattringer:2010zz}%
  \BibitemOpen
  \bibfield  {author} {\bibinfo {author} {\bibfnamefont {C.}~\bibnamefont
  {Gattringer}}\ and\ \bibinfo {author} {\bibfnamefont {C.~B.}\ \bibnamefont
  {Lang}},\ }\href {\doibase 10.1007/978-3-642-01850-3} {\emph {\bibinfo
  {title} {Quantum chromodynamics on the lattice: An Introductory
  Presentation}}}\ (\bibinfo  {publisher} {Lect. Notes Phys.},\ \bibinfo {year}
  {Springer, Berlin Heidelberg 2010})\ p.\ \bibinfo {pages} {211}\BibitemShut
  {NoStop}%
\bibitem [{\citenamefont {Wurtz}\ and\ \citenamefont
  {Lewis}(2013)}]{Wurtz:2013ova}%
  \BibitemOpen
  \bibfield  {author} {\bibinfo {author} {\bibfnamefont {M.}~\bibnamefont
  {Wurtz}}\ and\ \bibinfo {author} {\bibfnamefont {R.}~\bibnamefont {Lewis}},\
  }\href {\doibase 10.1103/PhysRevD.88.054510} {\bibfield  {journal} {\bibinfo
  {journal} {Phys.Rev.}\ }\textbf {\bibinfo {volume} {D88}},\ \bibinfo {pages}
  {054510} (\bibinfo {year} {2013})},\ \Eprint {http://arxiv.org/abs/1307.1492}
  {arXiv:1307.1492 [hep-lat]} \BibitemShut {NoStop}%
\bibitem [{\citenamefont {Berg}\ and\ \citenamefont
  {Billoire}(1983)}]{Berg:1982kp}%
  \BibitemOpen
  \bibfield  {author} {\bibinfo {author} {\bibfnamefont {B.}~\bibnamefont
  {Berg}}\ and\ \bibinfo {author} {\bibfnamefont {A.}~\bibnamefont
  {Billoire}},\ }\href {\doibase 10.1016/0550-3213(83)90620-X} {\bibfield
  {journal} {\bibinfo  {journal} {Nucl. Phys.}\ }\textbf {\bibinfo {volume}
  {B221}},\ \bibinfo {pages} {109} (\bibinfo {year} {1983})}\BibitemShut
  {NoStop}%
\bibitem [{\citenamefont {Maas}\ and\ \citenamefont
  {Mufti}(2015)}]{Maas:2014pba}%
  \BibitemOpen
  \bibfield  {author} {\bibinfo {author} {\bibfnamefont {A.}~\bibnamefont
  {Maas}}\ and\ \bibinfo {author} {\bibfnamefont {T.}~\bibnamefont {Mufti}},\
  }\href {\doibase 10.1103/PhysRevD.91.113011} {\bibfield  {journal} {\bibinfo
  {journal} {Phys. Rev.}\ }\textbf {\bibinfo {volume} {D91}},\ \bibinfo {pages}
  {113011} (\bibinfo {year} {2015})},\ \Eprint {http://arxiv.org/abs/1412.6440}
  {arXiv:1412.6440 [hep-lat]} \BibitemShut {NoStop}%
\bibitem [{\citenamefont {Michael}(1985)}]{Michael:1985ne}%
  \BibitemOpen
  \bibfield  {author} {\bibinfo {author} {\bibfnamefont {C.}~\bibnamefont
  {Michael}},\ }\href {\doibase 10.1016/0550-3213(85)90297-4} {\bibfield
  {journal} {\bibinfo  {journal} {Nucl. Phys.}\ }\textbf {\bibinfo {volume}
  {B259}},\ \bibinfo {pages} {58} (\bibinfo {year} {1985})}\BibitemShut
  {NoStop}%
\bibitem [{\citenamefont {L\"uscher}\ and\ \citenamefont
  {Wolff}(1990)}]{Luscher:1990ck}%
  \BibitemOpen
  \bibfield  {author} {\bibinfo {author} {\bibfnamefont {M.}~\bibnamefont
  {L\"uscher}}\ and\ \bibinfo {author} {\bibfnamefont {U.}~\bibnamefont
  {Wolff}},\ }\href {\doibase 10.1016/0550-3213(90)90540-T} {\bibfield
  {journal} {\bibinfo  {journal} {Nucl. Phys.}\ }\textbf {\bibinfo {volume}
  {B339}},\ \bibinfo {pages} {222} (\bibinfo {year} {1990})}\BibitemShut
  {NoStop}%
\bibitem [{\citenamefont {Blossier}\ \emph {et~al.}(2009)\citenamefont
  {Blossier}, \citenamefont {Della~Morte}, \citenamefont {von Hippel},
  \citenamefont {Mendes},\ and\ \citenamefont {Sommer}}]{Blossier:2009kd}%
  \BibitemOpen
  \bibfield  {author} {\bibinfo {author} {\bibfnamefont {B.}~\bibnamefont
  {Blossier}}, \bibinfo {author} {\bibfnamefont {M.}~\bibnamefont
  {Della~Morte}}, \bibinfo {author} {\bibfnamefont {G.}~\bibnamefont {von
  Hippel}}, \bibinfo {author} {\bibfnamefont {T.}~\bibnamefont {Mendes}}, \
  and\ \bibinfo {author} {\bibfnamefont {R.}~\bibnamefont {Sommer}},\ }\href
  {\doibase 10.1088/1126-6708/2009/04/094} {\bibfield  {journal} {\bibinfo
  {journal} {JHEP}\ }\textbf {\bibinfo {volume} {04}},\ \bibinfo {pages} {094}
  (\bibinfo {year} {2009})},\ \Eprint {http://arxiv.org/abs/0902.1265}
  {arXiv:0902.1265 [hep-lat]} \BibitemShut {NoStop}%
\bibitem [{\citenamefont {Seiler}(1982)}]{Seiler:1982pw}%
  \BibitemOpen
  \bibfield  {author} {\bibinfo {author} {\bibfnamefont {E.}~\bibnamefont
  {Seiler}},\ }\href@noop {} {\emph {\bibinfo {title} {{Gauge Theories as a
  Problem of Constructive Quantum Field Theory and Statistical Mechanics}}}}\
  (\bibinfo  {publisher} {Lect. Notes Phys.},\ \bibinfo {year} {1982})\ p.\
  \bibinfo {pages} {192}\BibitemShut {NoStop}%
\bibitem [{\citenamefont {Morningstar}\ and\ \citenamefont
  {Peardon}(2004)}]{Morningstar:2003gk}%
  \BibitemOpen
  \bibfield  {author} {\bibinfo {author} {\bibfnamefont {C.}~\bibnamefont
  {Morningstar}}\ and\ \bibinfo {author} {\bibfnamefont {M.~J.}\ \bibnamefont
  {Peardon}},\ }\href {\doibase 10.1103/PhysRevD.69.054501} {\bibfield
  {journal} {\bibinfo  {journal} {Phys. Rev.}\ }\textbf {\bibinfo {volume}
  {D69}},\ \bibinfo {pages} {054501} (\bibinfo {year} {2004})},\ \Eprint
  {http://arxiv.org/abs/hep-lat/0311018} {arXiv:hep-lat/0311018 [hep-lat]}
  \BibitemShut {NoStop}%
\bibitem [{\citenamefont {Philipsen}\ \emph {et~al.}(1996)\citenamefont
  {Philipsen}, \citenamefont {Teper},\ and\ \citenamefont
  {Wittig}}]{Philipsen:1996af}%
  \BibitemOpen
  \bibfield  {author} {\bibinfo {author} {\bibfnamefont {O.}~\bibnamefont
  {Philipsen}}, \bibinfo {author} {\bibfnamefont {M.}~\bibnamefont {Teper}}, \
  and\ \bibinfo {author} {\bibfnamefont {H.}~\bibnamefont {Wittig}},\ }\href
  {\doibase 10.1016/0550-3213(96)00156-3} {\bibfield  {journal} {\bibinfo
  {journal} {Nucl.Phys.}\ }\textbf {\bibinfo {volume} {B469}},\ \bibinfo
  {pages} {445} (\bibinfo {year} {1996})},\ \Eprint
  {http://arxiv.org/abs/hep-lat/9602006} {arXiv:hep-lat/9602006 [hep-lat]}
  \BibitemShut {NoStop}%
\bibitem [{\citenamefont {Elitzur}(1975)}]{Elitzur:1975im}%
  \BibitemOpen
  \bibfield  {author} {\bibinfo {author} {\bibfnamefont {S.}~\bibnamefont
  {Elitzur}},\ }\href {\doibase 10.1103/PhysRevD.12.3978} {\bibfield  {journal}
  {\bibinfo  {journal} {Phys. Rev.}\ }\textbf {\bibinfo {volume} {D12}},\
  \bibinfo {pages} {3978} (\bibinfo {year} {1975})}\BibitemShut {NoStop}%
\bibitem [{\citenamefont {Ilgenfritz}\ \emph {et~al.}(2011)\citenamefont
  {Ilgenfritz}, \citenamefont {Menz}, \citenamefont {Muller-Preussker},
  \citenamefont {Schiller},\ and\ \citenamefont
  {Sternbeck}}]{Ilgenfritz:2010gu}%
  \BibitemOpen
  \bibfield  {author} {\bibinfo {author} {\bibfnamefont {E.-M.}\ \bibnamefont
  {Ilgenfritz}}, \bibinfo {author} {\bibfnamefont {C.}~\bibnamefont {Menz}},
  \bibinfo {author} {\bibfnamefont {M.}~\bibnamefont {Muller-Preussker}},
  \bibinfo {author} {\bibfnamefont {A.}~\bibnamefont {Schiller}}, \ and\
  \bibinfo {author} {\bibfnamefont {A.}~\bibnamefont {Sternbeck}},\ }\href
  {\doibase 10.1103/PhysRevD.83.054506} {\bibfield  {journal} {\bibinfo
  {journal} {Phys.Rev.}\ }\textbf {\bibinfo {volume} {D83}},\ \bibinfo {pages}
  {054506} (\bibinfo {year} {2011})},\ \Eprint {http://arxiv.org/abs/1010.5120}
  {arXiv:1010.5120 [hep-lat]} \BibitemShut {NoStop}%
\bibitem [{\citenamefont {Maas}(2013{\natexlab{b}})}]{Maas:2011se}%
  \BibitemOpen
  \bibfield  {author} {\bibinfo {author} {\bibfnamefont {A.}~\bibnamefont
  {Maas}},\ }\href@noop {} {\bibfield  {journal} {\bibinfo  {journal} {Phys.
  Rep.}\ }\textbf {\bibinfo {volume} {524}},\ \bibinfo {pages} {203} (\bibinfo
  {year} {2013}{\natexlab{b}})},\ \Eprint {http://arxiv.org/abs/1106.3942}
  {arXiv:1106.3942 [hep-ph]} \BibitemShut {NoStop}%
\bibitem [{\citenamefont {Cucchieri}\ and\ \citenamefont
  {Mendes}(1996)}]{Cucchieri:1995pn}%
  \BibitemOpen
  \bibfield  {author} {\bibinfo {author} {\bibfnamefont {A.}~\bibnamefont
  {Cucchieri}}\ and\ \bibinfo {author} {\bibfnamefont {T.}~\bibnamefont
  {Mendes}},\ }\href {\doibase 10.1016/0550-3213(96)00177-0} {\bibfield
  {journal} {\bibinfo  {journal} {Nucl. Phys.}\ }\textbf {\bibinfo {volume}
  {B471}},\ \bibinfo {pages} {263} (\bibinfo {year} {1996})},\ \Eprint
  {http://arxiv.org/abs/hep-lat/9511020} {arXiv:hep-lat/9511020} \BibitemShut
  {NoStop}%
\bibitem [{\citenamefont {Cabibbo}\ and\ \citenamefont
  {Marinari}(1982)}]{Cabibbo:1982zn}%
  \BibitemOpen
  \bibfield  {author} {\bibinfo {author} {\bibfnamefont {N.}~\bibnamefont
  {Cabibbo}}\ and\ \bibinfo {author} {\bibfnamefont {E.}~\bibnamefont
  {Marinari}},\ }\href {\doibase 10.1016/0370-2693(82)90696-7} {\bibfield
  {journal} {\bibinfo  {journal} {Phys. Lett.}\ }\textbf {\bibinfo {volume}
  {B119}},\ \bibinfo {pages} {387} (\bibinfo {year} {1982})}\BibitemShut
  {NoStop}%
\bibitem [{\citenamefont {Suman}\ and\ \citenamefont
  {Schilling}(1993)}]{Suman:1993mg}%
  \BibitemOpen
  \bibfield  {author} {\bibinfo {author} {\bibfnamefont {H.}~\bibnamefont
  {Suman}}\ and\ \bibinfo {author} {\bibfnamefont {K.}~\bibnamefont
  {Schilling}},\ }\href@noop {} {\  (\bibinfo {year} {1993})},\ \Eprint
  {http://arxiv.org/abs/hep-lat/9306018} {arXiv:hep-lat/9306018} \BibitemShut
  {NoStop}%
\bibitem [{\citenamefont {B\"ohm}\ \emph {et~al.}(2001)\citenamefont {B\"ohm},
  \citenamefont {Denner},\ and\ \citenamefont {Joos}}]{Bohm:2001yx}%
  \BibitemOpen
  \bibfield  {author} {\bibinfo {author} {\bibfnamefont {M.}~\bibnamefont
  {B\"ohm}}, \bibinfo {author} {\bibfnamefont {A.}~\bibnamefont {Denner}}, \
  and\ \bibinfo {author} {\bibfnamefont {H.}~\bibnamefont {Joos}},\ }\href@noop
  {} {\emph {\bibinfo {title} {{Gauge theories of the strong and electroweak
  interaction}}}}\ (\bibinfo  {publisher} {Teubner},\ \bibinfo {address}
  {Stuttgart},\ \bibinfo {year} {2001})\ p.\ \bibinfo {pages} {784},\ \bibinfo
  {note} {stuttgart, Germany: Teubner (2001) 784 p}\BibitemShut {NoStop}%
\bibitem [{\citenamefont {Maas}(2011{\natexlab{a}})}]{Maas:2010qw}%
  \BibitemOpen
  \bibfield  {author} {\bibinfo {author} {\bibfnamefont {A.}~\bibnamefont
  {Maas}},\ }\href@noop {} {\bibfield  {journal} {\bibinfo  {journal} {JHEP}\
  }\textbf {\bibinfo {volume} {02}},\ \bibinfo {pages} {076} (\bibinfo {year}
  {2011}{\natexlab{a}})},\ \Eprint {http://arxiv.org/abs/1012.4284}
  {arXiv:1012.4284 [hep-lat]} \BibitemShut {NoStop}%
\bibitem [{\citenamefont {von Smekal}\ \emph {et~al.}(1998)\citenamefont {von
  Smekal}, \citenamefont {Hauck},\ and\ \citenamefont
  {Alkofer}}]{vonSmekal:1997vx}%
  \BibitemOpen
  \bibfield  {author} {\bibinfo {author} {\bibfnamefont {L.}~\bibnamefont {von
  Smekal}}, \bibinfo {author} {\bibfnamefont {A.}~\bibnamefont {Hauck}}, \ and\
  \bibinfo {author} {\bibfnamefont {R.}~\bibnamefont {Alkofer}},\ }\href
  {\doibase 10.1006/aphy.1998.5806} {\bibfield  {journal} {\bibinfo  {journal}
  {Ann. Phys.}\ }\textbf {\bibinfo {volume} {267}},\ \bibinfo {pages} {1}
  (\bibinfo {year} {1998})},\ \Eprint {http://arxiv.org/abs/hep-ph/9707327}
  {arXiv:hep-ph/9707327} \BibitemShut {NoStop}%
\bibitem [{\citenamefont {von Smekal}\ \emph {et~al.}(2009)\citenamefont {von
  Smekal}, \citenamefont {Maltman},\ and\ \citenamefont
  {Sternbeck}}]{vonSmekal:2009ae}%
  \BibitemOpen
  \bibfield  {author} {\bibinfo {author} {\bibfnamefont {L.}~\bibnamefont {von
  Smekal}}, \bibinfo {author} {\bibfnamefont {K.}~\bibnamefont {Maltman}}, \
  and\ \bibinfo {author} {\bibfnamefont {A.}~\bibnamefont {Sternbeck}},\ }\href
  {\doibase 10.1016/j.physletb.2009.10.030} {\bibfield  {journal} {\bibinfo
  {journal} {Phys. Lett.}\ }\textbf {\bibinfo {volume} {B681}},\ \bibinfo
  {pages} {336} (\bibinfo {year} {2009})},\ \Eprint
  {http://arxiv.org/abs/0903.1696} {arXiv:0903.1696 [hep-ph]} \BibitemShut
  {NoStop}%
\bibitem [{\citenamefont {Maas}(2011{\natexlab{b}})}]{Maas:2010nc}%
  \BibitemOpen
  \bibfield  {author} {\bibinfo {author} {\bibfnamefont {A.}~\bibnamefont
  {Maas}},\ }\href {\doibase 10.1140/epjc/s10052-011-1548-y} {\bibfield
  {journal} {\bibinfo  {journal} {Eur. Phys. J.}\ }\textbf {\bibinfo {volume}
  {C71}},\ \bibinfo {pages} {1548} (\bibinfo {year} {2011}{\natexlab{b}})},\
  \Eprint {http://arxiv.org/abs/1007.0729} {arXiv:1007.0729 [hep-lat]}
  \BibitemShut {NoStop}%
\bibitem [{\citenamefont {Maas}(2016)}]{Maas:2016edk}%
  \BibitemOpen
  \bibfield  {author} {\bibinfo {author} {\bibfnamefont {A.}~\bibnamefont
  {Maas}},\ }\href {\doibase 10.1140/epjc/s10052-016-4216-4} {\bibfield
  {journal} {\bibinfo  {journal} {Eur. Phys. J.}\ }\textbf {\bibinfo {volume}
  {C76}},\ \bibinfo {pages} {366} (\bibinfo {year} {2016})},\ \Eprint
  {http://arxiv.org/abs/1603.07525} {arXiv:1603.07525 [hep-lat]} \BibitemShut
  {NoStop}%
\bibitem [{\citenamefont {Cucchieri}\ \emph {et~al.}(2005)\citenamefont
  {Cucchieri}, \citenamefont {Mendes},\ and\ \citenamefont
  {Taurines}}]{Cucchieri:2004mf}%
  \BibitemOpen
  \bibfield  {author} {\bibinfo {author} {\bibfnamefont {A.}~\bibnamefont
  {Cucchieri}}, \bibinfo {author} {\bibfnamefont {T.}~\bibnamefont {Mendes}}, \
  and\ \bibinfo {author} {\bibfnamefont {A.~R.}\ \bibnamefont {Taurines}},\
  }\href {\doibase 10.1103/PhysRevD.71.051902} {\bibfield  {journal} {\bibinfo
  {journal} {Phys. Rev.}\ }\textbf {\bibinfo {volume} {D71}},\ \bibinfo {pages}
  {051902} (\bibinfo {year} {2005})},\ \Eprint
  {http://arxiv.org/abs/hep-lat/0406020} {arXiv:hep-lat/0406020} \BibitemShut
  {NoStop}%
\bibitem [{\citenamefont {Fradkin}\ and\ \citenamefont
  {Shenker}(1979)}]{Fradkin:1978dv}%
  \BibitemOpen
  \bibfield  {author} {\bibinfo {author} {\bibfnamefont {E.~H.}\ \bibnamefont
  {Fradkin}}\ and\ \bibinfo {author} {\bibfnamefont {S.~H.}\ \bibnamefont
  {Shenker}},\ }\href {\doibase 10.1103/PhysRevD.19.3682} {\bibfield  {journal}
  {\bibinfo  {journal} {Phys. Rev.}\ }\textbf {\bibinfo {volume} {D19}},\
  \bibinfo {pages} {3682} (\bibinfo {year} {1979})}\BibitemShut {NoStop}%
\bibitem [{\citenamefont {Maas}(2012)}]{Maas:2012ct}%
  \BibitemOpen
  \bibfield  {author} {\bibinfo {author} {\bibfnamefont {A.}~\bibnamefont
  {Maas}},\ }\href@noop {} {\bibfield  {journal} {\bibinfo  {journal} {Mod.
  Phys. Lett.}\ }\textbf {\bibinfo {volume} {A27}},\ \bibinfo {pages} {1250222}
  (\bibinfo {year} {2012})}\BibitemShut {NoStop}%
\bibitem [{\citenamefont {Caudy}\ and\ \citenamefont
  {Greensite}(2008)}]{Caudy:2007sf}%
  \BibitemOpen
  \bibfield  {author} {\bibinfo {author} {\bibfnamefont {W.}~\bibnamefont
  {Caudy}}\ and\ \bibinfo {author} {\bibfnamefont {J.}~\bibnamefont
  {Greensite}},\ }\href {\doibase 10.1103/PhysRevD.78.025018} {\bibfield
  {journal} {\bibinfo  {journal} {Phys. Rev.}\ }\textbf {\bibinfo {volume}
  {D78}},\ \bibinfo {pages} {025018} (\bibinfo {year} {2008})},\ \Eprint
  {http://arxiv.org/abs/0712.0999} {arXiv:0712.0999 [hep-lat]} \BibitemShut
  {NoStop}%
\bibitem [{\citenamefont {Langacker}(1981)}]{Langacker:1980js}%
  \BibitemOpen
  \bibfield  {author} {\bibinfo {author} {\bibfnamefont {P.}~\bibnamefont
  {Langacker}},\ }\href {\doibase 10.1016/0370-1573(81)90059-4} {\bibfield
  {journal} {\bibinfo  {journal} {Phys. Rept.}\ }\textbf {\bibinfo {volume}
  {72}},\ \bibinfo {pages} {185} (\bibinfo {year} {1981})},\ \bibinfo {note}
  {gUT}\BibitemShut {NoStop}%
\bibitem [{\citenamefont {{Wolfram Research{,} Inc.}}()}]{Mathematica}%
  \BibitemOpen
  \bibfield  {author} {\bibinfo {author} {\bibnamefont {{Wolfram Research{,}
  Inc.}}},\ }\href@noop {} {\enquote {\bibinfo {title} {Mathematica, {V}ersion
  11.1},}\ }\bibinfo {note} {Champaign, IL, 2017}\BibitemShut {NoStop}%
\end{thebibliography}%

\end{document}